\documentclass[a4paper,11pt]{article}
\usepackage{jheppub}
\usepackage{graphicx}
\usepackage{epsfig}
\usepackage{rotating}
\usepackage{amssymb}
\usepackage{subfigure}
\usepackage{dsfont}
\usepackage{psfrag}
\usepackage{amsmath,euscript,array,mathrsfs}
\usepackage{slashed}
\usepackage{array}
\usepackage{youngtab}
\usepackage{color}
\usepackage{bbold}

\newcommand{\beq}{\begin{equation}}
\newcommand{\eeq}{\end{equation}}
\newcommand{\beqs}{\begin{eqnarray}}
\newcommand{\eeqs}{\end{eqnarray}}
\newcommand{\Tr}{{\rm Tr}}

\newcommand{\dd}{\mbox{d}}
\newcommand{\arctanh}{{\rm arctanh}}

\renewcommand{\arraystretch}{1.9}
\newcommand{\be}{\begin{equation}}
\newcommand{\ee}{\end{equation}}
\newcommand{\ba}{\begin{array}}
\newcommand{\ea}{\end{array}}
\newcommand{\eq}[1]{Eq.~(\ref{#1})}

\newcommand\figwidth{15.2cm}

\makeatletter
\gdef\@fpheader{}
\makeatother

\begin{document}

\title{Holographic models of composite Higgs in the Veneziano limit. Part II. Fermionic sector}

\author[a]{Daniel Elander,}
\author[a]{Michele Frigerio,}
\author[b]{Marc Knecht,}
\author[a]{Jean-Lo\"ic Kneur}

\affiliation[a]{Laboratoire Charles Coulomb (L2C), University of Montpellier, CNRS, Montpellier, France}
\affiliation[b]{Centre de Physique Th\'eorique, CNRS/Aix-Marseille Univ./Univ. de Toulon (UMR 7332), CNRS-Luminy Case 907, 13288 Marseille Cedex 9, France}

\date{\today}

\abstract{
We continue our study of strongly-coupled, approximately scale-invariant gauge theories with a large number of flavours, which provide a suitable ultraviolet completion of the composite-Higgs scenario. We identify the requisite operators to realise partial compositeness of the Standard-Model fermions. In order to compute the spectrum of composite fermionic states, we extend the bottom-up holographic models, which we previously introduced to capture the main features of the non-perturbative dynamics in the Veneziano limit, by adding fermion fields in the bulk. We identify regions in parameter space where some fermionic bound states become light, depending in particular on the number of flavours, the operator scaling dimensions, and the bulk Yukawa couplings. We also observe a dense spectrum of states, when multi-scale dynamics is induced by a large backreaction of bulk scalars on the geometry. Adapting the formalism of the holographic Wilsonian renormalisation group, we study the linear coupling between the composite and elementary fermions, as a function of energy scale. We find that, in some circumstances, the associated operators are dangerously irrelevant: the renormalisation-group flow gives rise to a large linear coupling in the infrared, even when it is irrelevant from the point of view of the ultraviolet fixed point. We finally compute the partially composite spectrum, correlate it with the analysis of the flow, and assess the potential phenomenological implications, e.g.~for the top-quark partners.
}

\maketitle


\section{Introduction}

We consider a hypercolour (HC) gauge theory which is asymptotically free in the ultraviolet (UV), approaches an infrared (IR) fixed-point, thus becoming approximately 
scale-invariant at some scale $\Lambda_{\rm UV}$, and eventually undergoes confinement and develops a mass gap $m_*$.  In the process, some HC flavour symmetries may undergo spontaneous symmetry breaking, with $f$ the decay constant of the associated Nambu-Goldstone bosons (NGBs). For generic strong dynamics, one expects
\be
m_* \sim g_* f \sim \frac{4\pi}{\sqrt{N_C}} f \equiv 4\pi \tilde f~,
\label{NDA}
\ee
where $g_*$ is the typical coupling between the bound states, $N_C$ the number of hypercolours, and we defined a reduced decay constant $\tilde f \equiv f/\sqrt{N_C}$.
We wish to consider theories with a number $N_F$ of HC fermions, and take the Veneziano limit of large $N_C$ and fixed $x_F\equiv N_F/N_C$ of order one, corresponding to a large flavour symmetry. 
There are at least two motivations for such choice: (i) we wish to
obtain a coset large enough to contain the Standard Model (SM)
gauge and global symmetries; (ii)
we want the theory to lie in the vicinity of the lower edge of the
conformal window, resulting in near-conformal dynamics, which requires
$x_F$ of order unity.

If the SM Higgs boson were identified with one of the composite NGBs, then the scale $f$ would characterise the deviations from the SM predictions for the electroweak precision parameters as well as for the Higgs couplings. In this scenario current data imply a lower bound $f\gtrsim 1$ TeV. While NGBs are massless in the exact-global-symmetry limit, the other composite resonances are expected to have masses of order $m_*$ or heavier, which lie in the multi-TeV range, possibly out of the reach of the Large Hadron Collider (LHC) even for $f\sim 1$ TeV and $N_C\sim 10$. 

We will employ dual holographic models aimed at capturing the salient features of the HC theory, in order to study the spectrum of resonances in the Veneziano limit. The main question of our investigation is whether some resonances could be parametrically lighter than $m_*$, and thus provide the first observable footprints of the new strong dynamics. While the spectrum of bosonic resonances was analysed in a companion paper \cite{Elander:2020nyd}, here we focus on the fermionic sector. We will see that, in the large-$N_C$ limit, the lightest fermionic resonances are expected to be meson-like objects, rather than baryon-like objects, so that the generic scaling of \eq{NDA} remains valid for their masses. However, the spectrum may include massless chiral (partially) composite fermions, as well as vector-like composite fermions, whose mass may become significantly smaller than $m_*$ in specific regions of the parameter space. 

The idea of partial compositeness, originally introduced in~\cite{Kaplan:1991dc}, posits that the SM fermions may mix with fermionic states of the strongly-coupled sector containing the composite Higgs. This mechanism may be used to explain the mass hierarchies of the SM fermions as being due to the mixing with operators of different scaling dimensions, leading in particular to a large Yukawa coupling for the top quark. While most work on this subject has been within the framework of effective field theory (for reviews, see~\cite{Contino:2010rs,Panico:2015jxa}), in recent years there has been an interest in building UV complete models that  contain the requisite fermionic states, with the right quantum numbers to couple to the SM fermions, by considering gauge theories with constituent fermions in one or more representations of HC~\cite{Barnard:2013zea,Ferretti:2013kya,Vecchi:2015fma,
Belyaev:2016ftv,Bizot:2016zyu,Gertov:2019yqo}, or supersymmetric gauge theories \cite{Caracciolo:2012je}. 
However, so far little is known about the spectrum of such fermionic states. On the lattice, the fermionic spectra of $SU(4)$ and $Sp(4)$ gauge theories have been studied in~\cite{latt_su4_chimera} and~\cite{Lucini:2021xke,Bennett:2021mbw,Bennett:2022yfa}, respectively, while consequences for partial compositeness have been further explored in~\cite{Ayyar:2018glg}.

Conversely, within the holographic approach to strongly coupled dynamics, the study of the fermionic sector of composite Higgs models goes back to the early papers on the subject~\cite{Contino:2003ve,Contino:2004vy,Agashe:2004rs,Agashe:2005dk,Contino:2006qr}, which analysed bottom-up models for which the background geometry in the bulk is a slice of anti-de Sitter (AdS). More elaborate models can be constructed by including scalar fields whose backreaction on the bulk geometry causes it to deviate from AdS in the deep IR~\cite{Karch:2006pv,Csaki:2006ji,Falkowski:2008fz,Cabrer:2009we,Cabrer:2011fb}, leading to a soft wall that ends the geometry, while dynamically inducing a mass gap, similar to what is the case in top-down confining models~\cite{Witten:1998zw,Klebanov:2000hb,Chamseddine:1997nm,Maldacena:2000yy,Sakai:2004cn}. The fermionic sector of such soft-wall models has been studied in the context of warped extra dimensions in~\cite{Batell:2008me,Delgado:2009xb,Gherghetta:2009qs,MertAybat:2009mk,Archer:2011bk,Ahmed:2019zxm}. Taking inspiration from top-down models such as~\cite{Sakai:2004cn}, in which extended objects (D-branes) that probe the background geometry are responsible for symmetry breaking, recently bottom-up models have been constructed and both their bosonic and fermionic spectra have been computed in~\cite{Erdmenger:2020flu,Erdmenger:2020lvq}. While composite Higgs models have also been studied within the fully top-down approach to holography~\cite{Elander:2021kxk}, so far the fermionic spectrum has not been computed, nor has partial compositeness been implemented.

The Veneziano limit, with $x_F \sim 1$, requires taking into account the backreaction of the flavour sector on the geometry. We focus here on the two Models~I and~II introduced in~\cite{Elander:2020nyd} within the bottom-up approach to holography. The first contains a single bulk scalar field that is responsible both for breaking the flavour-symmetry and for ending the space in the IR via its backreaction on the metric. The amount of backreaction depends on the number of flavours, such that in the limit $x_F \rightarrow 0$ the background geometry approaches AdS. Model~II additionally contains a second bulk scalar, that is a flavour singlet, and whose non-trivial radial profile governs the breaking of scale invariance. To both these models, we add a bulk fermion, in order to describe composite fermionic states in the dual field theory. Furthermore, in Model~II we allow this fermion to couple to the flavour singlet scalar via a Yukawa term in the bulk.

We compute the spectrum of fermionic resonances in Models I and II. As we shall see, both these models are rich enough to incorporate multi-scale dynamics in certain regions of the parameter space. Moreover, the effect of the Yukawa coupling may be thought of as the bulk fermion acquiring a radially dependent mass, due to the profile of the flavour-singlet scalar field, which leads to the scaling dimension of the dual fermionic operator effectively varying with energy scale. Both these features may result in richer scenarios compared to bottom-up models with a single characteristic scale $m_*$ as in \eq{NDA}, such as those based on a slice of AdS ~\cite{Contino:2003ve,Contino:2004vy,Agashe:2004rs,Agashe:2005dk,Contino:2006qr}.

Next, we couple the strong sector to an elementary fermion, which in the case of partial compositeness would be one of the SM quarks (or leptons). We investigate the running of the linear coupling between the elementary and strong sectors, by making use of the formalism of the holographic Wilsonian renormalisation group (RG)~\cite{Heemskerk:2010hk,Faulkner:2010jy}, originally proposed to study the RG flow of double-trace operators. We pay particular attention to situations in which our models may contain dangerously irrelevant operators, that is, deformations that are irrelevant from the point of view of the UV fixed point, but nevertheless have a large effect on the low energy dynamics due to a non-trivial RG flow.

Finally, we compute the partially composite spectrum, and correlate it with expectations from the analysis of the RG flow. We discuss how and when the spectrum differs from that of composite Higgs models based on the AdS background, in particular when a fermionic resonance can be parametrically light.

This paper is organised as follows. In section~IIA, we review the gauge theory, paying particular attention to the fermionic operators, while, in section~IIB, we review Models I and II of~\cite{Elander:2020nyd} and introduce the fermionic sector in the bulk. In section~III, we compute the spectrum of the strongly coupled sector in isolation, and briefly discuss how it compares to lattice simulations. We also present a toy model for which some of the salient features of the spectrum can be reproduced analytically. Section~IV first describes how to apply the formalism of holographic Wilsonian RG to the linear coupling of partial compositeness, and then makes use of it to study the RG flows in Models~I and~II. In section~V, we compute the fermionic spectrum in the  scenario of partial compositeness, which is the most relevant phenomenologically. Finally, in section~VI we conclude with a discussion of the various novel scenarios that our models make possible, as well as consequences for future model building. Appendix~A contains a detailed classification of the HC fermionic operators, appendix~B collects technical results on fermions in five dimensions, appendix~C recalls some features of the bosonic spectra for comparison, and appendix~D contains an explicit derivation of the partial-compositeness RG flow equation.

\section{Modelling compositeness via gauge-gravity duality}

\subsection{The hypercolour theory}\label{sec:HC}

The ultraviolet completion of the composite Higgs scenario is subject to a number of theoretical and phenomenological constraints.
A gauge theory of fermions can satisfy such constraints, for appropriate choices of symmetries and fields, as we discussed in detail
in section II of \cite{Elander:2020nyd}. Here we just recall the main features of the suitable gauge theory we will employ.

We consider a gauge group $G_{HC}=Sp(2N_C)$, where $N_C$ is the number of hypercolours. Such group has a unique invariant tensor $\Omega_{ij}=-\Omega_{ji}$ with indexes in the fundamental ($i,j=1,\dots,2N_C$).
Our choice of the matter content, motivated by \cite{Elander:2020nyd} and further justified below, amounts to $2N_F$ Weyl fermions in the fundamental, pseudoreal representation of $Sp(2N_C)$, 
and one Weyl fermion in a two-index, real representation of $Sp(2N_C)$, 
\be
\psi^a_i~,~~a=1,\dots,2N_F~,\qquad \chi_{ij} ~~{\rm or}~~ \chi'_{ij}~.
\label{HCfermions}\ee
There are two possible irreducible, two-index representations: traceless and antisymmetric, $\chi_{ij}\Omega_{ij}=0$ and $\chi_{ij}=-\chi_{ji}$,
or symmetric, $\chi'_{ij}=\chi'_{ji}$. In either case, the fermion global, anomaly-free symmetry is $G_F=SU(2N_F)\times U(1)$,
while its vector subgroup is $H_F=Sp(2N_F$). We assume that, when the HC sector confines and develops a mass gap $m_*$, also the flavour symmetry
is spontaneously broken, $G_F\to H_F$, with decay constant $f$ and the Higgs belonging to the associated set of Goldstone bosons. The coset $SU(2N_F)/Sp(2N_F)$
corresponds to a vacuum expectation value of $\langle\psi^a\psi^b\rangle$ proportional to an $Sp(2N_F)$ invariant and antisymmetric tensor, $\Sigma_{ab}=-\Sigma_{ba}$.

In the confined phase the observables correspond to 
composite, HC-invariant operators. The bosonic operators were analysed in \cite{Elander:2020nyd}.
Here we focus on fermionic operators, relevant for the computation of the spectrum of composite fermions.
In this section we limit ourselves to operators (i) with canonical dimension $\le 9/2$, likely to be the most relevant even in the strongly-coupled regime,
and (ii) containing spin-$1/2$ components only,
with no spin-$3/2$ or larger components.
We refer the Reader to appendix \ref{zoo} for a more general list of fermionic operators, and for the explicit contractions of HC and spinor indexes, that here we drop for simplicity.

The minimal fermionic operator is made by two constituents only,
\be
\hat F \equiv F^{\mu\nu}\chi'\sigma_{\mu\nu}~,
\label{Fchi'}\ee
where $F^{\mu\nu}$ is the HC field strength, which transforms in the adjoint representation.
For an $Sp(2N_C)$ group, the adjoint coincides with the two-index symmetric representation of $\chi'$, allowing to form a HC-invariant
combination. The operator $\hat F$ has canonical dimension $7/2$, which is the smallest possible value for a composite fermion, in theories with no scalar constituents.

The next-to-minimal fermionic operators are fermion trilinears, with canonical dimension $9/2$.
There are four independent operators involving two $\psi$ and one $\chi$ constituents, 
\be 
F_1^{ab} = \psi^{[a}\chi \psi^{b]}~,\quad
F_2^{ab} = \psi^{\{a}\chi \psi^{b\}}~,\quad
F_3^{ab} = \psi^{[a} \overline{\chi} \psi^{b]}~,\quad
F_{4 \, b}^a = \psi^a \chi \overline{\psi}_b~,\quad  
\label{psipsichi}\ee
where $[~]$ and $\{~\}$ indicate antisymmetrisation and symmetrisation in the flavour $SU(2N_F)$ indexes, respectively. Note that $F_3$ has no symmetric counterpart, see appendix \ref{zoo} for details. In the model with $\chi$ replaced by $\chi'$, that is symmetric in its HC indexes,
the only difference is a reversed flavour symmetry,
\be 
{F'_1}^{ab} = \psi^{\{a}\chi' \psi^{b\}}~,\quad
{F'_2}^{ab} = \psi^{[a}\chi' \psi^{b]}~,\quad
{F'_3}^{ab} = \psi^{\{a} \overline{\chi}' \psi^{b\}}~,\quad
{F'_4}^a_{\ b} = \psi^a \chi' \overline{\psi}_b~.\quad 
\label{psipsichi'}\ee
Concerning the flavour of these operators, one should notice that 
(i) the two-index symmetric representation of $SU(2N_F)$ and $Sp(2N_F)$ coincide; (ii) the two-index antisymmetric  of $SU(2N_F)$ contains an
$Sp(2N_F)$ singlet, obtained by tracing with the tensor $\Sigma_{ab}$, plus a traceless antisymmetric of $Sp(2N_F)$; (iii)
the operators $F_4$ and $F'_4$  reduce to an $SU(2N_F)$ singlet, obtained by tracing with $\delta^b_{\ a}$, 
plus an $SU(2N_F)$ adjoint, 
and the latter decomposes under $Sp(2N_F)$ into a symmetric plus a traceless antisymmetric.

There are also two independent operators involving three $\chi$ constituents,
\be 
F_\chi = \chi\chi\chi~,\qquad 
\tilde{F}_\chi = \chi \overline{\chi}\chi~.
\label{chichichi}\ee
The analogous operators obtained by replacing $\chi$ with $\chi'$ vanish identically, see appendix \ref{zoo} for the explicit index contractions. 

We will be interested in two-point correlators of fermionic operators, as their poles correspond to the masses of composite fermion resonances.
At leading order in the number of hypercolours, one can check diagrammatically that
\be
\langle F \overline{F} \rangle \sim N_C^2~,
\label{FFcorr}
\ee
when $F$ is any of the operators in Eqs.~(\ref{Fchi'})--(\ref{psipsichi'}), with fixed flavour indices. Note that the large-$N_C$ behaviour is due to the presence of the field $\chi$, which transforms under a two-index representation of the HC gauge group. This is the case we will analyse in the dual holographic theory. 
For the operators in \eq{chichichi}, the two-point correlator grows with $N_C^3$, a behaviour that could also be described holographically in an analogous fashion. Note that one may also consider mixed correlators, between two different fermionic operators, which may be non-vanishing when the operators belong to the same Lorentz and flavour representations.
The non-standard scalings with $N_C$ are essentially due to the two HC indexes of $\chi$ ($\chi'$), to be contrasted with QCD-like theories with only fermions in the (anti-)fundamental representation. 
Nonetheless, it is important to notice that the fermion operators $F\sim\psi\chi\psi$ share some important properties with meson operators $M\sim\psi\psi$: their mass gap does not scale with $N_C$ (contrary to e.g.~baryon masses in QCD-like theories), so that generically one expects $m_F\sim m_M\sim m_*$, and their trilinear coupling to mesons scales with $N_C^{-1/2}$, namely $g_{\bar FFM}\sim g_{MMM} \sim g_*$.

We will be especially interested in the scaling dimension of the operators $F$ when the HC theory lies close to an interacting fixed point. While such fixed point is required to be strongly coupled in order for anomalous dimensions to be large, perturbative estimates may still provide some guidance. The one-loop anomalous dimension of three-fermion operators, defined by $[F]=9/2+\gamma$, were computed in general in
\cite{BuarqueFranzosi:2019eee}. Particularising these results to the operators of \eq{psipsichi}, we find
\be
\gamma_{1,2,3,4}  = \dfrac{g_C^2N_C}{16\pi^2}\Big[(0 , \, -4 , \, 0 , \, -3)+  {\cal O}(1/N_C) \Big] ~.
\ee
The operators of \eq{psipsichi'} have the same anomalous dimensions, up to  different ${\cal O}(1/N_C)$ corrections. Note that $\gamma_{1,3}$ do not grow with $N_C$, however this 
may well be a one-loop accident: 
higher-loop contributions to $\gamma_i$ have not been evaluated to date, but the contribution to $\gamma_i$ at $n$ loops is generically expected to be of order $\lambda^n/(16\pi^2)^n$, where 
$\lambda \equiv g_C^2N_C$ is the 't Hooft coupling.
We also remark that ${\cal O}(1/N_C)$ corrections may be quantitatively relevant even for $N_C\sim 10$.
These anomalous dimensions should be evaluated at the fixed point, where the HC gauge coupling takes some value  $\bar g_{C}$.
The perturbative two-loop (renormalisation-scheme independent) estimate for the latter \cite{Banks:1981nn} is given in our model by \cite{Elander:2020nyd} 
\be
\frac{\bar\lambda}{16\pi^2}\equiv\frac{\bar g_C^2 N_C}{16\pi^2} = \frac{9-2x_F}{13x_F-18} + {\cal O}(1/N_C)~.
\label{FP}\ee
Let us consider e.g. a partial compositeness operator $\overline{F_2} f_{SM}$, which mixes $F_2$ with a SM fermion $f_{SM}$.  Such operator becomes marginal when $\bar\gamma_2 \simeq -2$, and this requires $x_F\simeq 2.1$. However marginality corresponds to a large 't Hooft coupling, $\bar\lambda/(16\pi^2) \simeq 1/2$, so that the above perturbative result should be taken at most as an 
order-of-magnitude estimate.
In particular, the position of the fixed point is very sensitive to higher orders---for example, by replacing  the two-loop HC beta-function with the four-loop beta-function in the $\overline{\rm MS}$ scheme, the fixed point is reached for significantly smaller values of $x_F$: the condition $\bar\gamma_2 \simeq -2$ is realised for $x_F\simeq 0.4$. 
Higher order beta and gamma functions are in general scheme dependent; perturbative scheme-independent
approaches to higher loops have been developed for some QCD-like anomalous dimensions 
(see e.g.~\cite{Ryttov:2017dhd,Gracey:2018oym}), but not for the $F$ operators relevant here.
On the other hand, as $x_F$ grows, the fixed point rapidly becomes weakly coupled and the perturbative estimate of \eq{FP} can be trusted, e.g. $\bar\lambda/(16\pi^2) \simeq 1/34$ for $x_F=4$. This regime appears to be far from the floor of the conformal window, it corresponds to small anomalous dimensions, and moreover it does not admit a simple holographic description in terms of gravity.

Having completed the survey of the most relevant fermionic operators, let us discuss their SM quantum numbers. The SM symmetries are embedded in the unbroken flavour group $H_F$, as they should be preserved down to scales well below $f$.
The minimal model for composite quark partners has $N_F=5$, with
\be\begin{array}{c}
H_F=Sp(10)\supset SU(3)_c \times SU(2)_L \times SU(2)_R \times U(1)_B~,\\
\psi^a \sim 10_{Sp(10)} = [(3,1,1)_{1/3}+(\bar 3,1,1)_{-1/3}+(1,2,1)_0+(1,1,2)_0]_{SU_{3221}}~.
\end{array}
\label{1010}\ee
Here $SU(3)_c$ is identified with ordinary colour, $SU(2)_L\times SU(2)_R$ with weak interactions including custodial symmetry, and $U(1)_B$ with baryon number, and there are two possible embeddings of hypercharge, $Y=\pm T_{3R}+B/2$.
The bosonic operator $(\psi^a\psi^b)$ contains a Higgs component, $h\sim(1,2,2)_0$.  Composite partners for the quark doublets, $Q_L\sim(3,2,1)_{1/3}$, as well as for the up- and down-quark singlets, ${Q_R}^c = ({T_R}^c,{B_R}^c) \sim (\bar 3,1,2)_{-1/3}$,
can be found in any of the fermionic operators of \eq{psipsichi} or \eq{psipsichi'}. Indeed,  such components are contained in both the two-index symmetric and antisymmetric representations of $Sp(10)$, see \eq{4455}. There is also an alternative embedding 
for the down-quark singlet partner, ${B_R}^c\sim (\bar 3,1,1)_{2/3}$. Such component is present only in the $Sp(10)$ two-index antisymmetric operators. 

Therefore, the model allows to implement partial compositeness for any SM quark, by coupling it with one or more composite operators with the corresponding quantum numbers.
One expects that the composite-elementary mixing is the largest for the most relevant operator, thus it is sufficient to consider a single operator for each SM quark.
This allows, in turn, to induce the quark Yukawa couplings to the composite Higgs. 
In fact a single $Sp(10)$ two-index fermion is sufficient, in principle, to embed a whole quark family.
However, the Yukawa hierarchies, such as the small ratio $y_b/y_t$, are more easily explained by introducing couplings to operators with different scaling dimensions for the different quarks.
Note that the fermion operators in Eqs.~(\ref{psipsichi}) and (\ref{psipsichi'}) contain, beside SM quark partners, several other components which do not mix with the SM quarks,
and may also have an interesting phenomenology, in particular if they result in states significantly lighter than the HC mass gap, $m_*\sim 4\pi\tilde f$. 

The above discussion focused on models for a composite NGB Higgs boson, accompanied by composite partners for the SM quarks. However, composite fermions may have a number of different phenomenological applications, and we will see that the holographic theory, described in the following, is largely independent from the specific embedding of the SM symmetries within $G_F$, and from the specific couplings of the 
various SM fields to the HC sector. 
One example is provided by composite twin Higgs models \cite{Chacko:2005pe,Barbieri:2015lqa,Low:2015nqa}, engineered to protect the Higgs mass in the absence of new coloured states:
in this case the role of top-quark partners is played by composite fermions with only electroweak charges, whose mass is significantly less constrained. Such scenarios correspond to different cosets $G_F/H_F$, but typically they also require a large $N_F$.
Another possibility is that the lightest composite fermion may be stable, for example due to a non-trivial baryon number: it may then provide an interesting candidate for dark matter, carrying electroweak charges or even being a SM singlet \cite{Agashe:2004bm,Agashe:2010gt,Frigerio:2011zg}.
Such dark matter candidates may already be present in models for Higgs and top partners, or they may motivate a HC sector on their own.
Finally, note that a composite fermion may be arbitrarily light when it is a SM singlet, i.e.~a sterile neutrino,
with various, associated phenomenological implications \cite{Arkani-Hamed:1998wff,Robinson:2014bma,Chacko:2020zze}.
Our holographic analysis of the spectrum of composite fermions 
may be of interest in all these contexts.

\subsection{The holographic theory}
\label{sec:models}

\subsubsection{Bosonic sector}
\label{sec:bosonicsector}

Three models of composite Higgs, referred to as Models~I,~IIA ,and~IIB, were introduced in~\cite{Elander:2020nyd}, which focused on their bosonic sectors. In this paper, we restrict our attention to Model~I and~IIB, the latter of which we henceforth simply refer to as Model~II. The main results that we will use in our analysis of the fermionic sector concern the form of the background solutions, as well as the calculation of the Goldstone decay constant, which we will use to normalise the fermionic spectra. Both models consist of gravity, an $SU(2N_F)$ gauge field $\mathcal A_M$,\footnote{For the reasons discussed in \cite{Elander:2020nyd} (see the two paragraphs around Eq.~(4.3) therein), we do not include the $U(1)_\psi$ and $U(1)_\chi$ sectors in our analysis.} as well as a scalar sector. The scalar sector contains the antisymmetric complex scalar $\Phi_{ab}$, dual to $\psi^a \psi^b$ and responsible for flavour-symmetry breaking. In addition, Model~II also contains a flavour singlet scalar field $\phi$, responsible for breaking scale invariance. The action can be written on the form
\beq
	\mathcal S = \mathcal S_{grav} + \mathcal S_{gauge} + \mathcal S_{scalar}  \,,
\eeq
where the gravity and gauge field parts, respectively $\mathcal S_{grav}$ and $\mathcal S_{gauge}$, are the same for all the models:\footnote{We correct a typographical error, relative to~\cite{Elander:2020nyd}, which appeared in the numerical factor in front of the kinetic term for the gauge field.}
\beq
	\mathcal S_{grav} = N_C^2 \int \dd^5x \sqrt{-g} \, \frac{R}{4} \,, \ \ \
	\mathcal S_{gauge} = - N_C \int \dd^5x \sqrt{-g} \, \Tr \left[ \frac{1}{2} g^{MP} g^{NQ} \mathcal F_{MN} \mathcal F_{PQ} \right] \,,
\eeq
with $R$ the five-dimensional Ricci scalar, and $\mathcal F_{MN}$ the field strength associated with $\mathcal A_M$. The scalar part $\mathcal S_{scalar}$ of the action is model dependent, and for Model I it is given by
\beq
\label{eq:ActionModel1}
	\mathcal S_{scalar}^{(I)} = - \int \dd^5x \sqrt{-g} \, \bigg\{ N_C \Tr \left[ g^{MN} (D_M \Phi)^\dag D_N \Phi \right] + N_C^2 \mathcal V(\Phi) \bigg\} \,,
\eeq
while for Model~II, it takes the form
\begin{align}
\label{eq:ActionModel2}
	\mathcal S_{scalar}^{(II)} = - \int \dd^5x \sqrt{-g} \, \bigg\{ &N_C \Tr \left[ g^{MN} (D_M \Phi)^\dag D_N \Phi \right] \nonumber \\
	&+ N_C^2 \frac{1}{2} g^{MN} \partial_M \phi \partial_N \phi + N_C^2 \mathcal V(\Phi, \phi) \bigg\} \,.
\end{align}
In both models, the potential $\mathcal V$ is chosen such that it allows for solutions that in the UV approach that of AdS with unit scale. The background solutions for the metric and the anti-symmetric complex scalar are taken to be on the form
\beq
\label{eq:DWmetric}
	\dd s^2 = \dd r^2 + e^{2A(r)} \dd x_{1,3}^2 \,, \ \ \ \Phi(r) = \frac{\sigma(r)}{2} \Sigma \,,
\eeq
where the warp factor is given by $A(r)$, while the radial profile of $\Phi(r)$ is parametrised by the function $\sigma(r)$ and is proportional to the real, antisymmetric matrix $\Sigma$ introduced after Eq.~(\ref{HCfermions}). All background functions are assumed to only depend on the radial coordinate $r$, including the background profile of the scalar $\phi(r)$, present in Model II.

Focusing our attention on Model I, the scalar potential $\mathcal V$ is given in terms of a superpotential $\mathcal W$ as
\beqs
	\mathcal V(\Phi) &=& \frac{1}{2x_F} \mathcal W'(\mathcal I)^2 - \frac{4}{3} \mathcal W(\mathcal I)^2 \,, \\
\label{eq:Model1W}
	\mathcal W(\mathcal I) &=& - \frac{3}{2} \left[ 1 + x_F \sinh^2\left( \sqrt{\frac{\Delta}{3}} \, \mathcal I \right) \right] \,.
\eeqs
Here, $\mathcal I$ is an $SU(2N_F)$ invariant built from $\Phi$, the precise form of which does not presently concern us; we only mention that it evaluates to $\mathcal I = \sigma(r)$ on the background solutions (for details, see~\cite{Elander:2020nyd}). The parameter $\Delta$ is related to the scaling dimension of the operator dual to $\Phi$, while $x_F = N_F/N_C$ is the ratio between the number of flavours and colours. With this choice of potential, we find the background solution
\beqs
\label{eq:Model1solutions}
	\sigma(r) = \sqrt{\frac{3}{\Delta}} \arctanh \left(e^{-\Delta r} \right) \,, \ \ \ A(r) = r + \frac{x_F}{2\Delta} \log \left(1 - e^{-2\Delta r} \right) \,.
\eeqs
The end of space, conventionally chosen to be at $r = 0$, is generated dynamically by the backreaction of the scalar $\Phi$ on the metric. As can be seen from the form of the warp factor $A(r)$, this backreaction is large when the number of flavours $x_F$ is large, while in the limit $x_F \rightarrow 0$ the background approaches that of an AdS geometry. The scaling dimension of the operator $\mathcal O_\sigma = \Tr (\Sigma_{ab} \psi^a \psi^b)$ dual to $\sigma$ is given by $[\mathcal O_\sigma] = 2+ |2 - \Delta|$. Furthermore, for $\Delta \geq 2$, the radial profile of $\sigma(r)$ has the dual interpretation in terms of $\mathcal O_\sigma$ developing a VEV and breaking the flavour symmetry (purely) spontaneously.

Coming to Model II, the scalar potential $\mathcal V$ is again given in terms of a superpotential $\mathcal W$ as
\beq
	\mathcal V(\Phi,\phi) = \frac{1}{2x_F} \left( \frac{\partial \mathcal W(\mathcal I,\phi)}{\partial \mathcal I} \right)^2 + \frac{1}{2} \left( \frac{\partial \mathcal W(\mathcal I,\phi)}{\partial \phi} \right)^2 - \frac{4}{3} \mathcal W(\mathcal I,\phi)^2 \,,
\eeq
where
\beq
	\mathcal W(\mathcal I,\phi) =  - \frac{3}{2} \left[ 1 + x_F \sinh^2\left( \sqrt{\frac{\Delta}{3}} \mathcal I \right) + \sinh^2\left( \sqrt{\frac{\Delta_\phi}{3}} \phi \right) \right]  \,,
\eeq
leading to the background solution
\begin{align}
\label{eq:Model2Bsolutions}
	\phi(r) &= \sqrt{\frac{3}{\Delta_\phi}} \arctanh \left( \phi_c \sqrt{\frac{\Delta_\phi}{3}} \, e^{-\Delta_\phi r} \right) \,, \nonumber \\
	A(r) &= r + \frac{x_F}{2\Delta} \log \left( 1 - e^{-2\Delta r} \right) + \frac{1}{2\Delta_\phi} \log \left( 1 - \phi_c^2 \frac{\Delta_\phi}{3} e^{-2\Delta_\phi r} \right) \,,
\end{align}
with $\sigma(r)$ the same as in Eq.~\eqref{eq:Model1solutions} for Model I. The parameter $\Delta_\phi$ governs the scaling dimension of the operator $\mathcal O_\phi$ dual to $\phi$, such that $[\mathcal O_\phi] = 2 + |2 - \Delta_\phi|$. For $0 < \Delta_\phi < 2$, the profile of $\phi(r)$ has the interpretation as introducing the relevant deformation $\mathcal O_\phi$ in the dual field theory, thus explicitly breaking conformal invariance. The integration constant $\phi_c$ governs the size of this deformation.\footnote{Comparing to the notation of~\cite{Elander:2020nyd}, we have that $\phi_c = \phi_B \sqrt{\frac{3}{\Delta_\phi}}$.} Expanding $\phi$ asymptotically in the UV, one obtains
\beq
	\phi = \phi_c \, e^{-\Delta_\phi r} + \cdots \,,
\eeq
such that, for $\Delta_\phi < 2$, $\phi_c$ is equal to the source for $\mathcal O_\phi$. In order for the flavour-symmetry breaking scale to become dynamically related to the mass gap, we require that it is the profile of $\sigma(r)$, rather than $\phi(r)$, that is responsible for the end of the geometry in the IR, which implies that $\phi_c < \sqrt{\frac{3}{\Delta_\phi}}$. In particular, it follows that for a nearly marginal deformation $\Delta_\phi \simeq 0$, the maximum allowed value of the source $\phi_c$ is large.

In addition to knowing the background solutions themselves, we will also need to compute the Goldstone decay constant $f$. Decomposing the gauge field as $\mathcal A_M = A_{\hat A M} T^{\hat A} + V_{AM} T^A$, where $T^{\hat A}$ ($T^A$) are the broken (unbroken) generators of $SU(2N_F)$, the axial-vector $A_M(q,r)$ satisfies the equation of motion
\beq
\label{eq:eomA1}
	\Big[ \partial_r^2 + 2A' \partial_r - \left( q^2 e^{-2A} + g_5^2 \sigma^2 \right) \Big] P^{\mu\nu} A_\nu(q,r) = 0 \,,
\eeq
where $g_5$ is the bulk gauge coupling, and $q^\mu$ is the four-momentum. After writing $P^{\mu\nu} A_\nu(q,r) = \tilde A^\mu(q) a(q,r)$ and imposing the IR boundary condition $\partial_r a |_{r_1} = 0$, we obtain the decay constant as
\beq
\label{eq:decayconstant}
	f^2 = \lim\limits_{r \rightarrow \infty} \bigg\{ 2N_C \frac{e^{2A}}{g_5^2} \frac{\partial_r a}{a} \Big|_{q^2 = 0} \bigg\} \,.
\eeq
Taking out the overall factor of $N_C$, we define $\tilde f \equiv N_C^{-1/2} f$.

As usual, the holographic radial direction is related to energy scale on the field theory side. One may estimate the relation between the two by defining the scale $\Lambda(r)$ through~\cite{Csaki:2000cx}
\beq
\label{eq:Lambda}
	\Lambda(r)^{-1} = \int_r^{\infty} \dd \tilde r \, e^{-A(\tilde r)} \,.
\eeq
This expression is motivated by the fact that $\Lambda(r)^{-1}$ is the time it takes for a massless particle to reach from the boundary to the value of the radial coordinate $r$. We further define the characteristic IR scale as
\beq
	\Lambda_{\rm IR} \equiv \lim\limits_{r \rightarrow 0} \Lambda(r) \,.
\label{laIR}\eeq
Since in the IR the integrand appearing in Eq.~\eqref{eq:Lambda} scales as
\beq
	e^{-A} \sim \tilde r^{-\frac{x_F}{2\Delta}} + \cdots \,,
\eeq
one obtains that $\Lambda_{\rm IR} = 0$ when $x_F \geq 2\Delta$. In proximity to the limiting case, when $x_F \simeq 2\Delta$, one obtains a parametrically small $\Lambda_{\rm IR}$. In Model~II, there is an additional way for this to happen. Close to the upper bound $\phi_c < \sqrt{\frac{3}{\Delta_\phi}}$, in a range of the radial coordinate, one has the following scaling of the integrand appearing in Eq.~\eqref{eq:Lambda}:
\beq
\label{eq:scalinglargephic}
	e^{-A} \sim \tilde r^{-\frac{x_F}{2\Delta} -\frac{1}{2\Delta_\phi}} \,, \hspace{1cm} \frac{3 - \phi_c^2 \Delta_\phi}{6 \Delta_\phi} \lesssim \tilde r \lesssim \rm min \left( \frac{1}{2\Delta}, \frac{1}{2\Delta_\phi} \right) \,.
\eeq
When $\Delta_\phi \leq \frac{\Delta}{2\Delta - x_F}$, this may lead to a suppressed $\Lambda_{\rm IR}$, provided $\phi_c$ is sufficiently close to saturating its upper bound. We will see later that the multi-scale dynamics that results from a small $\Lambda_{\rm IR}$ has interesting effects  both on the spectrum (sections~\ref{sec:fermionicspectrum} and \ref{sec:partiallycompositespectrum}), as well as on the RG flow arising in the context of partial compositeness (section~\ref{sec:partialcompositeness}).

The relation between the two energy scales $\Lambda_{\rm IR}$ and ${\tilde f}$ is non-trivial, and in general needs to be determined numerically. When presenting our results for the fermionic spectrum in sections~\ref{sec:fermionicspectrum} and~\ref{sec:partiallycompositespectrum}, we will mostly normalise the masses in units of ${\tilde f}$. The choice of ${\tilde f}$ is suitable when discussing phenomenology, since the (non-)observation of experimental deviations from the SM predictions directly provides information about the ratio $v/f$, with $v$ the electroweak scale, and thus about ${\tilde f}$ given a choice of $N_C$. However, in some instances, we will find it more convenient to present our results in units of $\Lambda_{\rm IR}$. This includes the case when $\Delta < 2$, such that the flavour symmetry is explicitly broken, and there are no exact NGBs. Moreover, in some regions of parameter space, associated with multi-scale dynamics, there is a large separation between $\Lambda_{\rm IR}$ and ${\tilde f}$. The mass of the lightest fermionic composite state may be governed by the former, making it the more natural unit, in particular when discussing light states in the partially composite spectrum of section~\ref{sec:partiallycompositespectrum}.

\subsubsection{Fermionic sector}
\label{sec:fermionicsector}

In order to model fermionic states in the field theory, we supplement the bosonic part of the action by introducing a Dirac fermion $\Psi$ in the bulk. We have in mind that the field theory operator dual to $\Psi$ is one of those listed in Eq.~\eqref{psipsichi} or \eq{psipsichi'}, that may describe potential top partners. This leads to $\Psi$ being in the symmetric, anti-symmetric or adjoint representation of $SU(2N_F)$. In all of these cases, in order to reproduce the $N_C$ scaling of \eq{FFcorr}, the overall normalisation of the action should be given by $\mathcal N_\Psi = N_C^2$. Considering simultaneously several fermionic fields in the bulk is of course possible, but at the expense of making the model more complicated. Also, as discussed in section~\ref{sec:HC}, considering a single field $\Psi$ is not a severe restriction: on the one hand, a two-index representation of $SU(2N_F)$ is already large enough to provide partners for both the top and bottom quarks and, on the other hand, we expect one of the fermionic composite operators to be more relevant than the others and dominate the mixing with the elementary fermions at low energies. If one is interested in flavour-singlet fermionic operators (for example, in the $\chi'$ model, $\hat F$ or ${F'_4}^a_a$), the dual fermion $\Psi$ should be taken to be a singlet under $SU(2N_F)$, a possibility that we will mention where relevant. With these considerations, we choose the fermionic action $\mathcal S_\Psi$ to be given by
\beqs
\label{eq:SPsi}
	\mathcal N_\Psi^{-1} \mathcal S_\Psi &=& - \int \dd^5 x \sqrt{-g} \left[ \frac{1}{2} \left( \overline \Psi \Gamma^M D_M \Psi - \overline{ D_M \Psi} \Gamma^M \Psi \right) + H_\Psi \overline \Psi \Psi \right] \nonumber \\
	&&- \sum_{i=1,2} \frac{s_i}{2} \int \dd^4 x \sqrt{-\tilde g} \, \overline \Psi \Psi \Big|_{r_i} \,,
\eeqs
where summation over flavour indices are implied, and $H_\Psi$ is a function of the scalar fields of the model, the precise form of which is model dependent. We have introduced two boundaries, $r_1$ in the IR and $r_2$ in the UV, which will later serve as regulators in our computation of the spectrum, and $\tilde g$ is the determinant of the metric induced on these boundaries. We will discuss the possible choices of the signs $s_i = \pm 1$ of the boundary actions in a moment. The covariant derivative is defined in terms of the spin connection $\omega_M$ as
\beq
	D_M = \mathcal D_M + \omega_M \,, \hspace{1cm}
	\omega_M = \frac{1}{8} \omega_{M A B} [\gamma^A,\gamma^B] \,,
\eeq
where $\mathcal D_M$ is the covariant derivative corresponding to the representation of the flavour symmetry under which $\Psi$ transforms. We have that $\overline \Psi = \Psi^\dag i \gamma^0$, and that the flat space-time gamma matrices $\gamma^A$ satisfy the identities $\gamma^{0\dag} = -\gamma^0$,  $\gamma^{i\dag} =\gamma^i$,  $\gamma^{5\dag} = \gamma^5$. The curved space-time gamma matrices $\Gamma^M$ are given by
\beq
	\Gamma^M = e^M{}_A \gamma^A \,, \hspace{1cm} \{\Gamma^M,\Gamma^N\} = 2g^{MN} \,,
\eeq
where we used the vielbeins $e^M{}_A$, in terms of which the metric can be written as $g^{MN} = e^M{}_A  e^N{}_B \eta^{AB}$. The components of the spin connection are equal to
\beq
	\omega_{MAB} = e_{NA} \nabla_M e^N{}_B = e_{NA} \left( \delta^N{}_P \partial_M + \Gamma^N{}_{MP} \right) e^P{}_B  \,,
\eeq
where $\Gamma^N{}_{MP} = \frac{1}{2} g^{NQ} \left( \partial_{M} g_{PQ} + \partial_{P} g_{QM} - \partial_{Q} g_{MP} \right)$ is the Christophel symbol associated with the bulk metric $g_{MN}$. Moreover, we decompose $\Psi$ into its left- and right-handed components according to
\beq
	P_{L,R} = \frac{1 \mp \gamma^5}{2} \,, \hspace{1cm} \Psi = \Psi_L + \Psi_R \,, \hspace{1cm} \Psi_{L,R} = P_{L,R} \Psi \,,
\eeq
and use the notation
\beq
	\overline \Psi_{L,R} \equiv \overline{\Psi_{L,R}} = \left( P_{L,R} \Psi \right)^\dagger i \gamma^0 = \overline \Psi P_{R,L} \,.
\eeq
Finally, we note that the action Eq.~\eqref{eq:SPsi} is invariant under the transformations
\beqs
\label{eq:SPsiInvariance}
	\Psi_L &\rightarrow& \Psi_R \,, \ \ \ \nonumber
	\Psi_R \rightarrow - \Psi_L \,, \\
	H_\Psi &\rightarrow& - H_\Psi \,, \ \ \
	s_i \rightarrow - s_i \,.
\eeqs

Having established our notation, let us now discuss the specific models we will study in the following, in particular our choice of the function $H_\Psi$. We decompose $H_\Psi$ into a constant piece, given by the fermion mass $M_\Psi$, and $h_\Psi$ which is a function of the scalar fields of the model:
\beq
	H_\Psi = M_\Psi + h_\Psi \,,
\eeq
where it is assumed that $h_\Psi$ is chosen such that it vanishes asymptotically in the UV when evaluated on the background solutions. We restrict ourselves to choices of $H_\Psi$ giving rise to a mass term and/or a Yukawa interaction. It turns out that writing an $SU(2N_F)$ invariant Yukawa interaction involving the scalar $\Phi$ (in the antisymmetric representation) is forbidden. In Model I, we therefore make the choice $h_\Psi = 0$, such that $H_\Psi = M_\Psi$. In Model II, on the other hand, it is possible to write a Yukawa interaction involving the flavour-singlet scalar $\phi$, and we hence choose $h_\Psi(\phi) = y_5 \phi$, where $y_5$ is the bulk Yukawa coupling.

In our applications, it is sufficient to consider the fermion as propagating on top of a given background solution, which we will choose as in Eq.~\eqref{eq:Model1solutions} or Eq.~\eqref{eq:Model2Bsolutions}, corresponding to Models~I and~II, respectively. Making use of the domain-wall form of the metric in Eq.~\eqref{eq:DWmetric}, we choose the vielbein to be
\beq
\label{eq:vielbein}
	e^M{}_A = {\rm diag}(e^{-A(r)},e^{-A(r)},e^{-A(r)},e^{-A(r)}, 1) \,,
\eeq
after which the spin connection takes the simple form
\beq
	\omega_M = \left( \frac{1}{2} \partial_r A e^A \gamma_\mu \gamma^5 , 0 \right) \,.
\eeq
Using this, the Dirac equation, projected on its left- and right-handed components, becomes
\beqs
	(\partial_r + 2 \partial_r A + H_\Psi ) \Psi_R + e^{-A} \gamma^\mu \partial_\mu \Psi_L &=& 0 \,, \nonumber\\
	- (\partial_r + 2 \partial_r A - H_\Psi ) \Psi_L + e^{-A} \gamma^\mu \partial_\mu \Psi_R &=& 0 \,. \label{eq:DiracEq}
\eeqs
We will also write $h_\Psi(r)$, with the understanding that $h_\Psi$ has been evaluated on the background solution.

In gauge-gravity duality with fermions in the bulk, it is either the boundary value of $\Psi_L$ or $\Psi_R$ that becomes the source of the dual operator in the field theory. Because of the invariance of the bulk action under the transformations of Eq.~\eqref{eq:SPsiInvariance}, it is sufficient to consider the case of $\Psi_L$ being the source for the field theory operator $\mathcal O_R$, without loss of generality. By requiring that the variation of the action $\mathcal S_\Psi$ vanishes (for details, see appendix~\ref{sec:variationalproblem}), it can be shown that this corresponds to the choice $s_2 = -1$, and we hence restrict ourselves to this case in the following. Similar reasoning relates the choice of $s_1$ to the IR boundary condition imposed on $\Psi$, such that $s_1 = -1$ implies $\Psi_L |_{r_1} = 0$ while $s_1 = 1$ implies that $\Psi_R |_{r_1} = 0$. We will refer to these boundary conditions as $(-)$ and $(+)$, respectively. The former (latter) implies that the corresponding sector of the dual field theory is chiral (vector-like), that is, it does (not) include a massless chiral fermion. In both cases, the scaling dimension of $\mathcal O_R$ at the UV fixed point is given by $\Delta_R \equiv [\mathcal O_R] = 2 + M_\Psi$. Note that, although our prototype HC theory presented in section~\ref{sec:HC} is a vector-like gauge theory, the associated spectrum of confined states may or may not contain some massless chiral fermions. Besides, we leave open the possibility that our holographic analysis is applicable to different classes of HC theories. Thus, in the following we will consider both cases $(-)$ and $(+)$ on equal footing.

\section{Spectrum of composite fermions}
\label{sec:fermionicspectrum}

In this section, we show the numerical results for the spectrum of fermionic resonances, obtained by studying a bulk Dirac fermion $\Psi$ propagating on a background geometry which we will take to be either that of Model I or Model II given in section~\ref{sec:models}. The results we present apply to the strongly coupled sector in isolation; in section~\ref{sec:partiallycompositespectrum}, we will consider the effect on the spectrum in the case of coupling the strong sector to an additional elementary fermion. We also study a toy model that can be solved analytically, and which captures some of the essential features of the numerical results for Model~II. The details regarding the formalism can be found in appendix~\ref{sec:fermions}, the main results of which we summarise here.

The spectrum of fermionic resonances can be extracted from the poles of the two-point function $\langle \mathcal O_R(q) \overline{\mathcal O}_R(-q) \rangle$, which is computed following the procedure of holographic renormalisation~\cite{Bianchi:2001kw,Bianchi:2001de,Skenderis:2002wp} in the following way. First, we note that for the purposes of computing two-point functions, it is sufficient to retain only terms up to quadratic order in the fluctuations around a given background solution. In particular, one does not need to consider the fermion gauge coupling to vector bosons, nor the Yukawa coupling to scalar fluctuations. Second, one evaluates the action $\mathcal S_\Psi$ given in Eq.~\eqref{eq:SPsi} on shell, obtaining the regularised on-shell action $\mathcal S_{\Psi,{\rm reg}}$. In order to cancel divergences that may arise in the $r_2 \rightarrow \infty$ limit, one then defines $\mathcal S_{\Psi, \rm sub} \equiv \mathcal S_{\Psi, \rm reg} + \mathcal S_{\Psi, \rm ct}$, where the counter-terms take the form
\beq
	\mathcal S_{\Psi, \rm ct} = - \mathcal N_\Psi \int \dd^4 x \sqrt{-\tilde g} \, \overline\Psi_L \Big( \mathcal F(-\tilde g^{\mu\nu} \partial_\mu \partial_\nu) i \Gamma^\sigma \partial_\sigma \Big) \Psi_L \Big|_{r_2} \,,
\eeq
and locality dictates that $\mathcal F$ be polynomial. After taking into account an overall normalisation\footnote{\label{eq:defpsiR}
For future reference, we also define $\Psi_R = N_R(r) \psi_R$, where $N_R(r) \equiv e^{ -2A(r) - M_\Psi r + \int_r^\infty \dd \tilde r \, h_\Psi(\tilde r)}$.
}
\beq
\label{eq:defpsiL}
	\Psi_L = N_L(r) \psi_L \,, \ \ \
	N_L(r) \equiv \exp \left( -2A(r) + M_\Psi r - \int_r^\infty \dd \tilde r \, h_\Psi(\tilde r) \right) \,,
\eeq
the two-point function is obtained by differentiating twice, and taking the limit $r_2 \rightarrow \infty$, according to
\beq
\label{eq:OOR}
	\langle \mathcal O_R(q) \overline{\mathcal O}_R(-q) \rangle = \lim\limits_{r_2 \rightarrow \infty} \bigg\{ \frac{i \, \delta^2 \mathcal S_{\Psi, \rm sub}}{\delta \overline\psi_L(-q,r_2) \delta \psi_L(q,r_2)} \bigg\} \,,
\eeq
where the Fourier transform is defined according to the convention $\Psi(x) = \int \frac{\dd^4q}{(2\pi)^2} e^{i q_\mu x^\mu} \Psi(q)$ and we use the notation $\overline \Psi(q,r) \equiv \Psi^\dagger(-q,r) i \gamma^0$. After writing $\psi_L(q,r) = b(q,r) \tilde \psi_L(q)$, this results in (for details, see appendix~\ref{sec:twopointfunctions})
\beq
\label{eq:OORexplicit}
	\langle \mathcal O_R(q) \overline{\mathcal O}_R(-q) \rangle = \lim\limits_{r_2 \rightarrow \infty} \bigg\{ - \mathcal N_\Psi N_L^2 \, e^{5A}\frac{1}{\slashed{q}} \left( e^{-2A} q^2 \mathcal F(e^{-2A} q^2) + \frac{\partial_r b}{b} \right) \Big|_{r_2} \bigg\} \,.
\eeq
Here, the scalar $b$ satisfies the equation of motion, 
\beq
\label{eq:eomb}
	\Big[ \partial_r^2 + (\partial_r A + 2 H_\Psi) \partial_r - q^2 e^{-2A} \Big] b = 0 \,,
\eeq
derived from the Dirac equation~\eqref{eq:DiracEq},
with IR boundary condition given by either $b|_{r_1} = 0$ $(-)$ or $\partial_r b|_{r_1} = 0$ $(+)$. We remind the Reader that $H_\Psi = M_\Psi$ for Model I, while $H_\Psi = M_\Psi + y_5 \phi$ for Model II.

The first comment that we make about the expression for $\langle \mathcal O_R(q) \overline{\mathcal O}_R(-q) \rangle$ given in Eq.~\eqref{eq:OORexplicit} is that, while the counter-terms are important for cancelling UV divergences, since $\mathcal F$ is a polynomial its precise form does not affect the location of the poles. Second, the massive spectrum can be extracted by imposing either the $(-)$ or $(+)$ IR boundary condition together with the UV boundary condition $b|_{r_2} = 0$, and identifying those masses $m^2 = -q^2$ for which solutions to Eq.~\eqref{eq:eomb} exist, while finally recovering the physical result by taking the limits $r_1 \rightarrow 0$ (the IR regulator approaching the end of space) and $r_2 \rightarrow \infty$ (the UV regulator approaching the boundary at infinity). Third, the presence of a massless pole depends on the behaviour of the fraction $\frac{\partial_r b}{b}$ in the vicinity of $q^2 = 0$. In appendix~\ref{sec:masslesspoles}, we show that when imposing the $(-)$ IR boundary condition, $\langle \mathcal O_R(q) \overline{\mathcal O}_R(-q) \rangle$ contains a massless pole, while for the $(+)$ case, no such massless state is present. These results are in agreement with those for the special case of an AdS background studied in~\cite{Contino:2004vy}. The interpretation is that the strongly coupled sector described in the  $(-)$ case is chiral (i.e.~it contains an unpaired chiral fermion), while the $(+)$ case describes a vector-like sector. 

As mentioned above, the physical spectrum is obtained after taking the limits $r_1 \rightarrow 0$ and $r_2 \rightarrow \infty$. In order to improve the convergence of the numerics for computing spectra, one may make use of the asymptotic expansions of the fluctuations under study. Specifically, we implement such a procedure when setting up the IR and UV boundary conditions for $b$, as follows. Asymptotically in the UV, the general solution for $b$ can be expanded in powers of $e^{-r}$ as
\beq
\label{eq:bUVexp}
	b(q,r) = b_1(q) \Big( 1 + \cdots \Big) + b_2(q) \Big( e^{- (1 + 2M_\Psi) r} + \cdots \Big) \,.
\eeq
This implies that in the limit $r_2 \rightarrow +\infty$, the UV boundary condition $b|_{r_2} = 0$ selects $b_1(q) = 0$. Hence, in order to minimize the cutoff effects,
in the numerical computations we impose directly
\beq
	b|_{r_2} = b_2(q) e^{- (1 + 2M_\Psi) r_2} \,,
	\hspace{1cm} \partial_r b|_{r_2} = -(1+2M_\Psi) b_2(q) e^{- (1 + 2M_\Psi) r_2} \,,
\eeq
where $b_2(q)$ amounts to an unimportant overall normalisation of $b$. Conversely, in the IR, we set up the boundary condition for $b$ according to either the $(-)$ or $(+)$ case, by making use of the general IR expansion
\beq
\label{eq:bIR}
	b(q,r) = \tilde b_1(q) \bigg( 1 + \frac{(2\Delta)^{2-\frac{x_F}{\Delta}} q^2}{2(x_F - 2\Delta)^2 (1 - \phi_c^2 \frac{\Delta_\phi}{3})^{\frac{1}{\Delta_\phi}}} \, r^{2-\frac{x_F}{\Delta}} + \cdots \bigg) + \tilde b_2(q) \bigg( r^{1 - \frac{x_F}{2\Delta}} + \cdots \bigg)
\eeq
for Model~II. The corresponding IR expansion for Model~I is given by Eq.~\eqref{eq:bIR} after putting $\phi_c = 0$, thus eliminating the second factor of the denominator. The spectrum is then extracted by using, for each value of $m^2 = -q^2$, the equation of motion~\eqref{eq:eomb} to numerically evolve the solutions from the IR and from the UV towards an intermediate value of $r$ where the two are matched.

\subsection{AdS background}
\label{sec:AdS}

\begin{figure}[t]
\begin{center}
\includegraphics[width=\figwidth]{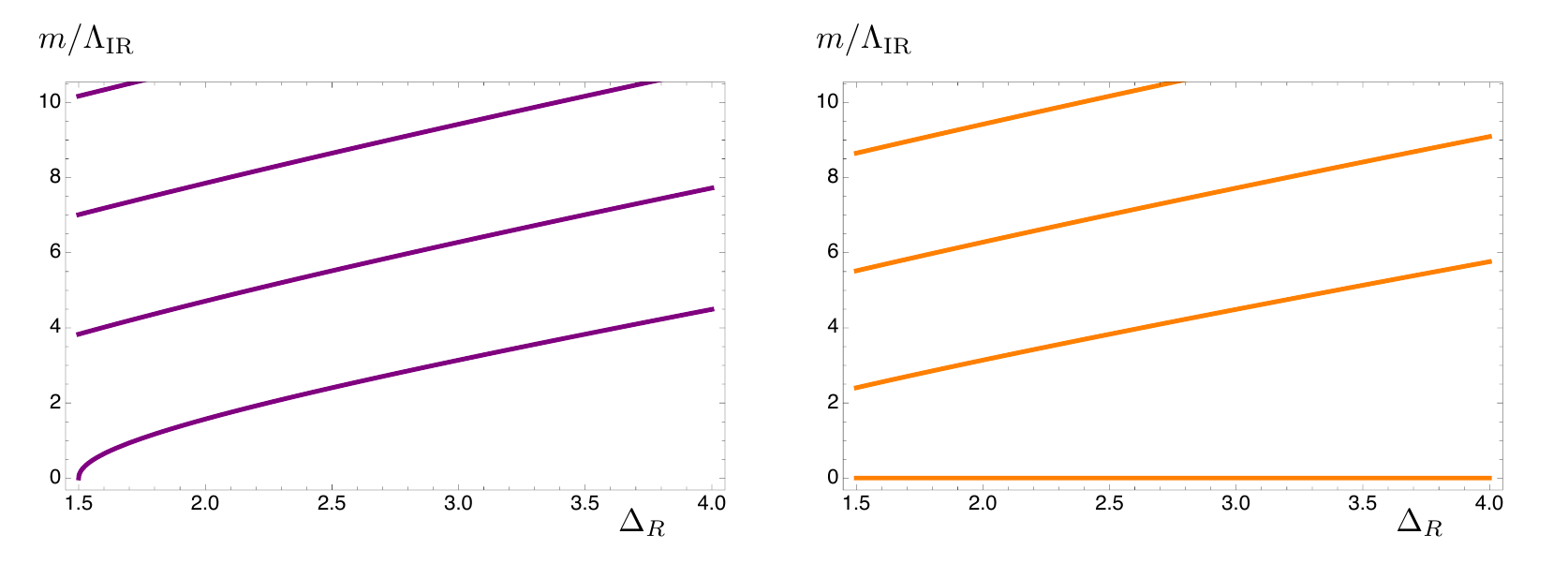}
\caption{AdS background. Spectrum of fermionic resonances as a function of $\Delta_R$ normalised to the IR scale $\Lambda_{\rm IR}$. The left panel shows the poles extracted from the correlator $\langle \mathcal O_R(q) \overline{\mathcal O}_R(-q) \rangle^+$ while the right panel shows the poles of $\langle \mathcal O_R(q) \overline{\mathcal O}_R(-q) \rangle^-$.}
\label{fig:AdSmp}
\end{center}
\end{figure}

Before discussing the fermionic spectrum for the models presented in section~\ref{sec:models}, let us first recall the results for an AdS background (for further details, see appendix~\ref{AdSapp}). In order to obtain a mass gap, we introduce a hard-wall IR cutoff that we take to be at $r = 0$ without loss of generality. In the case of the $(+)$ IR boundary condition, one obtains that the spectrum is given by those $m$ for which the
Bessel function $J_{\Delta_R - \frac{5}{2}}(m) = 0$. Similarly, for $(-)$ IR boundary condition, the spectrum is given by those $m$ for which  $J_{\Delta_R - \frac{3}{2}}(m) = 0$. Figure~\ref{fig:AdSmp} shows the resulting spectra, in units of the IR scale $\Lambda_{\rm IR} = 1$, as a function of the scaling dimension $\Delta_R = [\mathcal O_R]$. The $(+)$ case corresponds to the dual strongly-coupled sector being vector-like, and as a consequence the spectrum is gapped for generic values of $\Delta_R$. However, as the scaling dimension approaches $\Delta_R \rightarrow 3/2$, i.e. that of a free fermion, there is a state that becomes parametrically light. The interpretation is that, in this limit, a chiral fermion (in this case right-handed) decouples from the rest of the composite sector and becomes a free, elementary-like fermion, and as a consequence its (left-handed) chiral partner also becomes massless.\footnote{More precisely, as shown in appendix~\ref{AdSapp}, the two-point functions $\langle \mathcal O_R \overline{\mathcal O}_R \rangle^\pm$ both vanish exactly in the limit $\Delta_R \rightarrow 3/2$. After introducing the normalised operator $\tilde{\mathcal O} _R \equiv \mathcal O_R / \sqrt{2\Delta_R - 3}$, one can show that close to $\Delta_R \simeq 3/2$ its two-point functions approach the result for a canonically normalised free massless fermion, together with an infinite tower of poles whose residues become vanishingly small.} Conversely, the $(-)$ case corresponds to a chiral strongly-coupled sector, and hence a massless state is present for generic values of $\Delta_R$.

\subsection{Model I}

\begin{figure}[t]
\begin{center}
\includegraphics[width=\figwidth]{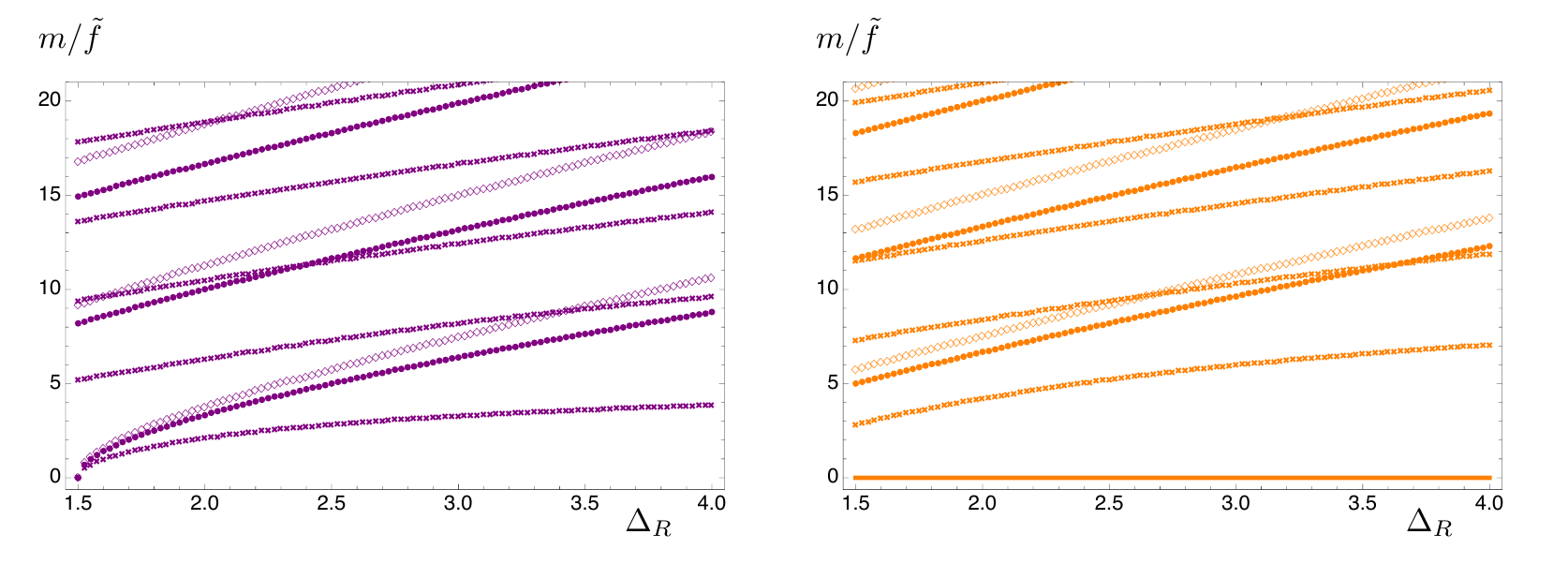}
\caption{Model I. Fermionic spectrum as a function of $\Delta_R$ for $x_F = 0.5, 2, 4$ (diamonds, dots, crosses), $\Delta = 2.5$, $g_5 = 8$, $r_1 = 10^{-12}$, $r_2 = 15$. The left and right panels correspond to imposing $(+)$ and $(-)$ IR boundary conditions, respectively. All plots are normalised to the decay constant $\tilde f$ (notice that $\Lambda_{\rm IR}/\tilde f \simeq 2.4, 2.1, 1.3$ for $x_F = 0.5, 2, 4$). The solid line in the right panel indicates the presence of a massless state.}
\label{fig:Model1F_1}
\vspace{0.4cm}
\includegraphics[width=\figwidth]{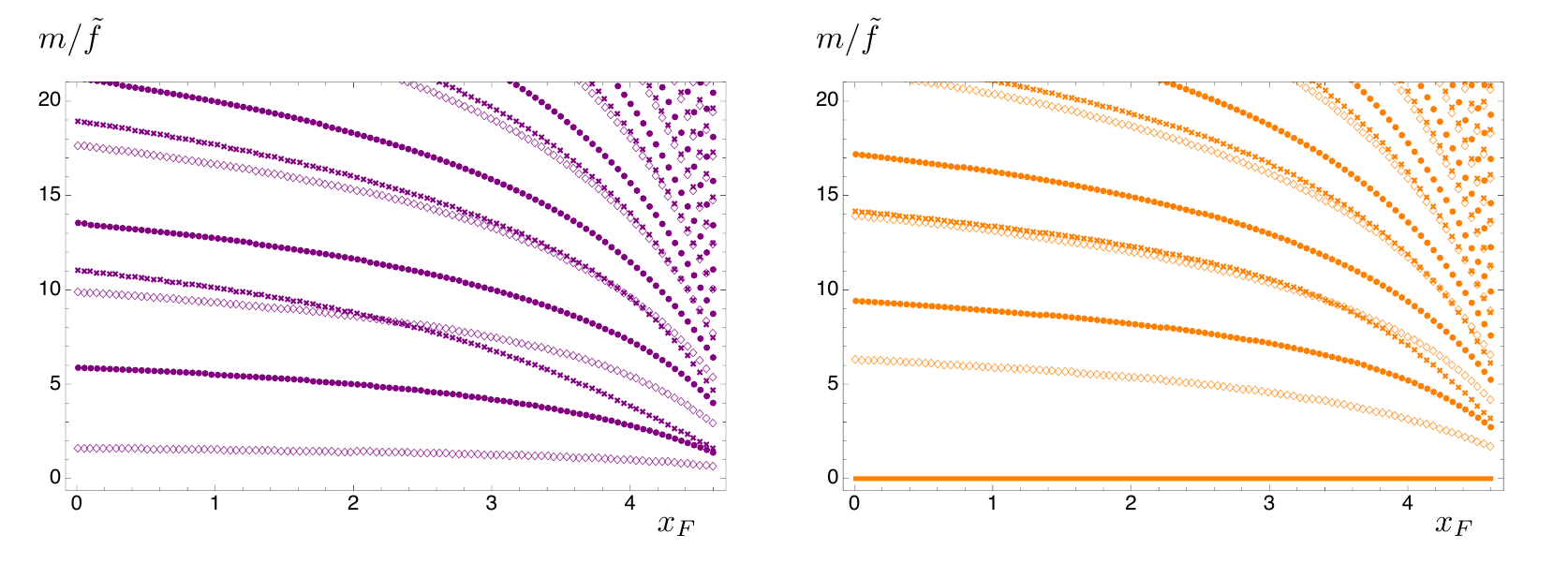}
\caption{Model I. Fermionic spectrum as a function of $x_F$ for $\Delta_R = 1.6, 2.5, 4$ (diamonds, dots, crosses), $\Delta = 2.5$, $g_5 = 8$. In the numerics, we used the IR regulator $r_1 = 10^{-12}$ for $x_F \leq 4.3$, while for $x_F > 4.3$ we used $r_1 = 10^{-30}$ in order to minimize cutoff effects; in both cases, the UV regulator was chosen to be $r_2 = 15$. The left and right panels correspond to imposing $(+)$ and $(-)$ IR boundary conditions, respectively. All plots are normalised to the decay constant $\tilde f$. The solid line in the right panel indicates the presence of a massless state.}
\label{fig:Model1F_2}
\end{center}
\end{figure}

In Figure~\ref{fig:Model1F_1}, we show the spectrum of fermionic states in Model I, as a function of the scaling dimension $\Delta_R$ of $\mathcal O_R$ at the UV fixed point, for different values of the number of flavours $x_F$. We have chosen the bulk gauge coupling to be equal to $g_5 = 8$, motivated by the comparison with lattice simulations, as discussed in~\cite{Elander:2020nyd} for the bosonic spectrum. As can be seen, the results are qualitatively the same as for the case of an AdS background (shown in Figure~\ref{fig:AdSmp}): for the $(+)$ boundary condition, there is one state that becomes parametrically light close to the free fermion case $\Delta_R \rightarrow \frac{3}{2}$, while for the $(-)$ boundary condition, there is always a massless state, as the dual strongly-coupled sector is chiral.

The effect of increasing the number of flavours is to bring the scale of the mass gap down relative to that of the decay constant $\tilde f$. 
There exists indeed a longstanding argument 
\cite{Chivukula:1992nw,Georgi:1992dw} for $f^2$ to grow proportionally to the largest between $N_C$ and $N_F$: this would imply that for large $N_F$ all boson and fermion masses should decrease 
relatively to $\tilde f$,
according to $m/\tilde{f}\sim x_F^{-1/2}$.
In our context, the dependence on $x_F$ appears more peculiar, as we illustrate in Figure~\ref{fig:Model1F_2}.
Interestingly, close to the bound at $x_F = 2\Delta$, when the characteristic IR scale $\Lambda_{\rm IR}$ tends towards zero (in units of the AdS radius), the spectrum approaches that of a continuum (in units of $\tilde f$). This differs from the behaviour of the bosonic spectrum discussed in appendix~\ref{sec:bosonicspectrum}, in which case instead there is a gapped continuum (see the left panel of Figure~\ref{fig:SpectrumBosons_1}). At the same time, we caution the Reader that our holographic models may be less trustable for large number of flavours, $x_F\gg 1$:
see discussion in \cite{Elander:2020nyd} as well as below \eq{FP}. Previous studies of holographic models with, or approaching, a (gapped) continuum include e.g.~\cite{Falkowski:2008yr,Csaki:2018kxb,Megias:2019vdb} as well as \cite{Berg:2005pd,Berg:2006xy,Elander:2017cle,Elander:2017hyr}, within the bottom-up and top-down approaches, respectively.

\begin{figure}[t]
\begin{center}
\includegraphics[width=\figwidth]{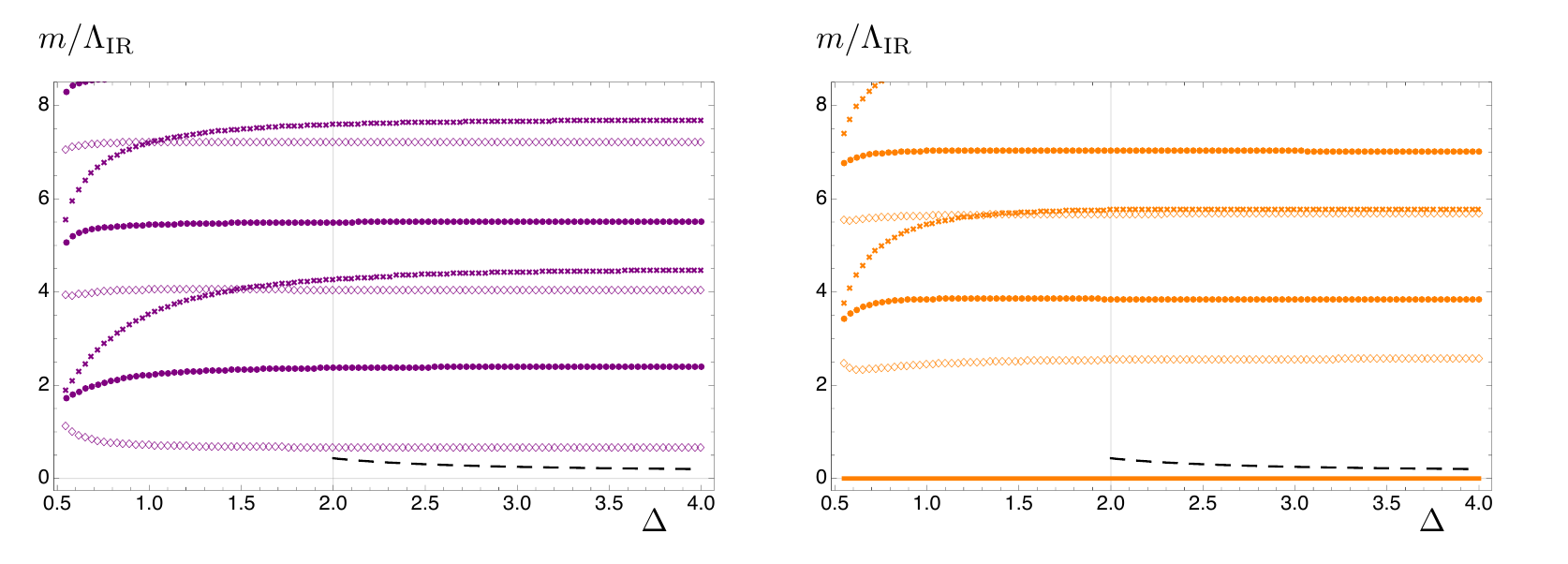}
\caption{Model I. Fermionic spectrum as a function of $\Delta$ for $\Delta_R = 1.6, 2.5, 4$ (diamonds, dots, crosses), $x_F = 1$, $g_5 = 8$, $r_1 = 10^{-12}$, $r_2 = 15$. The decay constant $\tilde f$ is represented by the dashed black line. We indicated by a vertical line the value $\Delta = 2$ below which the flavour symmetry is explicitly broken. The left and right panels correspond to imposing $(+)$ and $(-)$ IR boundary conditions, respectively. These plots are normalised to the IR scale $\Lambda_{\rm IR}$. The solid line in the right panel indicates the presence of a massless state.}
\label{fig:Model1F_3}
\end{center}
\end{figure}

Finally, Figure~\ref{fig:Model1F_3} shows the spectrum as a function of $\Delta$, related to the scaling dimension of the flavour-symmetry-breaking operator, including values $\Delta < 2$ for which explicit breaking is present. Again, the decrease of the mass gap (in units of $\tilde f$) for smaller values of $\Delta$ can be understood in terms of approaching the bound at $x_F = 2\Delta$.

\subsection{Model II}

\begin{figure}[t]
\begin{center}
\includegraphics[width=\figwidth]{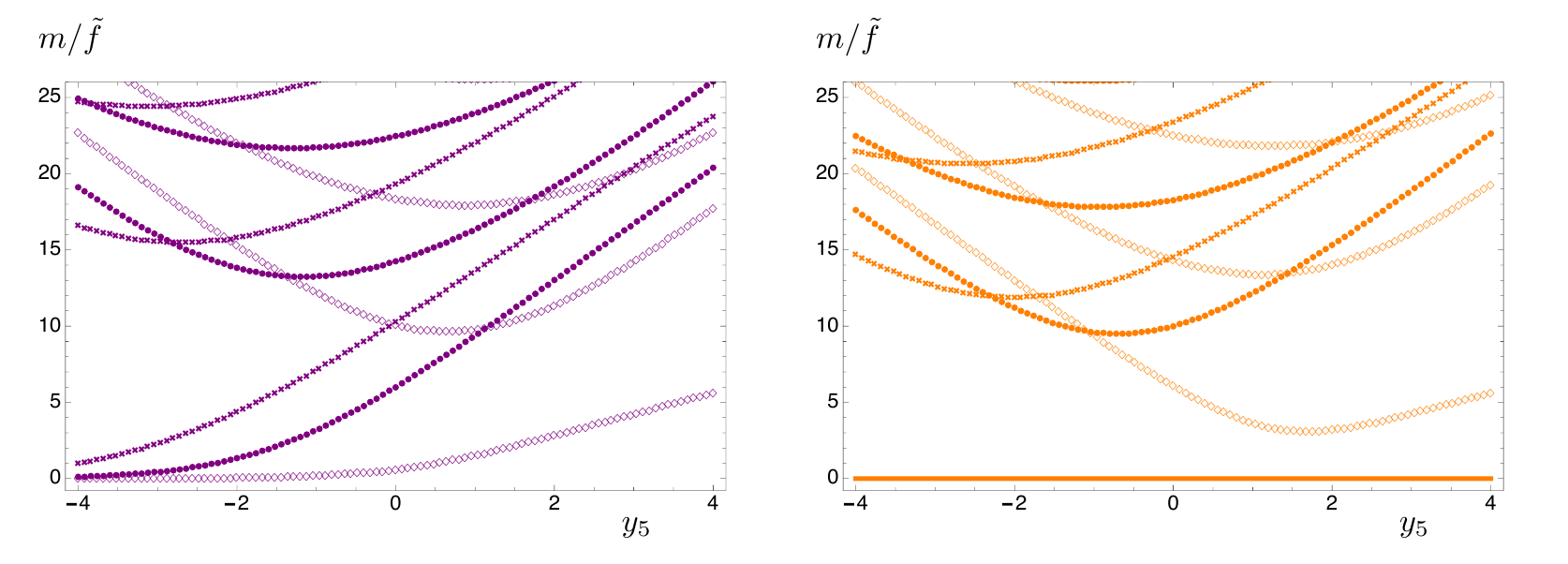}
\caption{Model II. Fermionic spectrum as a function of $y_5$ for $\Delta_R = 1.51, 2.5, 4$ (diamonds, dots, crosses), $\Delta = 3$, $x_F = 1$, $g_5 = 8$, $\Delta_\phi = 1$, $\phi_c = 1.5$, $r_1 = 10^{-12}$, $r_2 = 15$. The left and right panels correspond to imposing $(+)$ and $(-)$ IR boundary conditions, respectively. All plots are normalised to the decay constant $\tilde f$. The solid line in the right panel indicates the presence of a massless state.}
\label{fig:Model2BF_1}
\end{center}
\end{figure}

In Model~II, the inclusion of a bulk Yukawa coupling $y_5$ between the bulk fermion $\Psi$ and the flavour singlet scalar field $\phi$ allows for a fermionic spectrum with richer features than that of Model I. In Figure~\ref{fig:Model2BF_1}, we show the spectrum as a function of $y_5$. While for both large negative and positive values of $y_5$, the masses of the heavy states increase, they may be accompanied by light states:
\begin{itemize}
	\item[(i)] For large negative $y_5$ and $(+)$ IR boundary condition, there is a state that becomes parametrically light. 
	The lower the scaling dimension $\Delta_R$, the more pronounced this effect.
	\item [(ii)] For large positive $y_5$ and $(-)$ IR boundary condition, there is a light state as long as the scaling dimension $\Delta_R \simeq 3/2$ is close to that of a free fermion. As $y_5$ is further increased, the mass of this light state is however lifted.
\end{itemize}
While the first of these two mechanisms for obtaining a light state works regardless of the scaling dimension $\Delta_R$, the second requires being in the vicinity of the free fermion case. Hence, we may regard (i) as being more realistic of a strongly coupled theory than (ii), in the absence of a mechanism in strongly coupled field theories that generates scaling dimensions close to the free fermion case. In the next subsection~\ref{sec:toymodel}, we will describe a simplified toy model designed to capture the qualitative features of the spectrum shown in Figure~\ref{fig:Model2BF_1}.

Let us comment that a non-zero Yukawa coupling can roughly be thought of as a bulk mass with radial dependence determined by the background profile of $\phi$, and hence on the dual field theory side as a scaling dimension that varies with energy scale (we will elaborate further on this point in section~\ref{sec:partialcompositeness} where we introduce an effective scaling dimension of $\mathcal O_R$). One may worry that, for large negative $y_5$, this causes the scaling dimension of $\mathcal O_R$ to effectively be lower than that of a free fermion, in apparent contradiction with unitarity bounds~\cite{Mack:1975je}. Such an interpretation should however be taken with caution since, strictly speaking, scaling dimensions can be given a precise meaning only at conformal fixed points, which in our case implies that the unitarity bound is applicable only at the UV fixed point.

\begin{figure}[t]
\begin{center}
\includegraphics[width=\figwidth]{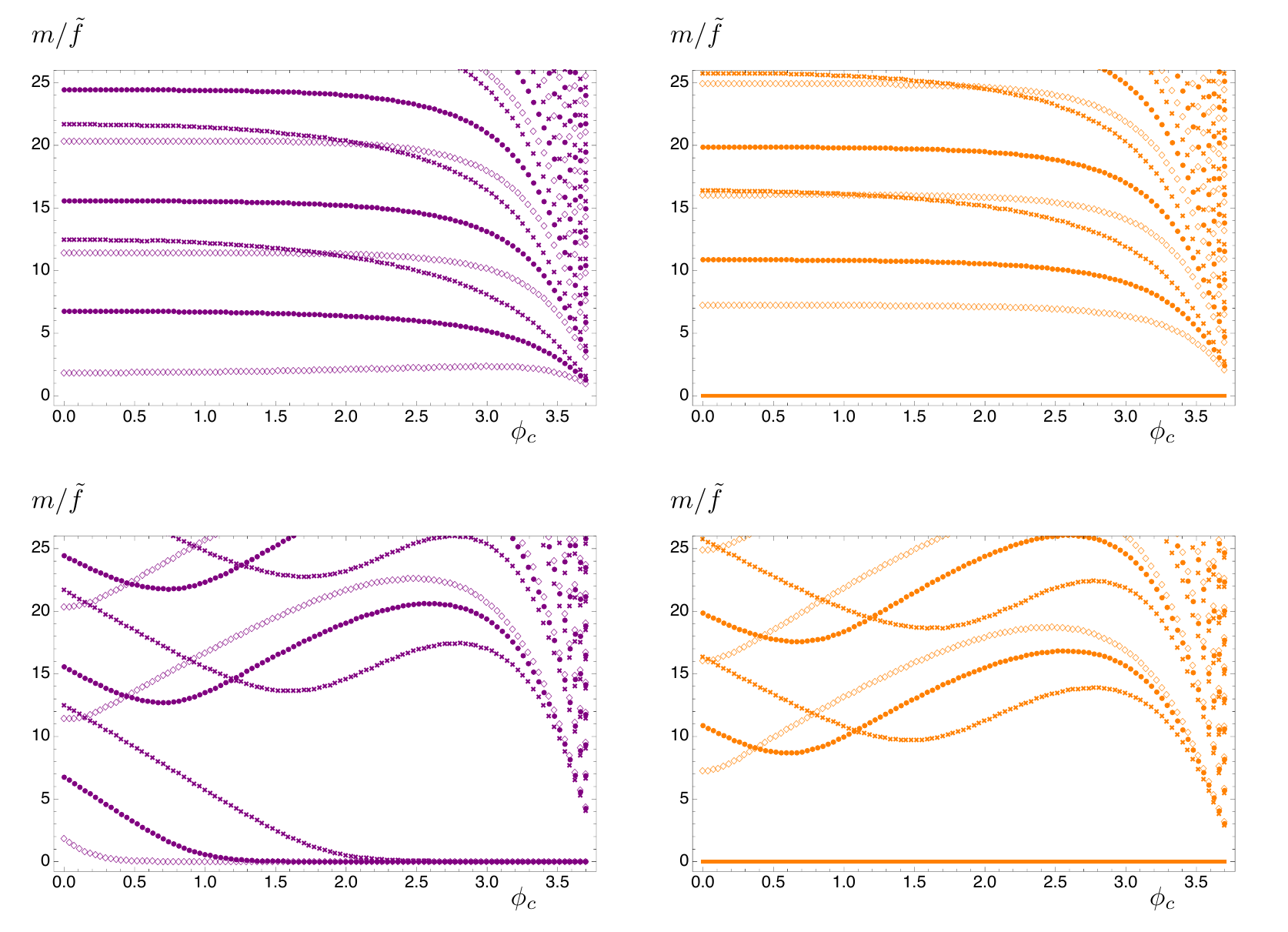}
\caption{Model II. Fermionic spectrum as a function of $\phi_c$ for $\Delta_R = 1.6, 2.5, 4$ (diamonds, dots, crosses), $\Delta = 3$, $x_F = 1$, $g_5 = 8$, $\Delta_\phi = 0.2$, $r_1 = 10^{-12}$, $r_2 = 15$, and $y_5 = 0, -2$ (top, bottom panels). The maximum allowed value of $\phi_c$ is given by $\sqrt{3/\Delta_\phi} \approx 3.87$. The left and right panels correspond to imposing $(+)$ and $(-)$ IR boundary conditions, respectively. All plots are normalised to the decay constant $\tilde f$. The solid lines in the right panels indicates the presence of a massless state.}
\label{fig:Model2BF_2}
\end{center}
\end{figure}

As discussed in section~\ref{sec:bosonicsector}, when $\Delta_\phi \leq \frac{\Delta}{2\Delta - x_F}$, it is possible to induce a small $\Lambda_{\rm IR}$ by increasing the source $\phi_c$ of the flavour singlet operator $\mathcal O_\phi$. We illustrate this effect in Figure~\ref{fig:Model2BF_2} for two values of the Yukawa coupling $y_5$. Close to the upper bound $\phi_c < \sqrt{\frac{3}{\Delta_\phi}}$, the spectrum approaches that of a continuum, similarly to what happens in Figure~\ref{fig:Model1F_2} for large number of flavours $x_F$. However, there is an important difference in that the bosonic spectrum (shown in the right panel of Figure~\ref{fig:SpectrumBosons_1}) now approaches a continuum that is no longer gapped. Furthermore, as is apparent from the bottom panels of Figure~\ref{fig:Model2BF_2}, for which $y_5 = -2$, the effect of the Yukawa coupling on the spectrum becomes more pronounced as $\phi_c$ is increased, eventually resulting in a light state for the case of $(+)$ IR boundary condition. We also note that for $\phi_c = 0$, the fermionic spectrum of Model~II is the same as for Model~I, since in this case the background geometries of the two models coincide while $\phi(r)$ vanishes identically such that the Yukawa coupling has no effect.

\begin{figure}[t]
\begin{center}
\includegraphics[width=\figwidth]{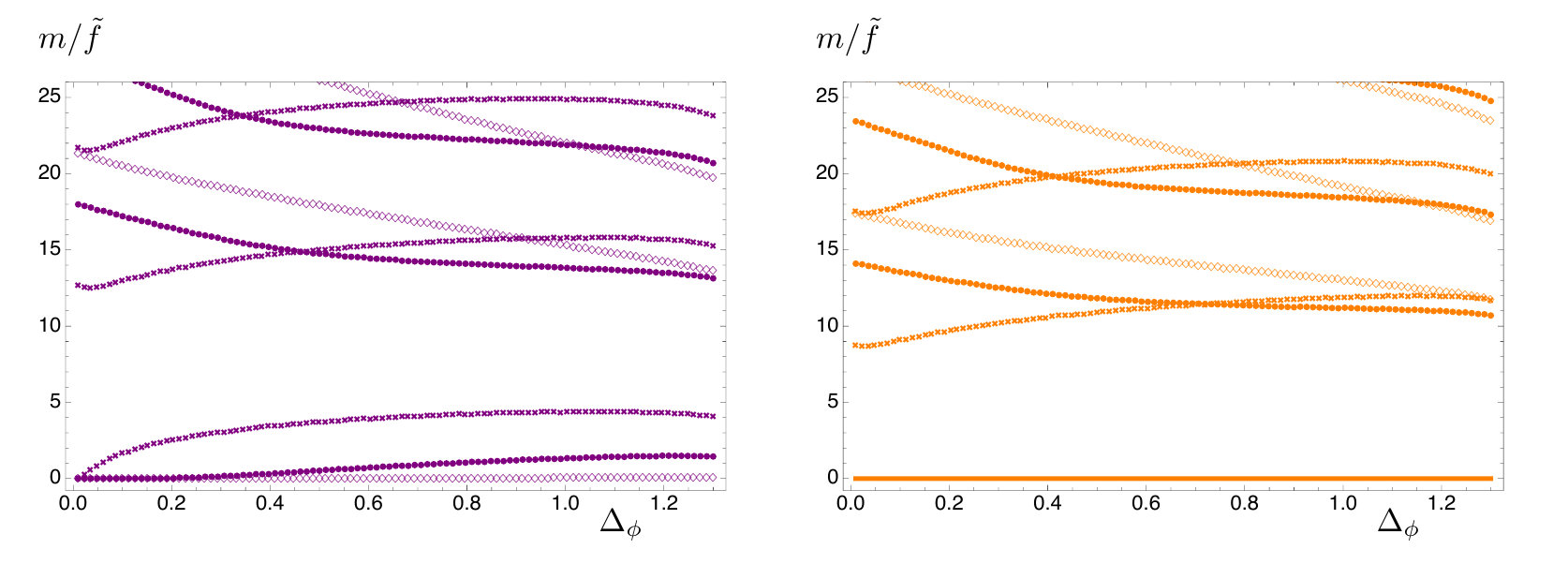}
\caption{Model II. Fermionic spectrum as a function of $\Delta_\phi$ for $\Delta_R = 1.51, 2.5, 4$ (diamonds, dots, crosses), $\Delta = 3$, $x_F = 1$, $g_5 = 8$, $y_5 = -2$, $\phi_c = 1.5$, $r_1 = 10^{-12}$, $r_2 = 15$. The left and right panels correspond to imposing $(+)$ and $(-)$ IR boundary conditions, respectively. All plots are normalised to the decay constant $\tilde f$. The solid line in the right panel indicates the presence of a massless state.}
\label{fig:Model2BF_3}
\end{center}
\end{figure}

Finally, in Figure~\ref{fig:Model2BF_3} we show the spectrum as a function of $\Delta_\phi$ for a particular value of the Yukawa coupling $y_5 = -2$. The parameters are chosen such that for $\Delta_\phi = 1$, the spectrum is the same as that of Figure~\ref{fig:Model2BF_1}. Interestingly, the effect of the Yukawa coupling becomes more pronounced as $\Delta_\phi \rightarrow 0$, when the scaling dimension of $\mathcal O_\phi$ approaches that of the marginal case. In particular, this leads to a smaller mass for the light state present in the vector-like case, corresponding to the $(+)$ IR boundary condition.

\subsection{Toy model}
\label{sec:toymodel}

In order to gain insight on the fermionic spectrum in models where the bulk fermion mass varies radially, such as Model II, let us consider a toy model that can be solved analytically. Take a Dirac fermion $\Psi$ propagating on an AdS geometry, introduce a hard-wall cutoff at $r=0$, and assume as usual a left-handed source $\Psi_L$, dual to a right-handed operator in the field theory. The AdS case with a constant bulk-fermion mass is analysed in appendix \ref{AdSapp}. To capture the effect of a bulk mass that varies radially, we let $\Psi$ have mass $M_*$ in the region $0 \leq r < r_*$ and mass $M_\Psi$ in the rest of the geometry $r \geq r_*$. In other words, we have
\beq
	H_\Psi = M_\Psi + M_\Delta \cdot \Theta(r_* - r) \,,\qquad M_\Delta \equiv M_* - M_\Psi \,.
\eeq
In comparing with Model II, $r_*$ is of order $\Delta_\phi^{-1}$, while $M_\Delta$ is analogous to $y_5 \Delta_\phi \int_{0}^\infty \dd r \, \phi(r)$.

This model is solvable: the equation of motion for $b$, given in Eq.~\eqref{eq:eomb}, is satisfied by
\beq
	b =
\begin{cases}
	e^{-\left(M_* + \frac{1}{2}\right) r} \big[ c_1(Q) Y_{M_* + \frac{1}{2}}(e^{-r} Q) + c_2(Q) J_{M_* + \frac{1}{2}}(e^{-r} Q) \big] & \mbox{for } 0 \leq r < r_* \,, \\
	e^{-\left(M_\Psi + \frac{1}{2}\right) r} \big[ d_1(Q) Y_{M_\Psi + \frac{1}{2}}(e^{-r} Q) + d_2(Q) J_{M_\Psi + \frac{1}{2}}(e^{-r} Q) \big] & \mbox{for } r \geq r_* \,,
\end{cases}
\eeq
where $Q^2=-q^2$ and $Y_\nu,J_\nu$ are Bessel functions.
The ratio of the integration constants $c_1$ and $c_2$ is determined by imposing either the $(+)$ or $(-)$ IR boundary conditions,
\beq
	\frac{c_1(Q)}{c_2(Q)} = - \frac{J_{M_* \mp \frac{1}{2}}(Q)}{Y_{M_* \mp \frac{1}{2}}(Q)}  \hspace{1cm} (\pm) \,,
\eeq
where we used $dC_\nu(z)/dz=C_{\nu-1}(z)-\nu C_\nu(z)/z$ for $C=Y,J$.
In addition, $d_1$ and $d_2$ are determined by requiring continuity of both $\Psi_L$ and $\Psi_R$ at $r = r_*$, which in turn implies continuity of both $b$ and $\partial_r b$. This determines $b$ up to an overall normalisation that plays no role in the following.

As usual, following the argument given in appendix~\ref{sec:masslesspoles}, there is a massless mode for $(-)$ IR boundary condition, while in the  $(+)$ case no such state exists. In order to compute the massive spectrum, one should first impose the UV boundary condition $b |_{r_2} = 0$, which implies
\beq
	\frac{d_1(Q)}{d_2(Q)} = - \frac{J_{M_\Psi+\frac 12}(e^{-r_2} Q)}{Y_{M_\Psi+\frac 12}(e^{-r_2} Q)} \,,
\eeq
and then take the limit $r_2 \rightarrow +\infty$. Since the right-hand side vanishes in this limit, one finds that the (massive) spectrum is determined by those $Q = m$ which satisfy $d_1(Q)=0$. Having required continuity at $r_*$, this condition can be written as 
\begin{align}
	& Y_{M_*+\frac{1}{2}}\left(e^{-r_*}Q\right) J_{M_{\Psi}-\frac{1}{2}}\left(e^{-r_*} Q\right) - Y_{M_*-\frac{1}{2}}\left(e^{-r_*}Q\right) J_{M_{\Psi}+\frac{1}{2}}\left(e^{-r_*} Q\right) \nonumber \\
	+ & \frac{c_2(Q)}{c_1(Q)} \Big[ J_{M_*+\frac{1}{2}}\left(e^{-r_*}Q\right) J_{M_{\Psi}-\frac{1}{2}}\left(e^{-r_*} Q\right) -
    J_{M_*-\frac{1}{2}}\left(e^{-r_*}Q\right) J_{M_{\Psi}+\frac{1}{2}}\left(e^{-r_*} Q\right)\Big] = 0  \,.
\end{align}
In order to study light states,  one may expand in powers of $e^{-r_*} Q$, obtaining the condition
\beqs
\label{eq:exprstar}
	&& \left[ \frac{\Gamma\left(M_* + \frac{1}{2}\right)}{\Gamma\left(M_\Psi + \frac{1}{2}\right)} + \frac{\Gamma\left(M_* - \frac{1}{2}\right) (M_\Psi - M_*)}{4\Gamma\left(M_\Psi + \frac{3}{2}\right)} e^{-2r_*} Q^2 + \mathcal O\left(e^{-4r_*}Q^4\right) \right] + \\
	&& \left(e^{-r_*} Q \right)^{2M_* + 1} \left(\tan\left(\pi M_*\right) - \frac{c_2(Q)}{c_1(Q)}\right) \left[ \frac{2^{-(2M_* + 1)} \pi (M_\Psi - M_*)}{\Gamma\left(M_\Psi + \frac{3}{2}\right) \Gamma\left(M_* + \frac{3}{2}\right)} + \mathcal O\left(e^{-2r_*}Q^2\right) \right] = 0 \,. \nonumber
\eeqs

\begin{figure}[t]
\begin{center}
\includegraphics[width=\figwidth]{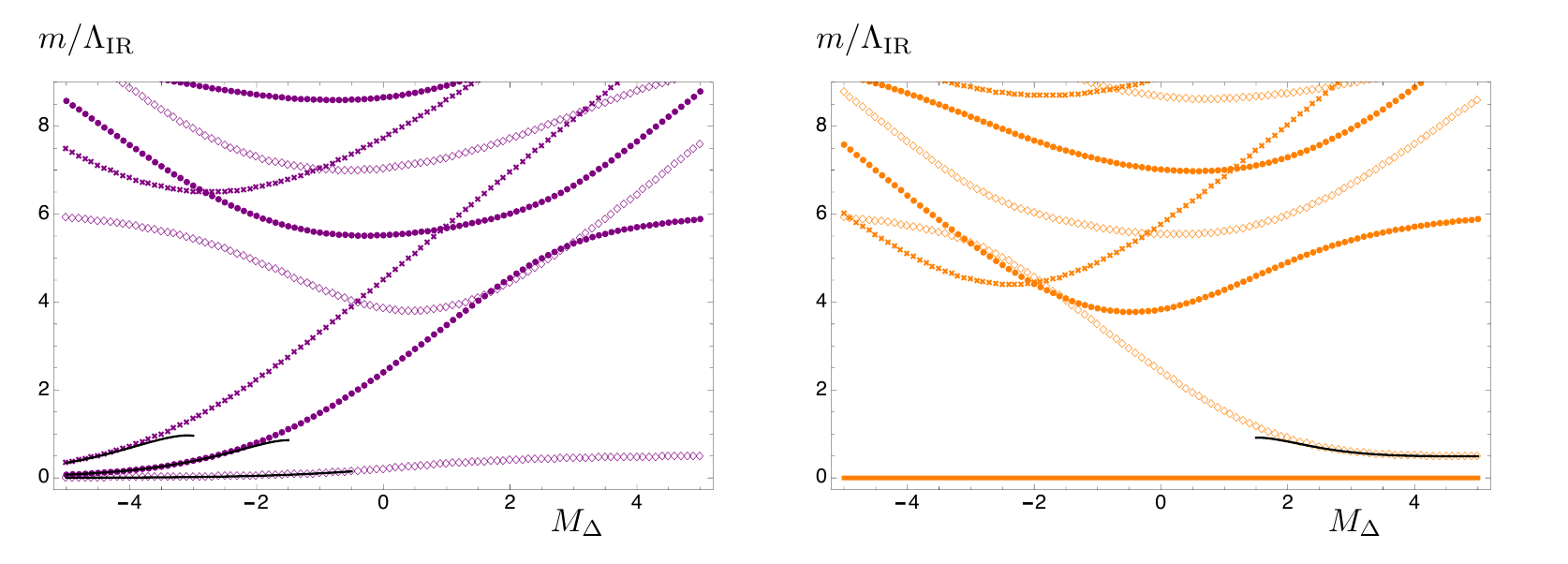}
\caption{Toy model. Fermionic spectrum as a function of $M_\Delta = M_* - M_\Psi$ for  $r_* = 1$ and $\Delta_R = 1.51, 2.5, 4$ (diamonds, dots, crosses). The left and right panels correspond to imposing $(+)$ and $(-)$ IR boundary conditions, respectively. All plots are normalised to the IR scale $\Lambda_{\rm IR}$. The solid line in the right panel indicates the presence of a massless state. Rather than using the analytic expressions, we found it convenient to use the same numerical procedure in extracting the spectrum as for Models~I and~II, and in so doing we used the UV cutoff $r_2 = 15$. Finally, the black lines are obtained from the approximations written in Eq.~\eqref{eq:ToyYapprox1} and Eq.~\eqref{eq:ToyYapprox2}.}
\label{fig:ModelToyY_1}
\end{center}
\end{figure}

Starting from this expression, we will now argue that a parametrically light state is present in the two following cases:
\begin{itemize}
\item[(i)] Consider $(+)$ IR boundary condition, in which case
\beq
	\frac{c_2(Q)}{c_1(Q)} - \tan\left(\pi M_*\right) = Q^{1-2M_*} \bigg[ \frac{2^{2M_* - 1}}{\pi} \Gamma\left(M_* - \frac{1}{2}\right)\Gamma\left(M_* + \frac{1}{2}\right) + \mathcal O\left(Q^2\right)\bigg] \,.
\eeq
Taking $M_\Psi$ to be constant and $M_*$ to be large and negative, the two lines of Eq~\eqref{eq:exprstar} can be made to be of the same order if $Q^2 \sim e^{(2M_* +1)r_*}$. Solving for $Q^2$, we obtain the estimate for the mass of the light state,
\beq
\label{eq:ToyYapprox1}
	m^2 = \frac{(2M_\Psi + 1)(4M_*^2 - 1)}{2(M_\Psi - M_*)} e^{(2M_* + 1) r_*} \,,
\eeq
which is exponentially suppressed.
\item[(ii)]
Consider $(-)$ IR boundary condition, in which case
\beq
	\frac{c_2(Q)}{c_1(Q)} - \tan\left(\pi M_*\right) = Q^{-(2M_*+1)} \bigg[ \frac{2^{2M_* + 1}}{\pi} \Gamma\left(M_* + \frac{1}{2}\right)\Gamma\left(M_* + \frac{3}{2}\right) + \mathcal O\left(Q^2\right)\bigg] \,,
\eeq
and furthermore take the scaling dimension $\Delta_R = M_\Psi + 2$ to be close to the free fermion case, namely $M_\Psi = - \frac{1}{2} + \epsilon$. For large positive $M_*$, the second term on the first line of Eq~\eqref{eq:exprstar} is of the same order as the second line, if $Q^2 \sim e^{(1-2M_*)r_*}$. Keeping also the first term on the first line, which is of order $\mathcal O(\epsilon)$, and solving for $Q^2$, we can estimate the mass of the light state as
\beq
\label{eq:ToyYapprox2}
	m^2 = \frac{8M_* - 4}{1 + 2M_*} e^{2r_*} \epsilon + 2 (2 M_* - 1) e^{(1-2M_*) r_*} \,,
\eeq
such that asymptotically, for large $M_*$, one obtains the small finite value for the mass $m = 2 \sqrt{\epsilon} \, e^{r_*}$. Notice that while in case (i) it was possible to obtain a light state for any value of the UV scaling dimension $\Delta_R$, in this case the argument crucially relies on taking $\Delta_R \simeq 3/2$ to be close to that of a free fermion.
\end{itemize}
We show the spectrum of this simplified model, as a function of $M_\Delta$, in Figure~\ref{fig:ModelToyY_1}. It can be seen that the approximations of Eq.~\eqref{eq:ToyYapprox1} and Eq.~\eqref{eq:ToyYapprox2} work well as long as $|M_\Delta|$ is sufficiently large. 
Note that the dependence of the spectrum on $M_\Delta$ is qualitatively similar to the dependence on $y_5$ of the spectrum of Model~II in Figure~\ref{fig:Model2BF_1}.

\subsection{Comparison with lattice results}\label{latt_comp}

Let us briefly discuss how our results may compare with existing lattice
simulations. For gauge theories with
symmetry breaking patterns related to composite Higgs scenarios,
lattice studies of the fermionic
bound states remain scarce to date, in contrast with simulations
addressing the properties of the corresponding
mesonic spectra, for which several studies are available, see the
references and discussion in \cite{Elander:2020nyd}.
One elaborate
analysis is provided by Ref. \cite{latt_su4_chimera},  
considering an $SU (4)$ gauge theory
and, for technical reasons, a simplified fermion content with a symmetry
pattern that does not accommodate the SM Higgs: two Dirac
fermions $q$ in the fundamental (dim-4) and two Dirac fermions $Q$ in the
two-index antisymmetric (dim-6) representations.
The fact that two
dynamical fermion species are simulated simultaneously provides a
realisation, within this specific model, of the fermionic states
described by the operators in Eq.~(\ref{psipsichi}) above.

The masses of the lightest fermionic bound states obtained after taking
the continuum and chiral limits are displayed in Table II and Figures~8 and
10 of \cite{latt_su4_chimera},
and they are also compared with the masses of the mesonic
bound states obtained previously for the same theory \cite{latt_su4}. 
Restricting
our attention to the spin-1/2 states, we
approximately find, in our normalisation,\footnote{We identify $F_4$ of \cite{latt_su4,latt_su4_chimera} with
$\sqrt{2} {\tilde f}$, 
and moreover use $F_6=\sqrt{2} F_4$~\cite{latt_su4}.}
\be
\frac{m_{Qqq}}{\tilde f} = 12.6(8)\,, \qquad
\frac{m_{V}}{\tilde f} = 8.6(6) \,,
\label{su4_baryon}
\ee
where we also quote the mass of the lightest vector meson for
comparison.

Quite recently, results for fermionic bound states were also obtained
for an $Sp(4)$ gauge symmetry \cite{Lucini:2021xke,Bennett:2021mbw}, 
with two dynamical, Dirac fermions
in the fundamental and three in the antisymmetric representations. From
Figure~4 of \cite{Lucini:2021xke}, one extracts
\be
m_{Qqq} \sim 2 m_{V}~.
\label{sp4_baryon}
\ee

Given that in these papers the gauge \cite{latt_su4_chimera} and flavour \cite{latt_su4_chimera,Lucini:2021xke,Bennett:2021mbw}
groups are different from the ones considered here, and that
moreover both simulations may not be close to
near-conformal dynamics,\footnote{We note that, in different contexts, such as QCD with a large number of flavours, many elaborate lattice analyses have been performed, with results approaching near-conformal dynamics  (see \cite{DeGrand:2015zxa} for a review).} a quantitative comparison with our results is not pertinent. We can
nevertheless examine if some qualitative features reasonably match,  
by considering our Model~II, with sizeable values of $\Delta_\phi$ and $\phi_c$, controlling the explicit breaking of conformal symmetry,
$x_F\sim 1$, and fermionic anomalous dimension presumably not too far from its canonical value, $\Delta_R\lesssim 9/2$. 
With all such caveats, a rather generic trend is that, for $\Delta_R\sim 4$,
the Model~II results for the fermionic states appear roughly consistent with the masses of Eq.~(\ref{su4_baryon}) or 
Eq.~(\ref{sp4_baryon}), 
see Figure~6 (top-left panel). In our analysis such masses have a milder sensitivity to $\Delta_\phi$ and $\phi_c$, as long as $\phi_c$ is not very large. 
Note that for comparable parameter choices we had obtained $m_V/\tilde f \sim 7-8$ \cite{Elander:2020nyd}.

\section{Partial compositeness and holographic Wilsonian RG flows}
\label{sec:partialcompositeness}

Fermion partial compositeness is realised by coupling linearly an elementary fermion $\chi_L$ to a fermionic composite operator $\mathcal{O}_R$, issued from the strongly-coupled sector: 
$\int \dd^4x ( \lambda \, \overline{\mathcal O}_R \chi_L + {\rm h.c.})$.\footnote{Here, $\chi_L$ is not to be confused with the two-index representation fermion $\chi_{ij}$ of the gauge theory discussed in section~\ref{sec:HC}.} In holography, such a coupling is provided between the boundary value of a bulk fermion, and the operator $\mathcal{O}_R$ for which it is a source in the dual field theory. In order to describe partial compositeness, we also need to make the source dynamical on the boundary, that is, add a four-dimensional kinetic term for $\chi_L$. Furthermore, we would like to describe how the coupling $\lambda$ runs with energy scale. In particular, we will identify scenarios where the IR value of $\lambda$ can be enhanced, despite $\lambda$ being irrelevant in the UV.

One expects the coupling $\lambda$ to be generated by some UV physics at a scale $\Lambda_{\rm UV}$ well above the HC mass gap $m_*$, while observations are sensitive to its infrared value after RG evolution, $\lambda(m_*)$. Let us briefly discuss the phenomenological constraints on the size of $m_*$ and $\Lambda_{\rm UV}$, in turn. In partial compositeness one needs to introduce one coupling $\lambda$ for each of the SM fermions $\chi_L$, in order to reproduce the SM Yukawa couplings. This implies a lower bound on $\lambda(m_*)$, in particular in the case of the top-quark Yukawa coupling, $y_t\simeq \lambda_{t_L} \lambda_{t_R} / g_* \simeq 1$. In turn, via the couplings $\lambda$, a generic strong dynamics would induce all sorts of higher-dimension operators suppressed by powers of $m_*$, especially flavour and CP violating ones. This sets a lower bound on $m_*$ which, in the case of generic strong dynamics, is of the order of $\sim 10^2$ ($\sim 10^3$) TeV for partial compositeness of quarks \cite{Keren-Zur:2012buf} (leptons \cite{Frigerio:2018uwx}). In order to keep $m_*$ as low as $\sim 10$ TeV, one needs global symmetries of the composite sector to forbid $e-\mu$ flavour transitions, as well as electric dipole moments of quarks and leptons. Let us remark that these estimates hold assuming a single mass scale $m_*$ for the lightest composite states. 
When a SM fermion mixes with a composite fermion with mass $m_F$ significantly smaller than $m_*$ (the typical mass of a composite vector), then some flavour-violating operators may be enhanced by powers of $m_*/m_F$, thus strengthening the constraints \cite{Panico:2015jxa}.

Beside the $\overline{\mathcal O}_R \chi_L$ operators, that are linear in the elementary fermions, the UV physics could typically induce also four-fermion operators involving two or more elementary fermions (as in e.g. extended technicolour models), suppressed by powers of $\Lambda_{\rm UV}$. For an anarchic flavour structure, this leads to a lower bound $\Lambda_{\rm UV} \gtrsim 10^5$ TeV \cite{UTfit:2007eik}, which can be relaxed to $\Lambda_{\rm UV} \gtrsim 10^3$ TeV by flavour selection rules suppressing the first-family couplings \cite{Luty:2004ye}.
One thus concludes that flavour constraints require at least $m_*/\Lambda_{\rm UV}\lesssim 10^{-2}$.
In the following we will study the RG evolution of $\lambda$ between $\Lambda_{\rm UV}$ and $\Lambda_{\rm IR}$. The latter is defined by \eq{laIR} and
is a good proxy for the vector resonance mass $m_*$, as long as the strong dynamics is characterised by a single scale. We will discuss interesting limits where this is not the case, and $\Lambda_{\rm IR} \sim m_F \ll m_*$ is realised instead. In such case, the Yukawa couplings are obtained by evaluating $\lambda(m_F)$ at the lower energy scale $m_F$ (rather than $m_*$). We will find that the IR region of the geometry may significantly modify the running of $\lambda$, and lead to a significant mixing between the elementary fermion and the strongly coupled sector, even for irrelevant scaling dimensions $\Delta_R$, provided that $\Lambda_{\rm UV}/\Lambda_{\rm IR}$ is not too large.

We will show here how to use the formalism of the holographic Wilsonian RG \cite{Heemskerk:2010hk,Faulkner:2010jy} in order to derive a beta function for $\lambda$. In particular, the treatment of fermionic operators was developed in \cite{Laia:2011wf,Elander:2011vh}.
We first consider the case when the bulk Dirac fermion $\Psi$ is uncharged, so that it does not couple to gauge fields. We also assume that no Yukawa interaction with bulk scalar fluctuations is present, although we allow the fermion mass term $H_\Psi(r) = M_\Psi + h_\Psi(r)$ to depend on the radial coordinate. Furthermore, we treat the background geometry as fixed. In this way the action is quadratic in the fermionic fields, allowing for Gaussian integration. Later we will comment on the possible impact of cubic or higher order interactions, whose treatment is considerably more involved.

\subsection{Holography at finite cutoff}

An initial assumption of holographic Wilsonian RG is that the GKPW relation \cite{Gubser:1998bc,Witten:1998qj}, relating the boundary values of bulk fields to sources for composite operators in the field theory, continues to hold also at generic finite values $\tilde r$ of the radial coordinate. The finite radial cutoff in the bulk corresponds to the introduction of a finite UV cutoff on the field theory side. We hence write the bulk partition function restricted to values of the radial coordinate $r \leq \tilde r$:
\beq
\label{eq:Zbulk}
	Z_{\rm bulk}[\overline \psi_L(\tilde r),\psi_L(\tilde r);\tilde r] = \int \mathcal D \overline \psi \mathcal D \psi |_{r<\tilde r} \, e^{i \mathcal S_\Psi[r_1,\tilde r;s_1, -1]} \,.
\eeq
Here, the action $S_\Psi$ restricted to an interval $r_1 \leq r \leq r_2$ is defined by
\begin{align}
\label{eq:SPsiFiniteCutoff}
	\mathcal S_\Psi[r_1,r_2;s_1,s_2] \equiv & - \int_{r_1}^{r_2} \dd r \int \dd^4 x \sqrt{-g} \left[ \frac{1}{2} \left( \overline \Psi \Gamma^M D_M \Psi - \overline{D_M \Psi} \Gamma^M \Psi \right) + H_\Psi \overline \Psi \Psi \right] \nonumber \\ & - \frac{s_1}{2} \int \dd^4 x \sqrt{-\tilde g} \, \overline \Psi \Psi \Big|_{r_1} - \frac{s_2}{2} \int \dd^4 x \sqrt{-\tilde g} \, \overline \Psi \Psi \Big|_{r_2} \,,
\end{align}
where $\Psi_L = N_L \psi_L$ and $\Psi_R = N_R\psi_R$ with the normalisations given by Eq.~\eqref{eq:defpsiL} together with the accompanying footnote~\ref{eq:defpsiR}, while $s_{1,2}$ encode the signs of the boundary actions at $r=r_{1,2}$. Note that, as a consequence of choosing $s_2= -1$ in Eq.~\eqref{eq:Zbulk}, the value of the left-handed component $\psi_L$ and its conjugate $\overline \psi_L$ are kept fixed at the radial cutoff $r=\tilde r$.\footnote{For a detailed derivation of this correspondence, see appendix~E of~\cite{Elander:2011vh}.} The reason for this is that we will consider the case when $\psi_L$ is the source of a composite operator in the dual field theory (as explained before, the case of a right-handed source can be treated analogously). Furthermore, we have kept the IR regulator $r_1$, with the understanding that physical quantities are obtained in the limit $r_1 \rightarrow r_o$ where $r_o$ is the value of the radial coordinate at the end of the space. Finally, as before, we choose the IR boundary condition $(\pm)$ on $\psi_L$ corresponding to the sign $s_1$, which we keep general in the following discussion.

The GKPW relation at finite cutoff, postulated to hold at any value of $\tilde r$, now becomes
\beq
\label{eq:GKPW-HP}
	Z_{\rm bulk}[\overline \psi_L(\tilde r),\psi_L(\tilde r);\tilde r]
	= \int \mathcal D {\mathbb M}_{\Lambda(\tilde r)} \exp \bigg\{i \mathcal S_{\rm QFT}[{\mathbb M};\Lambda(\tilde r)]+ i \int \dd^4x \Big( \overline{\mathcal O}_R \psi_L(\tilde r) + \overline \psi_L(\tilde r) \mathcal O_R \Big) \bigg\} \,,
\eeq
where we have collectively denoted by $\mathbb M$ all the fields appearing in the field theory, and introduced a field theory cutoff $\Lambda$. The details regarding how the radial coordinate is related to $\Lambda$ and in particular which regularisation scheme is used in the field theory are difficult questions that have not been fully answered (although see~\cite{McGough:2016lol,Taylor:2018xcy,Hartman:2018tkw}). We leave both the functional dependence of $\Lambda$ on the radial coordinate, as well as the precise form of the field theory action $\mathcal S_{\rm QFT}[{\mathbb M};\Lambda]$, unspecified for the moment. Later on, we will make the assumption that $\Lambda$ is given by Eq.~\eqref{eq:Lambda}.

As a next step, consider the field theory with partition function given by
\beqs
\label{eq:ZQFT1}
	 Z_{\rm QFT}[\Lambda;\xi] &=& \int \mathcal D \overline \chi_L \mathcal D \chi_L \mathcal D {\mathbb M}_\Lambda e^{i \mathcal S[\overline \chi_L,\chi_L,\mathbb M;\Lambda;\xi]} \,, \nonumber\\
	 \mathcal S[\overline \chi_L,\chi_L,\mathbb M;\Lambda;\xi] &=& \mathcal S_{\rm QFT}[{\mathbb M};\Lambda] - \int \dd^4q \, \overline \chi_L(-q) i \slashed q \chi_L(q) \nonumber \\ &&
	 + \int \dd^4q \Big( \mathcal N_\Psi^{-1/2} \xi(q) \overline{\mathcal O}_R(-q) \chi_L(q) + {\rm h.c.} \Big) \,,
\eeqs
describing an elementary fermion $\chi_L$ with canonically normalised kinetic term, coupled to a strongly-coupled sector with strength $\xi$ that depends on the four-momentum squared $q^2$. Here it is understood that the composite operator $\mathcal O_R$ may refer to some specific component of a larger flavour multiplet. In general, the coupling $\xi$ can be expanded in powers of $q^2$ leading to higher derivative terms.\footnote{Even if not turned on initially, the RG flow will generate such terms, and hence it is necessary to include them in the analysis. Note that such higher derivative operators, accounting for the UV physics being integrated out, are still quadratic in the fields.} We also normalised $\xi$ to be $N_C$-independent in the large-$N_C$ limit, by including the factor $\mathcal N_\Psi^{-1/2}$ to compensate for the $N_C$-scaling of $\mathcal O_R$. As shown in section \ref{sec:HC}, the operators needed to realise partial compositeness of the SM fermions scale with $\mathcal N_\Psi = N_C^2$. After the change of variables\footnote{Since $\chi_L$ couples to a specific component $\mathcal O_R$ of a complete flavour multiplet, in a slight abuse of notation we refer to $\psi_L$ as the corresponding component of the source multiplet.}
\be
\psi_L(q,\tilde r) = \mathcal N_\Psi^{-1/2} \xi(q) \chi_L(q)~,
\ee
we can use Eq.~\eqref{eq:GKPW-HP} to rewrite $Z_{\rm QFT}[\Lambda;\xi]$ in terms of bulk quantities, as
\begin{align}
\label{eq:ZQFT2}
	Z_{\rm QFT}[\Lambda(\tilde r);\xi] = \int \mathcal D \overline \psi_L(\tilde r) \mathcal D \psi_L(\tilde r) \, & Z_{\rm bulk}[\overline \psi_L(\tilde r),\psi_L(\tilde r);\tilde r] \times \nonumber \\
	& \exp\left(- i \, \mathcal N_\Psi \int \dd^4q \, \overline \psi_L(-q,\tilde r) \frac{i \slashed q}{\xi^2(q)} \psi_L(q,\tilde r) \right) \,.
\end{align}
The usefulness of this relation stems from the fact that it holds at any $\tilde r$ and for any given coupling $\xi$. Finally, we comment that it is useful to keep in mind the scaling dimension of the various quantities as a function of the scaling dimension $\Delta_R$ of $\mathcal O_R(x)$ at a fixed point: one has in particular $[\psi_L(q)]=-\Delta_R$ and $[\xi]=5/2-\Delta_R$.

\subsection{Integrating out higher energy modes}

The main idea of the holographic Wilsonian RG is to introduce a cutoff surface at some finite value of the radial coordinate $r=\mathfrak{r}$, and divide the path integral in the bulk into two pieces corresponding to $r \leq \mathfrak{r}$ and $r>\mathfrak{r}$. Integrating out the latter part of the geometry is analogous to integrating out high momentum (UV) modes in the field theory. In this way, one can derive a bulk description of the four-dimensional Wilsonian action, as a function of the cutoff scale $\Lambda(\mathfrak{r})$. To this end, we consider the field theory defined at some UV scale $\Lambda_{\rm UV} = \Lambda(r_{\rm UV})$ with partition function given by $Z_{\rm QFT}[\Lambda_{\rm UV};\xi_{\rm UV}]$, and rewrite it in terms of bulk quantities, by making use of Eq.~\eqref{eq:ZQFT2} at $\tilde r = r_{\rm UV}$. Next, we divide the bulk partition function into IR and UV parts:
\beqs
\label{eq:ZIRandUV}
	Z_{\rm bulk}[\overline \psi_L(r_{\rm UV}),\psi_L(r_{\rm UV});r_{\rm UV}] &=& \int \mathcal D \overline \psi \mathcal D \psi |_{r<r_{\rm UV}} \, e^{i \mathcal S_\Psi[r_1,r_{\rm UV};s_1,-1]} \nonumber \\
	&=& \int \mathcal D \overline \psi_L(\mathfrak{r}) \mathcal D \psi_L(\mathfrak{r}) \Bigg[ \left( \int \mathcal D \overline \psi \mathcal D \psi |_{r<\mathfrak{r}} \, e^{i \mathcal S_\Psi[r_1,\mathfrak r;s_1,-1]} \right) \times \nonumber \\
	&& \left( \int \mathcal D \overline \psi_R(\mathfrak{r}) \mathcal D \psi_R(\mathfrak{r}) \int \mathcal D \overline \psi \mathcal D \psi|_{\mathfrak{r}<r<r_{\rm UV}} \, e^{i \mathcal S_\Psi[\mathfrak r,r_{\rm UV};+1,-1]} \right) \Bigg] \nonumber \\
	&=& \int \mathcal D \overline \psi_L(\mathfrak{r}) \mathcal D \psi_L(\mathfrak{r}) \, Z_{\rm bulk}[\overline \psi_L(\mathfrak{r}),\psi_L(\mathfrak{r});\mathfrak{r}] \times \nonumber \\
	&& \hspace{0.9cm} Z_{\rm UV}[\overline \psi_L(\mathfrak{r}),\psi_L(\mathfrak{r}), \overline \psi_L(r_{\rm UV}),\psi_L(r_{\rm UV});\mathfrak{r},r_{\rm UV}] \,,
\eeqs
where we introduced an intermediate value of the radial coordinate $\mathfrak{r}$, separating the IR and UV contributions. In the last equality, we used Eq.~\eqref{eq:Zbulk} and defined the UV functional
\begin{align}
	& Z_{\rm UV}[\overline \psi_L(\mathfrak{r}),\psi_L(\mathfrak{r}), \overline \psi_L(r_{\rm UV}),\psi_L(r_{\rm UV});\mathfrak{r},r_{\rm UV}] \equiv \nonumber \\
	& \hspace{3cm} \int \mathcal D \overline \psi_R(\mathfrak{r}) \mathcal D \psi_R(\mathfrak{r}) \int \mathcal D \overline \psi \mathcal D \psi|_{\mathfrak{r}<r<r_{\rm UV}} \, e^{i \mathcal S_\Psi[\mathfrak r,r_{\rm UV};+1,-1]} \,,
\end{align}
which encodes the effect of integrating out the UV degrees of freedom.

After plugging Eq.~\eqref{eq:ZIRandUV} into Eq.~\eqref{eq:ZQFT2} evaluated at $\tilde r = r_{\rm UV}$, we obtain\footnote{Cf. Eq.~(6.9) of~\cite{Elander:2011vh}.}
\beqs
\label{eq:ZQFTUV}
	Z_{\rm QFT}[\Lambda_{\rm UV};\xi_{\rm UV}] &=& \int \mathcal D \overline \psi_L(\mathfrak{r}) \mathcal D \psi_L(\mathfrak{r}) \, Z_{\rm bulk}[\overline \psi_L(\mathfrak{r}),\psi_L(\mathfrak{r});\mathfrak{r}] e^{i \mathcal S_{\rm UV}[\mathfrak{r},r_{\rm UV}]} \,,\\ \nonumber
	e^{i \mathcal S_{\rm UV}[\mathfrak{r},r_{\rm UV}]} &\equiv& \int \mathcal D \overline \psi_L(r_{\rm UV}) \mathcal D \psi_L(r_{\rm UV}) \, Z_{\rm UV}[\overline \psi_L(\mathfrak{r}),\psi_L(\mathfrak{r}), \overline \psi_L(r_{\rm UV}),\psi_L(r_{\rm UV});\mathfrak{r},r_{\rm UV}] \times \\ && \hspace{1.75cm} \exp\left(- i \, \mathcal N_\Psi \int \dd^4q \, \overline \psi_L(-q,r_{\rm UV}) \frac{i \slashed q}{\xi_{\rm UV}^2(q)} \psi_L(q,r_{\rm UV}) \right) \,.
\eeqs
Note that, since we assumed the bulk action $\mathcal S_\Psi$ contained in $Z_{\rm UV}$ to be quadratic in the fields, $\exp (i \mathcal S_{\rm UV}[\mathfrak{r},r_{\rm UV}])$ remains Gaussian as $\mathfrak{r}$ is varied. Let us parametrise it according to
\beq
\label{eq:SUVdefinition}
	\mathcal S_{\rm UV}[\mathfrak{r},r_{\rm UV}] = - \, \mathcal N_\Psi \int \dd^4q \, \overline \psi_L(-q,\mathfrak{r}) \frac{i \slashed q}{\xi^2(q,\mathfrak{r})} \psi_L(q,\mathfrak{r}) \,,
\eeq
where we have allowed $\xi$ to depend on $\mathfrak{r}$. In other words, after imposing the boundary condition that its initial value at $r_{\rm UV}$ is given by $\xi(q,r_{\rm UV}) = \xi_{\rm UV}(q)$, the full functional dependence of $\xi(q,\mathfrak{r})$ on $\mathfrak{r}$ is defined by Eq.~\eqref{eq:SUVdefinition}. After plugging Eq.~\eqref{eq:SUVdefinition} into Eq.~\eqref{eq:ZQFTUV}, and using Eq.~\eqref{eq:ZQFT2} at $\tilde r = \mathfrak{r}$, one finally obtains
\begin{align}
\label{eq:ZQFTrelation}
	Z_{\rm QFT}[\Lambda_{\rm UV};\xi_{\rm UV}] &= \int \mathcal D \overline \chi_L \mathcal D \chi_L \mathcal D {\mathbb M}_{\Lambda_{\rm UV}} e^{i \mathcal S[\overline \chi_L,\chi_L,\mathbb M;\Lambda_{\rm UV};\xi_{\rm UV}]} \nonumber \\
	&= \int \mathcal D \overline \chi_L \mathcal D \chi_L \mathcal D {\mathbb M}_{\Lambda(\mathfrak{r})} e^{i \mathcal S[\overline \chi_L,\chi_L,\mathbb M;\Lambda(\mathfrak{r});\xi(q,\mathfrak{r})]} = Z_{\rm QFT}[\Lambda(\mathfrak{r}),\xi(q,\mathfrak{r})] \,,
\end{align}
for any $\mathfrak{r}$. In other words, $\mathcal S[\overline \chi_L,\chi_L,\mathbb M;\Lambda(\mathfrak{r});\xi(q,\mathfrak{r})]$ defined in Eq.~\eqref{eq:ZQFT1} is to be interpreted as the Wilsonian action, where the $\mathfrak{r}$-dependence of the coupling $\xi(q,\mathfrak{r})$ encodes the effect of  integrating out higher energy modes corresponding to the UV part of the bulk partition function.

\subsection{RG flow equation}

In order to derive the beta-function for $\xi$, we note that since the left-hand side of Eq.~\eqref{eq:ZQFTUV} does not depend on $\mathfrak{r}$, neither should its right-hand side. In the large-$N_C$ limit, this implies that
\beq
	\partial_\mathfrak{r} \big( \mathcal S_\Psi[r_1,\mathfrak r;s_1,-1] + \mathcal S_{\rm UV}[\mathfrak{r},r_{\rm UV}] \big) = 0 \,.
\eeq
After also demanding that the variation $\delta ( \mathcal S_\Psi[r_1,\mathfrak r;s_1,-1] + \mathcal S_{\rm UV}[\mathfrak{r},r_{\rm UV}] )$ vanishes on-shell, one can derive the flow equation
\beq
\label{eq:flowxi}
	\partial_\mathfrak{r} \xi^2 = - N_L^{-2} e^{-5A} q^2 + N_L^2 e^{3A} \xi^4 \,,
\eeq
where $N_L(\mathfrak{r})$ is defined in \eq{eq:defpsiL} and, at a fixed point, has scaling dimension $[N_L]=\Delta_R-2$. We provide more details regarding the derivation of Eq.~\eqref{eq:flowxi} in appendix~\ref{sec:hWRGapp}.\footnote{We also note the correspondence with the flow equation given in Eq.~(3.26) of Ref.~\cite{Elander:2011vh} with the identification $F = - \frac{1}{i \slashed q} e^{4A} N_L^2 \xi^2$. Here, one also has to take into account a change in sign due to the assumed (opposite) direction of the radial coordinate.} Given a background geometry, determined by the warp factor $A(\mathfrak{r})$, 
and a choice for the bulk fermion mass $H_\Psi(\mathfrak{r})$, this allows us to study the RG flow of $\xi$.

However, we already note a problem with interpreting Eq.~\eqref{eq:flowxi} as describing the flow of a coupling $\xi$ of an elementary fermion to a strongly coupled sector: namely that $\xi(q,\mathfrak{r}) = 0$ is {\it not} a solution. In other words, even if one imposes that $\xi(q,r_{\rm UV}) = 0$, so that the elementary fermion completely decouples from the strong sector, the RG flow for $\xi$ induces higher derivative couplings at the lower cutoff $\Lambda(\mathfrak{r}) < \Lambda_{\rm UV}$. These should be interpreted as belonging to the strongly-coupled sector in isolation, by making the observation that if one integrates out the external fermion $\chi_L$ in Eq.~\eqref{eq:ZQFT1}, one generates a double-trace operator proportional to $\overline{\mathcal O}_R \slashed{\partial} \mathcal O_R$. The RG flow of the associated coupling was studied in Refs.~\cite{Laia:2011wf,Elander:2011vh}. Crucially, in a strongly-coupled theory considered in isolation, even if such double-trace coupling is turned off initially, it is generated by the RG flow as the cutoff $\Lambda$ is lowered. This can also be expected on general grounds from field theory considerations~\cite{Heemskerk:2010hk}. To see how this comes about, we start from Eq.~\eqref{eq:ZQFTrelation} and use Eq.~\eqref{eq:ZQFT1} to obtain
\begin{align}
	Z_{\rm QFT}[\Lambda_{\rm UV};\xi_{\rm UV} = 0] &= Z_{\rm QFT}[\Lambda(\mathfrak{r});\xi^{(0)}(q,\mathfrak{r})] = \int \mathcal D \overline \chi_L \mathcal D \chi_L \mathcal D {\mathbb M}_\Lambda e^{i \mathcal S[\overline \chi_L,\chi_L,\mathbb M;\Lambda(\mathfrak{r});\xi^{(0)}(q,\mathfrak{r})]} \nonumber \\ & \hspace{3.93cm} = \int \mathcal D {\mathbb M}_\Lambda \exp \left( i \mathcal S_{\rm QFT}^{(0)}[\mathbb M;\Lambda(\mathfrak{r})] \right) \,, \\
	\mathcal S_{\rm QFT}^{(0)}[\mathbb M;\Lambda] &\equiv \mathcal S_{\rm QFT}[{\mathbb M};\Lambda] - \mathcal N_\Psi^{-1} \int \dd^4q \, f_{\rm DT}(q,\mathfrak r) \overline{\mathcal O}_R(-q) (i \slashed{q}) \mathcal O_R(q) \,,
\end{align}
where we denote by the superscript $(0)$ that we are considering the RG flow generated by solving the flow equation Eq.~\eqref{eq:flowxi}, having imposed the boundary condition $\xi^{(0)}(q,r_{\rm UV}) = 0$. As anticipated, the integration over $\chi_L$ ($\overline \chi_L$) has generated a double-trace operator with coupling defined by
\beq
\label{eq:DTdefinition}
	f_{\rm DT}(q,\mathfrak r) \equiv \frac{\left( \xi^{(0)}(q,\mathfrak r) \right)^2}{q^2} \,,
\eeq
that satisfies the flow equation
\beq
\label{eq:flowDT}
	\partial_\mathfrak{r} f_{\rm DT} = - N_L^{-2} e^{-5A} + N_L^2 e^{3A} q^2 f^2_{\rm DT} \,.
\eeq
Even though, by definition, $f_{\rm DT}(q,r_{\rm UV}) = 0$, such that no double-trace coupling is present at the UV scale $\Lambda_{\rm UV}$, Eq.~\eqref{eq:flowDT} leads to a non-trivial RG flow for $f_{\rm DT}$ as the cutoff $\Lambda$ is lowered. We also note that the potentially troubling non-local nature of the factor $1/q^2$ in Eq.~\eqref{eq:DTdefinition} is offset by the fact that a derivative expansion of Eq.~\eqref{eq:flowxi} yields $( \xi^{(0)})^2 = \mathcal O(q^2)$, so that $f_{\rm DT} = \mathcal O(q^0)$ (we will return to this point later).

In the context of partial compositeness, it is necessary to separate the double-trace coupling $f_{\rm DT}$, which is present already in the RG flow of the strongly coupled theory in isolation, from the linear coupling to the external fermion. Hence, considering a general RG flow for $\xi$, we have that
\beqs
	Z_{\rm QFT}[\Lambda_{\rm UV};\xi_{\rm UV}] &=& Z_{\rm QFT}[\Lambda(\mathfrak{r});\xi(q,\mathfrak{r})] = \int \mathcal D \overline \chi_L \mathcal D \chi_L \mathcal D {\mathbb M}_\Lambda e^{i \tilde{\mathcal S}[\overline \chi_L,\chi_L,\mathbb M;\Lambda(\mathfrak{r});\lambda(q,\mathfrak{r})]} \,,
	\\ \tilde{\mathcal S}[\overline \chi_L,\chi_L,\mathbb M;\Lambda;\lambda] &\equiv& \mathcal S^{(0)}_{\rm QFT}[{\mathbb M};\Lambda] - \int \dd^4q \, \overline \chi_L(-q) i \slashed q \chi_L(q) \nonumber \\
	&&+ \int \dd^4q \Big( \mathcal N_\Psi^{-1/2} \lambda(q,\mathfrak r) \overline{\mathcal O}_R(-q) \chi_L(q) + {\rm h.c.} \Big) \,,
\label{eq:effectiveaction}
\eeqs
where the coupling $\lambda$ is defined by
\beq
\label{eq:lambdadef}
	\lambda^2(q,\mathfrak r) \equiv \xi^2(q,\mathfrak r) - q^2 f_{\rm DT}(q,\mathfrak r) \,.
\eeq
The effective action $\tilde{\mathcal S}[\overline \chi_L,\chi_L,\mathbb M;\Lambda;\lambda]$ given in Eq.~\eqref{eq:effectiveaction} then separates into a part that describes the RG flow of the strongly-coupled sector considered in isolation and another that describes the effect of its coupling $\lambda$ to the elementary fermion $\chi_L$. Note that, due to the factor of $q^2$ in Eq.~\eqref{eq:lambdadef}, the distinction between $\xi$ and $\lambda$ only matters for higher-derivative operators. Finally, the flow equation for the linear coupling $\lambda$ is given by
\beq
\label{eq:flowlambda}
	\partial_\mathfrak{r} \lambda^2 = N_L^2 e^{3A} \lambda^2 \left( \lambda^2 + 2 q^2 f_{\rm DT} \right) \,,
\eeq
and admits, as it should, $\lambda = 0$ as a solution.

We would like to write the flow equations~\eqref{eq:flowxi},~\eqref{eq:flowDT}, and~\eqref{eq:flowlambda} in terms of dimensionless couplings, $\tilde \xi$, $\tilde \lambda$, and $\tilde f_{\rm DT}$. However, scaling dimensions of operators are only strictly defined at fixed points, away from which they are subject to regularisation-scheme dependence. Furthermore, we need a precise relation between the radial cutoff and $\Lambda$, which we from now on assume takes the form given in Eq.~\eqref{eq:Lambda}. With these considerations in mind, we define the effective scaling dimension for $\mathcal O_R$ as
\beq
\label{eq:Deff}
	\Delta_R^{({\rm eff})}(\Lambda) \equiv (\Lambda \partial_\Lambda \mathfrak{r}) H_\Psi + 2  = \Lambda^{-1} e^A H_\Psi+2 \,,
\eeq
where we inverted \eq{eq:Lambda} to compute $\mathfrak{r}(\Lambda)$. Our definition (\ref{eq:Deff}) is indeed consistent with the expectation at fixed points: for an AdS geometry with radius $L$, such that $A(\mathfrak{r}) = \mathfrak{r}/L$ and $\Lambda(\mathfrak{r}) = \exp(\mathfrak{r}/L)$, and for constant mass $H_\Psi = M_\Psi$, one obtains $\Delta_R^{({\rm eff})} = L M_\Psi + 2$. We assumed this to be the case in the far UV, that is for $\Lambda\to\infty$, with $L=1$. However, our definition also works as expected for more general geometries that may flow close to an IR fixed point with a different AdS radius. We furthermore note that, under rather general assumptions, \eq{eq:Deff} implies that $\Delta_R^{({\rm eff})}$ becomes equal to two in the deep IR. More precisely, {\it if} one assumes the following to hold at the end of space: (i) the warp factor $A$ diverges to $-\infty$ (i.e. the end of space is dynamically generated), (ii) the IR scale $\Lambda_{\rm IR} > 0$ does not vanish, and (iii) the radially dependent mass $H_\Psi$ remains finite, {\it then} it follows that $\Delta_R^{({\rm eff})}(\Lambda_{\rm IR}) = 2$. While the first assumption is violated in the case of an AdS background with a hard-wall cutoff, all these assumptions are satisfied for Models I and II, given the background solutions that we consider.

We now define the dimensionless couplings
\beq
	\tilde \xi(q,\Lambda) \equiv F(\Lambda) \xi(q,\Lambda) \,, \qquad \tilde \lambda(q,\Lambda) \equiv F(\Lambda) \lambda(q,\Lambda) \,,
\eeq
and
\beq
	\tilde f_{\rm DT}(q,\Lambda) \equiv F^2(\Lambda) \Lambda^2 f_{\rm DT}(q,\Lambda) \,,
\eeq
where
\beqs
	F(\Lambda) &\equiv& \exp \left( \int_\Lambda^\infty \frac{\dd \tilde \Lambda}{\tilde \Lambda} \, \left[ \Delta_R - \Delta_R^{(\rm eff)}(\tilde \Lambda) \right] \right) \Lambda^{\Delta_R - \frac{5}{2}} \nonumber \\
	&=& \Lambda^{-1/2} \exp \left(M_\Psi \mathfrak{r} - \int_\mathfrak{r}^\infty \dd r \, h_\Psi(r) \right) \,,
\eeqs
and $\Delta_R \equiv M_\Psi + 2$ is the scaling dimension of $\mathcal O_R$ at the UV fixed point. We then obtain the flow equation for $\tilde\xi$ as
\beq
\label{eq:flowtildexi}
	\Lambda \partial_\Lambda \tilde \xi^2 = - \frac{q^2}{\Lambda^2} + 2 \left(\Delta_R^{(\rm eff)} - \frac{5}{2} \right) \, \tilde \xi^2 + \tilde \xi^4 \,,
\eeq
while the corresponding flow equations for $\tilde\lambda$ and $\tilde f_{\rm DT}$ are given by
\begin{align}
\label{eq:flowtildelambda}
	\Lambda \partial_\Lambda \tilde \lambda^2 &= 2 \left(\Delta_R^{(\rm eff)} - \frac{5}{2} + \frac{q^2}{\Lambda^2} \tilde f_{\rm DT} \right) \, \tilde \lambda^2 + \tilde \lambda^4 \,, \nonumber \\
	\Lambda \partial_\Lambda \tilde f_{\rm DT} &= - 1 + \left(2 \Delta_R^{(\rm eff)} - 3 \right) \, \tilde f_{\rm DT} + \frac{q^2}{\Lambda^2} \tilde f_{\rm DT}^2 \,,
\end{align}
where the double-trace coupling satisfies the UV boundary condition $\tilde f_{\rm DT}(q,\Lambda_{\rm UV}) = 0$.

When the four-momentum is small compared to the renormalisation scale, i.e. $q^2 \ll \Lambda^2$, one may consider the derivative expansion of the coupling $\tilde \lambda$:
\beq
\label{eq:lambdaexpansion}
	\tilde \lambda(q,\Lambda) = \tilde \lambda_0(\Lambda) +  \tilde \lambda_1(\Lambda) \Lambda^{-2} q^2 + \cdots \,,
\eeq
and similarly for the double-trace coupling $\tilde f_{\rm DT}$:
\beq
\label{eq:xiexpansion}
	\tilde f_{\rm DT} = \tilde f_{{\rm DT},0}(\Lambda) + \tilde f_{{\rm DT},1}(\Lambda) \Lambda^{-2} q^2 + \cdots \,.
\eeq
At the lowest order, we obtain
\beq
\label{eq:flowtildelambda0}
	\Lambda \partial_\Lambda \tilde \lambda_0^2 = 2 \left(\Delta_R^{(\rm eff)} - \frac{5}{2}  \right) \, \tilde \lambda_0^2 + \tilde \lambda_0^4 \,,
\eeq
which, as can be seen, does not require knowing the solution for $\tilde f_{\rm DT}$. As expected, since the lowest order coupling $\tilde \lambda_0$ describes the most relevant deformation, its flow is also independent from that of higher-order operators. The flow equation~\eqref{eq:flowtildelambda0} can be solved formally to give
\beq
\label{eq:lambda0flow}
	\tilde \lambda_0^2(\Lambda) = \frac{F^2(\Lambda) }{F^2(\Lambda_{\rm UV}) \tilde \lambda_{0,{\rm UV}}^{-2} + \int_\Lambda^{\Lambda_{\rm UV}} \frac{\dd \tilde \Lambda}{\tilde \Lambda} \, F^2(\tilde \Lambda)} \,,
\eeq
where $\tilde \lambda_{0,{\rm UV}} \equiv \tilde \lambda_0(\Lambda_{\rm UV})$ is the value of the coupling at the UV cutoff.

While at the lowest order in the derivative expansion there is no distinction between $\tilde \xi$ and $\tilde \lambda$, this is no longer true at subsequent orders. For the purpose of illustrating this, we therefore also write explicitly the flow equations at the next order:
\beq
\label{eq:lambda1flow}
	\Lambda \partial_\Lambda \tilde \lambda_1 = \left(\Delta_R^{(\rm eff)} - \frac{1}{2} +\frac{3}{2} \tilde \lambda_0^2 \right) \, \tilde \lambda_1 + \tilde \lambda_0 \tilde f_{{\rm DT},0} \,,
\eeq
where $\tilde f_{{\rm DT},0}$ satisfies
\beq
	\Lambda \partial_\Lambda \tilde f_{{\rm DT},0} = \left( 2 \Delta_R^{(\rm eff)} - 3 \right) \tilde f_{{\rm DT},0} - 1 \,,
\eeq
with boundary condition $\tilde f_{{\rm DT},0}(\Lambda_{\rm UV}) = 0$. As can be seen, the lowest order coupling $\tilde \lambda_0$ drives the flow of the coupling at the next order $\tilde \lambda_1$, such that even if $\tilde \lambda_1$ is put to zero as an initial condition, it becomes generated by the flow provided $\tilde \lambda_0$ is non-zero. Also, the flow of the double-trace coupling $\tilde f_{{\rm DT},0}$ enters non-trivially in the RG equations.

Finally, let us make a few comments regarding the assumptions we have made in our derivation of the RG flow equations. The first is that we considered a single fermion in the bulk, whereas in the field theory there may be several different fermionic composite operators of interest, and they may mix. Such mixing would be easily incorporated into the formalism by simply generalising it to $n$ bulk fermions and a non-diagonal matrix $H_\Psi$. The result would be a coupled system of RG flow equations for $n$ couplings that generalizes Eq.~\eqref{eq:flowxi} for $\xi$.

Secondly, as mentioned, the reason that the effective action of Eq.~\eqref{eq:ZQFT1}, and hence Eq.~\eqref{eq:effectiveaction}, remains of the same form along the RG flow is due the absence of interaction terms in the bulk action $S_\Psi$, which ensures that $\exp (i \mathcal S_{\rm UV}[\mathfrak{r},r_{\rm UV}])$ remains Gaussian along the flow. In other words, we have treated the bulk fermion as propagating on top of a given background, having neglected the fluctuations of all other fields. The full treatment of Models~I and~II should include fluctuations of the entire bosonic sector, in particular the gauge field associated with the part of the flavour symmetry under which $\mathcal O_R$ is charged. As a consequence, $\exp (i \mathcal S_{\rm UV}[\mathfrak{r},r_{\rm UV}])$ no longer remains Gaussian along the flow, and hence the Wilsonian effective action should contain all possible operators consistent with the symmetries, built from an arbitrary number of fields, as is usually the case. In this more general treatment, the RG flow equation~\eqref{eq:flowxi} for $\xi$ (and hence $\lambda$) would remain the same, but it would need to be supplemented by additional RG flow equations for the couplings associated with operators of higher order in the number of fields.

In the following, we will primarily focus on the RG flow for $\tilde \lambda_0$, as it describes the most relevant deformation, and being at the lowest order in the derivative expansion, it is the coupling that most affects the masses of light states. We will consider three explicit examples for the background: AdS, Model I, and Model II.

\subsection{AdS background}
\label{sec:RGAdS}

Let us specialise the general flow equations to the case of an AdS background with constant $H_\Psi = M_\Psi$. Without loss of generality we fix the end of space to be at $r_o= 0$. The field-theory energy scale is given by $\Lambda(r) = e^r$, which implies that $\Lambda_{\rm IR} = 1$, and the effective scaling dimension of $\mathcal O_R$ is constant, $\Delta_R^{({\rm eff})}(\Lambda) = \Delta_R  = 2 + M_\Psi$. In this case the dimensionless coupling $\tilde \xi$ reads
\beq
	\tilde \xi(q,\Lambda) \equiv \Lambda^{\delta_R} \xi(q,\Lambda) \,, \hspace{1cm} \delta_R \equiv \Delta_R - \frac{5}{2} \,,
\eeq
and the flow equation~\eqref{eq:flowtildexi} simply becomes 
\beq
\label{eq:betalambda}
	\Lambda \partial_\Lambda \tilde \xi^2  = - \Lambda^{-2} q^2 + 2 \delta_R \, \tilde \xi^2 + \tilde \xi^4 \,.
\eeq
Its general solution is given by\footnote{Note the similarity between \eq{eq:betalambda} and the differential equation for $x\equiv -\frac{\partial_r b}{b}$, that can be derived from the equation of motion \eqref{eq:eomb}: $\Lambda \partial_\Lambda x  = - \Lambda^{-2} q^2 - (1+2M_\Psi) x + x^2$. Hence the similarity between the solutions: \eq{eq:xisolAdS} and \eq{eq:dbLdivbL}, respectively.}
\beq
\label{eq:xisolAdS}
	\tilde \xi^2(Q,\Lambda) = - \frac{Q}{\Lambda} \frac{J_{\delta_R+1}(\Lambda^{-1}Q) - c_\xi(Q) Y_{\delta_R+1}(\Lambda^{-1}Q)}{J_{\delta_R}(\Lambda^{-1}Q) - c_\xi(Q) Y_{\delta_R}(\Lambda^{-1}Q)} \,,
\hspace{1cm} Q \equiv \sqrt{-q^2} \,,
\eeq
where $J_\alpha$ and $Y_\alpha$ are Bessel functions, and the integration constant $c_\xi(Q)$ determines the boundary value $\tilde\xi^2(Q,\Lambda_{\rm UV})$.

\begin{figure}[t]
\begin{center}
\includegraphics[width=\figwidth]{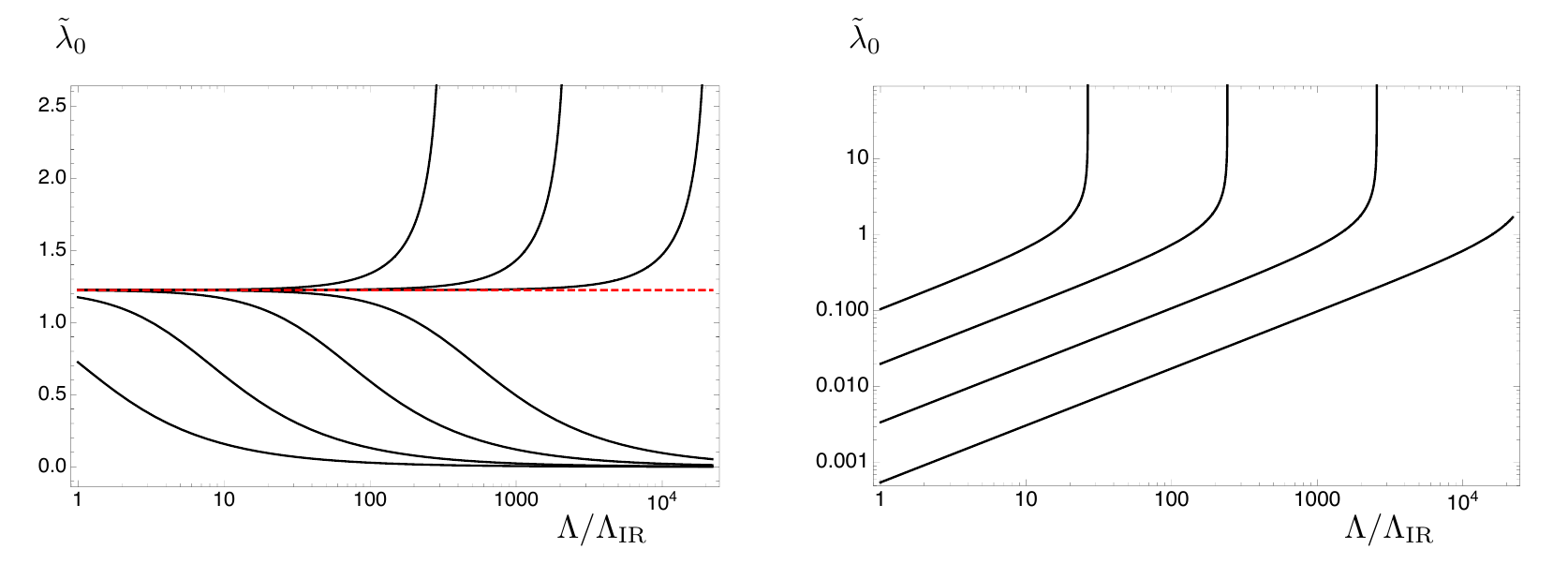}
\caption{The coupling $\tilde \lambda_0$ as a function of the energy scale $\Lambda$ for an AdS background. The two panels show a few possible RG flows for the cases of a relevant deformation with $\Delta_R = 1.75$ (left), and an irrelevant deformation with $\Delta_R = 3.25$ (right). The dashed red line indicates the position of the IR fixed point (left panel).}
\label{fig:RGAdS}
\end{center}
\end{figure}

Although Eq.~\eqref{eq:xisolAdS} gives the general solution for $\tilde \xi$, from which $\tilde \lambda^2 = \tilde \xi^2 - \frac{q^2}{\Lambda^2} \tilde f_{\rm DT}$ can be determined, we find it illustrative to study the derivative expansion of $\tilde \lambda$ given in Eq.~\eqref{eq:lambdaexpansion}. The flow equation~\eqref{eq:flowtildelambda0} for the lowest order coupling $\tilde \lambda_0$ becomes
\beq
	\Lambda \partial_\Lambda \tilde \lambda_0^2 = 2 \delta_R \, \tilde \lambda_0^2 + \tilde \lambda_0^4 \,,
\eeq
whose solution is
\beq
\label{eq:lambdasolution}
	\tilde \lambda_0^2(\Lambda) = \frac{2\delta_R}{\left( \dfrac{2\delta_R}{\tilde \lambda_{0,{\rm UV}}^2} + 1 \right) \left( \dfrac{\Lambda_{\rm UV}}{\Lambda} \right)^{2\delta_R} -1} \, ,
\eeq
where $\Lambda_{\rm UV} = e^{r_{\rm UV}}$ is the UV scale at which we specify the initial value 
$\tilde \lambda_0(\Lambda_{\rm UV}) \equiv \tilde \lambda_{0,{\rm UV}}$. For $\delta_R \geq 0$, the operator $\overline {\mathcal O}_R \chi_L$ represents an irrelevant deformation of the CFT, and the RG flow approaches an IR fixed point at $\tilde \lambda_0 = 0$. 
Conversely, for $\delta_R < 0$, one has that $\overline {\mathcal O}_R \chi_L$ is relevant, and the RG flow approaches a non-trivial IR fixed point, at $\tilde \lambda_0^2 = - 2\delta_R$. In the latter case, there is also a UV fixed point at $\tilde \lambda_0 = 0$. We illustrate a few examples of RG flows in Figure~\ref{fig:RGAdS}.

Finally, we note that if one makes the requirement that $\tilde \lambda_0$ does not reach a Landau pole at a scale $\Lambda < \Lambda_{\rm UV}$, then Eq.~\eqref{eq:lambdasolution} implies that its IR value $\tilde \lambda_{0,{\rm IR}}$ must be restricted to the range
\beq
\label{eq:lambdaIRmax}
	0 \leq | \tilde \lambda_{0,{\rm IR}}| \leq \lambda_{0,{\rm IR}}^{(\rm max)}  = \sqrt{\frac{2 \delta_R}{\left(\frac{\Lambda_{\rm UV}}{\Lambda_{\rm IR}}\right)^{2\delta_R} - 1}} \,.
\eeq
Suppose the UV cutoff $\Lambda_{\rm UV}$ is large compared to IR scale $\Lambda_{\rm IR}$. Then, for $\delta_R > 0$, the maximum allowed IR value of the coupling is suppressed as
\beq
\label{eq:lambdamaxAdSirrelevant}
	\lambda_{0,{\rm IR}}^{(\rm max)} \simeq \sqrt{2\delta_R} \left( \frac{\Lambda_{\rm IR}}{\Lambda_{\rm UV}} \right)^{\delta_R} \,,
\eeq
while for $\delta_R < 0$, it approaches that of the IR fixed point $\lambda_{0,{\rm IR}}^{(\rm max)} \simeq \sqrt{-2\delta_R}$.

\subsection{Model I}

\begin{figure}[t]
\begin{center}
\includegraphics[width=\figwidth]{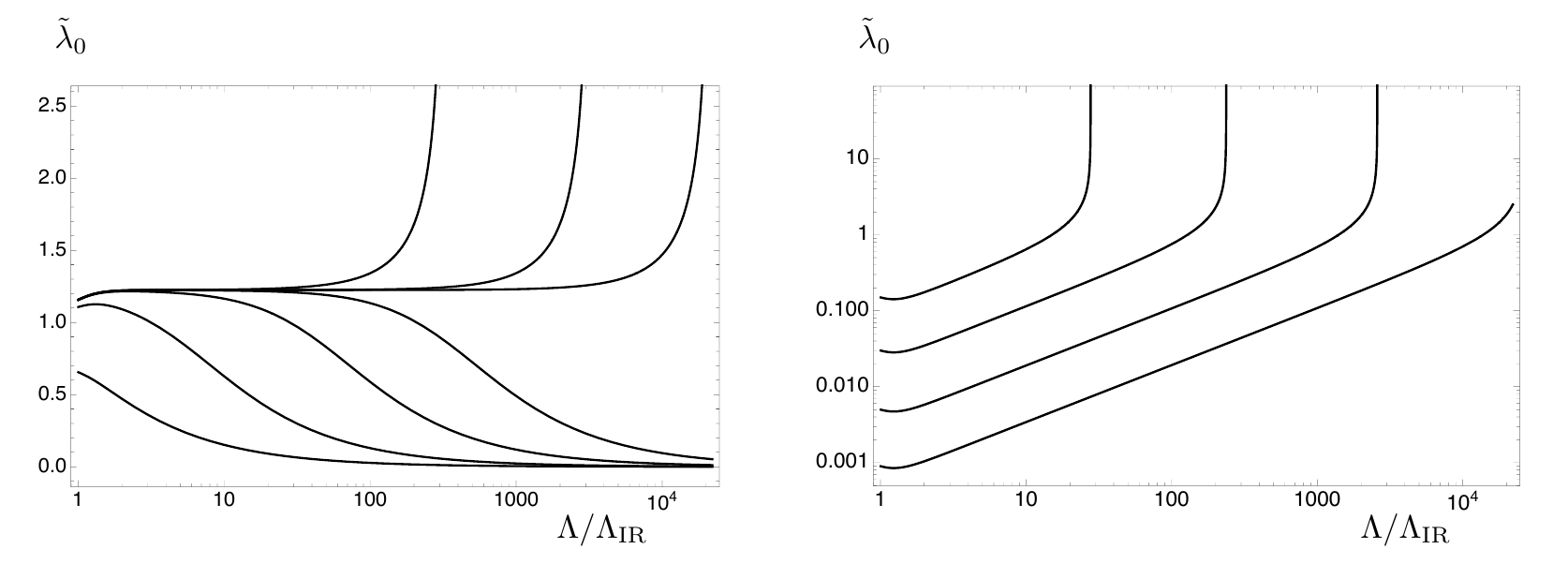}
\caption{Model I. The coupling $\tilde \lambda_0$ as a function of the energy scale $\Lambda$, for $\Delta = 2.5$ and $x_F = 3$. The two panels show a few possible RG flows for the cases of a relevant deformation with $\Delta_R = 1.75$ (left), and an irrelevant deformation with $\Delta_R = 3.25$ (right).
}
\label{fig:RGModel1_1}
\end{center}
\end{figure}

Let us now consider the flow of $\tilde\lambda_0$ in the background of Model I. The geometry is determined by the warp factor given in Eq.~\eqref{eq:Model1solutions}, i.e. $A(r) = r + x_F/(2\Delta) \log \left( 1 - e^{-2\Delta r} \right)$, and depends on the number of flavours $x_F$, as well as $\Delta$, related to the scaling dimension of the flavour-symmetry breaking operator, as explained in section~\ref{sec:models}. We recall that the end of space is located at $r_o = 0$. As before, we have constant $H_\Psi=M_\Psi$. Remarkably, it is possible to solve analytically the flow equation~\eqref{eq:flowtildelambda0} for $\tilde\lambda_0$, even for this non-trivial background. First, one  evaluates Eq.~\eqref{eq:Lambda} in terms of a hypergeometric function
\beq
	\Lambda(\mathfrak{r}) = \frac{e^\mathfrak{r}}{\, _2F_1\left(\frac{1}{2 \Delta},\frac{x_F}{2 \Delta};1+\frac{1}{2 \Delta};e^{-2 \Delta \mathfrak{r}}\right)} \,.
\eeq
Then, the solution of Eq.~\eqref{eq:flowtildelambda0} becomes
\beq
	\tilde \lambda^2_0(\mathfrak{r}) = \frac{{2\delta _R} \,  _2F_1\left(\frac{1}{2 \Delta
   },\frac{x_F}{2 \Delta };1+\frac{1}{2 \Delta };e^{-2 \Delta \mathfrak{r}
   }\right)}
   {\tilde \lambda_C e^{-2 \delta _R \mathfrak{r}}-\, _2F_1\left(\frac{x_F}{2 \Delta
   },-\frac{\delta _R}{\Delta };1-\frac{\delta _R}{\Delta };e^{-2
   \Delta \mathfrak{r}}\right)} \,,
\eeq
where $\tilde \lambda_C$ is an integration constant, that can be written in terms of the IR value of the coupling $\tilde \lambda_0(0)\equiv \tilde \lambda_{0,{\rm IR}}$ as
\beq
	\tilde \lambda_C = \Gamma \left(1-\frac{x_F}{2 \Delta }\right) \left[ \frac{\Gamma
   \left(1+\frac{1}{2 \Delta }\right)}{\Gamma \left(1-\frac{x_F-1}{2 \Delta
   }\right)} \frac{2\delta _R}{\tilde{\lambda
   }_{0,\text{IR}}^2} +\frac{\Gamma \left(1-\frac{\delta _R}{\Delta
   }\right)}{\Gamma \left(1 - \frac{x_F + 2 \delta _R}{2 \Delta
   }\right)}\right] \,.
\eeq
We illustrate a few examples of RG flows in Figure~\ref{fig:RGModel1_1}. In the far UV, $\Delta_R^{(\rm eff)} \simeq \Delta_R$, and hence the RG flow is close to that in the AdS case. The main qualitative difference with AdS occurs in the deep IR, where at the end of the space $\Delta_R^{(\rm eff)}(\Lambda_{\rm IR}) = 2$. For the case of a relevant deformation, this has the effect of diverting the RG flow of $\tilde\lambda_0$ from the would-be fixed point at $\tilde\lambda_0 = \sqrt{5-2\Delta_R}$, corresponding to the AdS case, towards $\tilde\lambda_0 \simeq 1$.

\begin{figure}[t]
\begin{center}
\includegraphics[width=\figwidth]{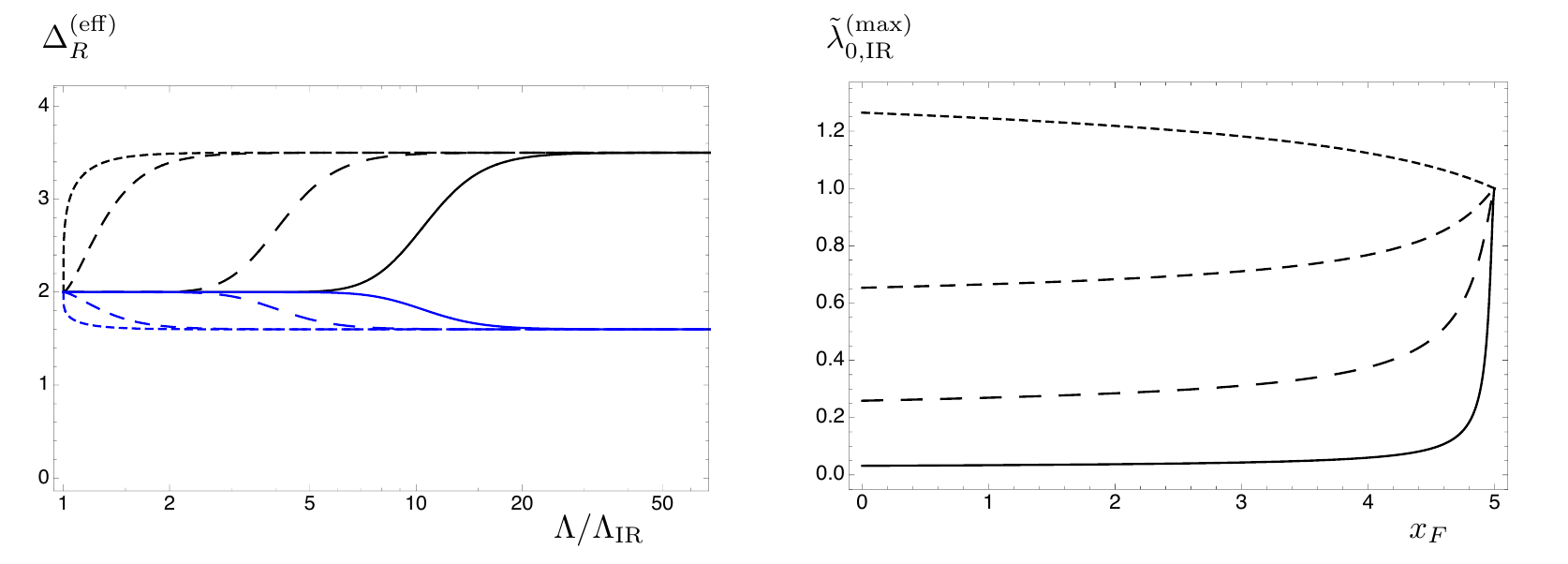}
\caption{Model I. The left panel shows the effective scaling dimension $\Delta_R^{({\rm eff})}$ as a function of the energy scale $\Lambda$, for $\Delta = 2.5$, $x_F = 1, 3, 4.7, 4.9$ (short-dashed, dashed, long-dashed, solid lines), and two choices $\Delta_R = 1.6, 3.5$ (in blue, black). The right panel shows the maximum IR value $\tilde \lambda_{0,{\rm IR}}^{({\rm max})}$ as a function of $x_F$, for $\Delta = 2.5$, $\frac{\Lambda_{\rm UV}}{\Lambda_{\rm IR}} = 10^3$, and $\Delta_R = 1.7, 2.3, 2.6, 3$ (short-dashed, dashed, long-dashed, solid lines).}
\label{fig:RGModel1_2}
\vspace{0.4cm}
\includegraphics[width=\figwidth]{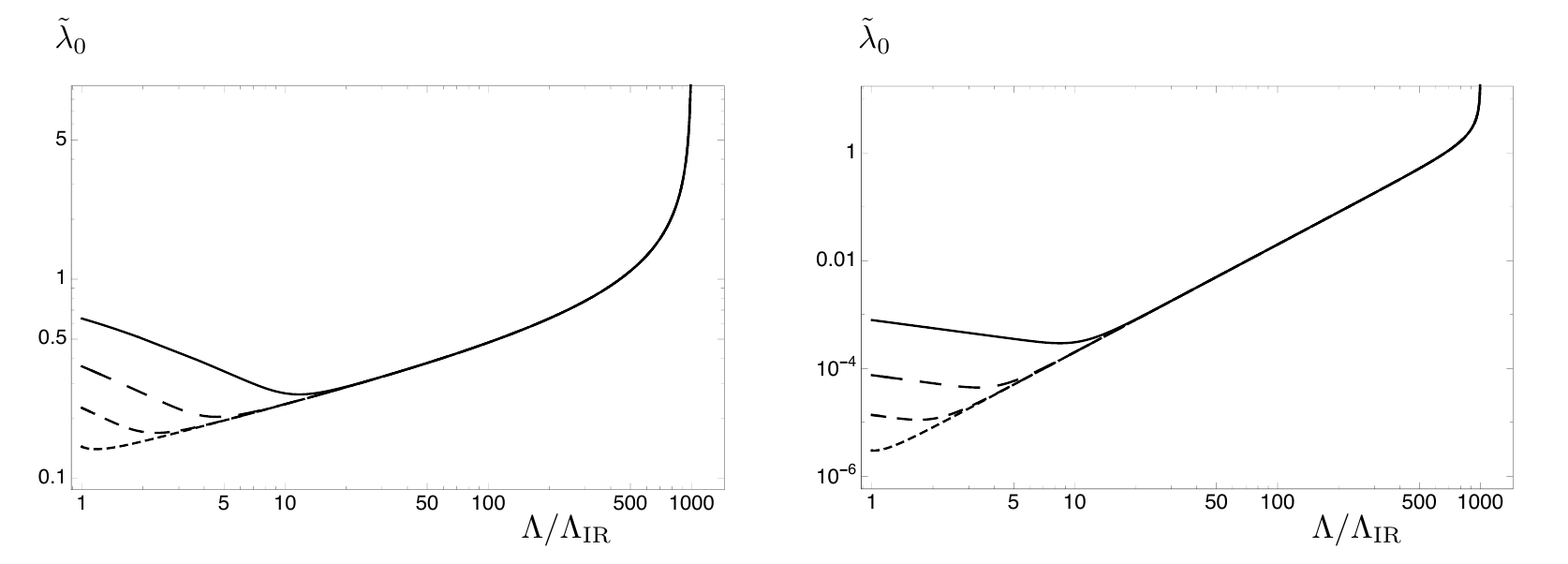}
\caption{Model I. The coupling $\tilde \lambda_0$ as a function of the energy scale $\Lambda$, for $\Delta = 2.5$ and $x_F = 2, 4.2, 4.7, 4.9$ (short-dashed, dashed, long-dashed, solid lines), and $\Delta_R = 2.75, 4.5$ (left, right panels). For each of the flows depicted, the parameter $\tilde \lambda_{0,\rm IR}$ is chosen such that the Landau pole is at $\frac{\Lambda_{\rm UV}}{\Lambda_{\rm IR}} = 10^3$.
}
\label{fig:RGModel1_3}
\end{center}
\end{figure}

As usual, when the flow is expressed as a function of the coupling in the IR, one must check whether the flow can be extrapolated up to the desired UV scale. In order to avoid a Landau pole at $\Lambda < \Lambda_{\rm UV}$, $\tilde \lambda_{0,{\rm IR}}$ is constrained to be in the range
$0 \leq |\tilde \lambda_{0,{\rm IR}}| \leq \tilde \lambda_{0,{\rm IR}}^{({\rm max})}$, with
\beqs
\label{eq:lambdamaxModelI}
	\tilde \lambda_{0,{\rm IR}}^{({\rm max})}{}^2 &=& \frac{2 \delta _R \ \Gamma \left(1+\frac{1}{2 \Delta }\right)
   \Gamma \left(1-\frac{x_F}{2 \Delta }\right) \Gamma \left(1 -\frac{x_F+2 \delta _R}{2 \Delta }\right)}
   {\Gamma\left(1-\frac{x_F-1}{2 \Delta }\right)}
   \Bigg[e^{2 \delta
   _R r_{\rm UV}} \Gamma \left(1 -\frac{x_F+2 \delta _R}{2 \Delta }\right)
   \times \nonumber \\
   && _2F_1\left(\frac{x_F}{2 \Delta },-\frac{\delta _R}{\Delta
   };1-\frac{\delta _R}{\Delta };e^{-2 \Delta r_{\rm UV}}\right)-\Gamma
   \left(1-\frac{x_F}{2 \Delta }\right) \Gamma \left(1-\frac{\delta
   _R}{\Delta }\right)\Bigg]^{-1} \,.
\eeqs
One may ask whether it is possible for the effective scaling dimension $\Delta_R^{(\rm eff)}$ to remain close to $2$ over a sufficiently large range of energies that it leads to significant deviations of the RG flow in the IR, compared that of the AdS case. It would be especially interesting if even for irrelevant deformations, when $\Delta_R > 5/2$, one can reach sizeable values of the coupling $\tilde \lambda_0$ in the IR. This leads us to consider the number of flavours close to the upper bound $x_F \simeq 2\Delta$, where the background solutions exhibit multiscale dynamics (see section~\ref{sec:models}). We have that the maximum IR value of $\tilde\lambda_0$ is
\beq
	\tilde \lambda_{0,{\rm IR}}^{({\rm max})} = 1+\frac{\psi ^{(0)}\left(-\frac{\delta _R}{\Delta
   }\right) -\psi ^{(0)}\left(\frac{1}{2 \Delta
   }\right)+B_{e^{-2 \Delta  r_{\text{UV}}}}\left(-\frac{\delta
   _R}{\Delta },0\right)}{4 \Delta } (2\Delta - x_F) + \mathcal O\left((2\Delta - x_F)^2\right) \,,
\eeq
where $\psi^{(0)}(x)$ is the digamma function and $B_\alpha(x,y)$ is the incomplete beta function. As can be seen, when the bound is saturated at $x_F = 2\Delta$, one has that $\tilde \lambda_{0,{\rm IR}}^{({\rm max})} = 1$, independent of the scaling dimension $\Delta_R$ of $\mathcal O_R$ at the UV fixed point. In the left panel of Figure~\ref{fig:RGModel1_2}, we show the effective scaling dimension $\Delta^{({\rm eff})}(\Lambda)$ as a function of the energy scale $\Lambda$ for a few different values of $x_F$. As anticipated, when $x_F \simeq 2\Delta$, one has that $\Delta^{({\rm eff})}(\Lambda) \simeq 2$ over a large range of energies. The maximum IR value $\tilde \lambda_{0,{\rm IR}}^{({\rm max})}$ of the coupling is illustrated in the right panel of Figure~\ref{fig:RGModel1_2}. It will be especially relevant in the next section, where we compute the spectrum as a function of $\tilde\lambda_{0,{\rm IR}}$. Figure~\ref{fig:RGModel1_3} shows a few examples of the RG flow for different values of $x_F$ with the parameter $\tilde \lambda_{0,\rm IR}$ chosen such that the RG flows coincide in the UV. As can be seen, proximity to the upper bound at $x_F = 2\Delta$ may lead to the multiscale dynamics responsible for generating sizeable values of $\tilde\lambda_0$ in the deep IR. While this is an interesting result, we remind the Reader that our models are less trustable when the number of flavours is large~\cite{Elander:2020nyd}.

\subsection{Model II}

The background solutions of Model~II are given in \eq{eq:Model2Bsolutions}, i.e. the warp factor is $A(r) = r + \frac{x_F}{2\Delta} \log \left( 1 - e^{-2\Delta r} \right) + \frac{1}{2\Delta_\phi} \log \left( 1 - \phi_c^2 \frac{\Delta_\phi}{3} e^{-2\Delta_\phi r} \right)$ while the profile for the flavour-singlet scalar is $\phi(r) = \sqrt{\frac{3}{\Delta_\phi}} \arctanh \left( \phi_c \sqrt{\frac{\Delta_\phi}{3}} \, e^{-\Delta_\phi r} \right)$. We remind the Reader that $H_\Psi = M_\Psi + y_5 \phi$. Hence, compared to Model~I, there are three additional parameters: $\Delta_\phi$, $\phi_c$ (governing the scaling dimension and strength of the relevant deformation $\mathcal O_\phi$, respectively), and $y_5$ (the bulk Yukawa coupling). In order to study the flow of $\tilde\lambda_0$ in the background of Model~II, we need to resort to numerics. We have confirmed that moderate values of $\Delta_\phi = 1$, $\phi_c = 1.5$, and $y_5 = 0$ leads to similar RG flows as those depicted in Figures~\ref{fig:RGModel1_1} and~\ref{fig:RGModel1_3}. In particular, as for Model~I, it is possible to obtain large deviations in the IR, compared to the AdS case, when $x_F \simeq 2\Delta$.

\begin{figure}[t]
\begin{center}
\includegraphics[width=\figwidth]{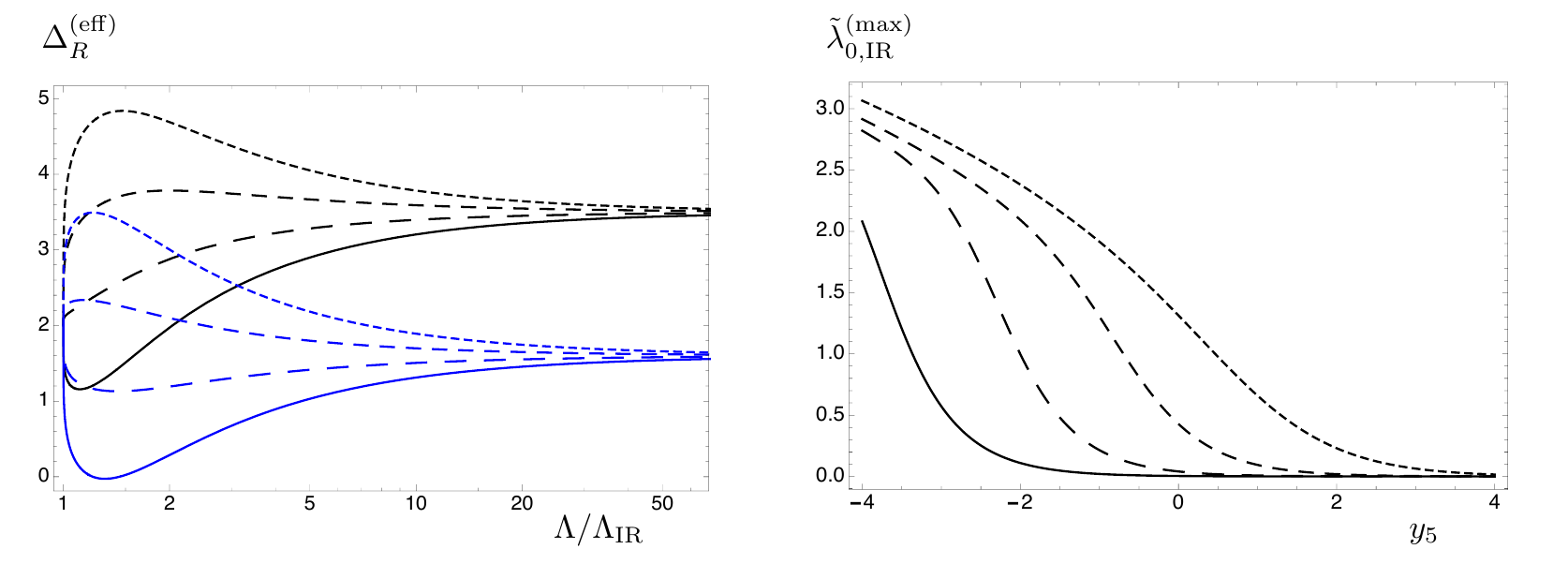}
\caption{Model II. Both panels have $\Delta = 3$, $x_F = 1$, $\Delta_\phi = 1$, and $\phi_c = 1.5$. The left panel shows the effective scaling dimension $\Delta_R^{({\rm eff})}$ as a function of the energy scale $\Lambda$, for $y_5 = 1.5, 0.5, -0.5, -1.5$ (short-dashed to solid lines), and two choices $\Delta_R = 1.6, 3.5$ (in blue, black). The right panel shows the maximum IR value $\tilde \lambda_{0,{\rm IR}}^{({\rm max})}$ as a function of $y_5$, for $\frac{\Lambda_{\rm UV}}{\Lambda_{\rm IR}} \simeq 10^3$, and $\Delta_R = 1.51, 2.5, 3, 3.4$ (short-dashed to solid lines).}
\label{fig:RGModel2B_3}
\vspace{0.4cm}
\includegraphics[width=\figwidth]{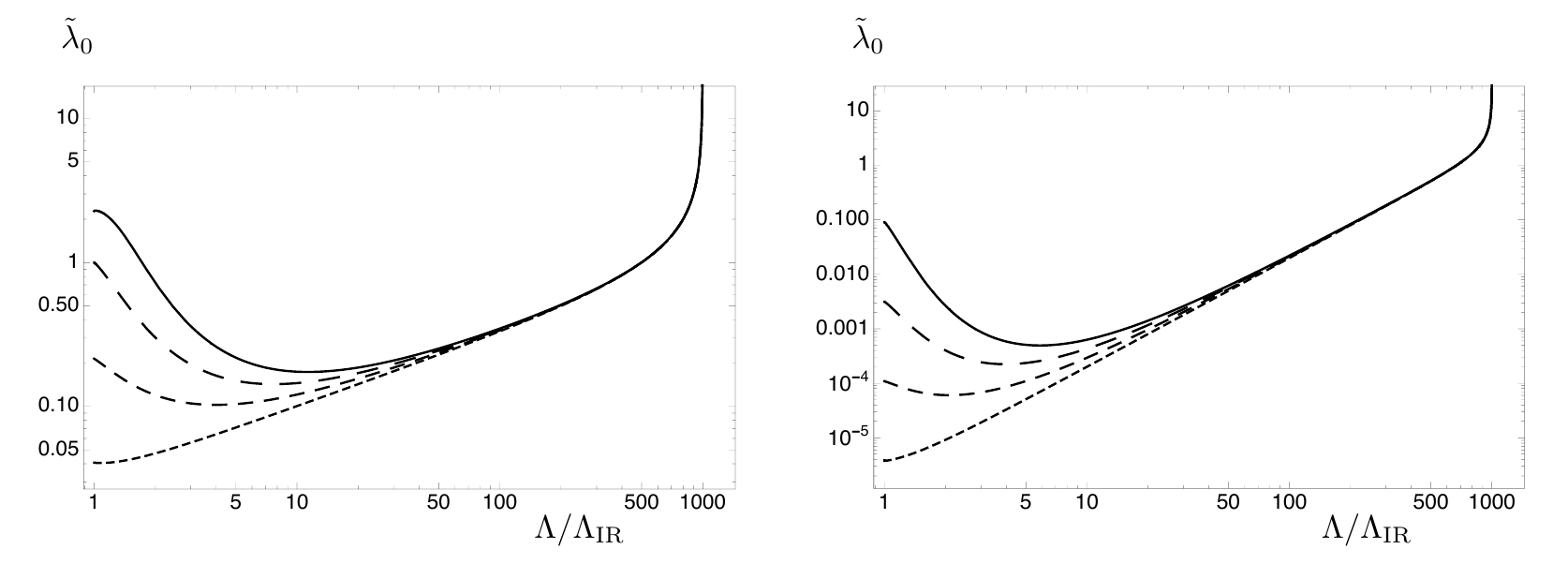}
\caption{Model II. The coupling $\tilde \lambda_0$ as a function of the energy scale $\Lambda$, for $\Delta = 3$, $x_F = 1$, $\Delta_\phi = 1$, and $\phi_c = 1.5$. In the left panel $\Delta_R = 3$ and $y_5 = 0, -1, -2, -3$ (short-dashed to solid lines), while in the right panel  $\Delta_R = 4.5$ and $y_5 = 0, -2, -4, -6$ (short-dashed to solid lines). The boundary condition for $\tilde \lambda_0$ is chosen such that, for all RG flows shown, the Landau pole is at $\frac{\Lambda_{\rm UV}}{\Lambda_{\rm IR}} \simeq 10^3$.
}
\label{fig:RGModel2B_1}
\end{center}
\end{figure}

\begin{figure}[t]
\begin{center}
\includegraphics[width=\figwidth]{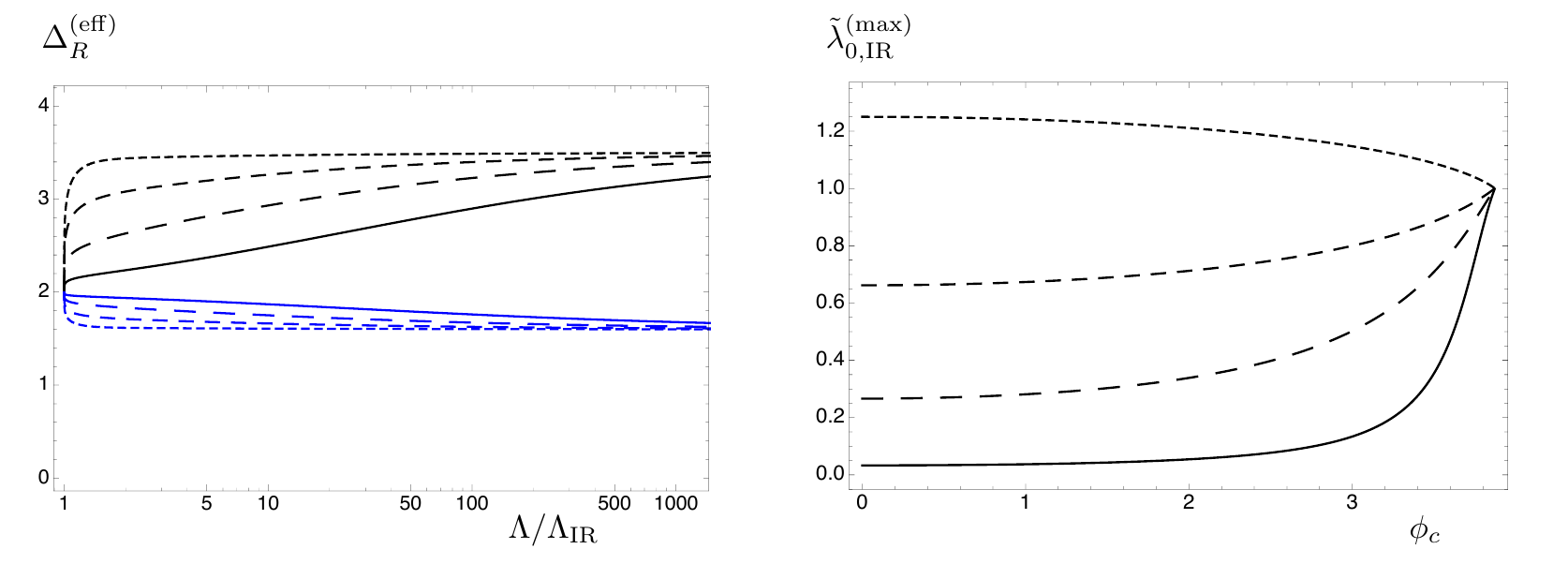}
\caption{Model II. Both panels have $\Delta = 3$, $x_F = 1$, $\Delta_\phi = 0.2$, and $y_5 = 0$. The left panel shows the effective scaling dimension $\Delta_R^{({\rm eff})}$ as a function of the energy scale $\Lambda$, for $\phi_c = 1, 2.5, 3.3, 3.7$ (short-dashed to solid lines), and two choices $\Delta_R = 1.6, 3.5$ (in blue, black). The right panel shows the maximum IR value $\tilde \lambda_{0,{\rm IR}}^{({\rm max})}$ as a function of $\phi_c$, for $\frac{\Lambda_{\rm UV}}{\Lambda_{\rm IR}} \simeq 10^3$, and $\Delta_R = 1.7, 2.3, 2.6, 3$ (short-dashed to solid lines).}
\label{fig:RGModel2B_4}
\vspace{0.4cm}
\includegraphics[width=\figwidth]{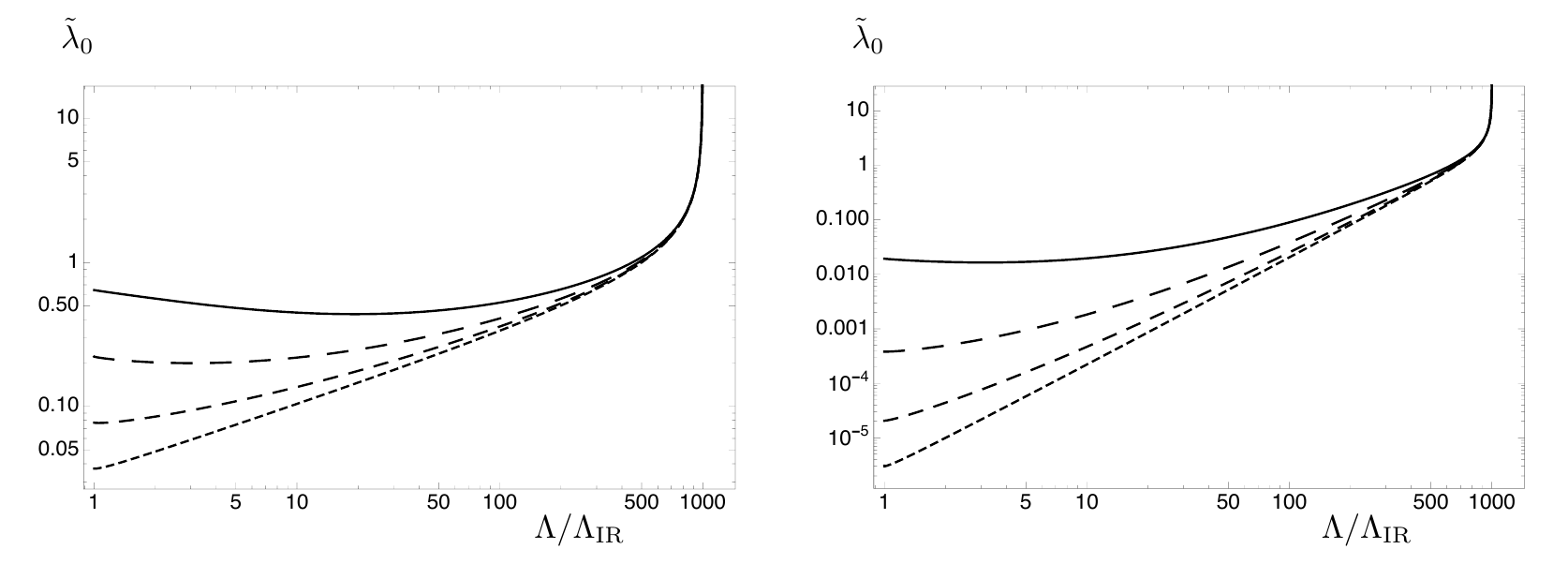}
\caption{Model II. The coupling $\tilde \lambda_0$ as a function of the energy scale $\Lambda$, for $\Delta = 3$, $x_F = 1$, $\Delta_\phi = 0.2$, $y_5 = 0$, and  $\Delta_R = 3, 4.5$ (left and right panels), for $\phi_c = 1, 2.5, 3.3, 3.7$ (short-dashed to solid lines). The maximum allowed value of $\phi_c$ is given by $\sqrt{3/\Delta_\phi} \approx 3.87$. The boundary condition for $\tilde \lambda_0$ is chosen such that, for all RG flows shown, the Landau pole is at $\frac{\Lambda_{\rm UV}}{\Lambda_{\rm IR}} \simeq 10^3$.
}
\label{fig:RGModel2B_2}
\end{center}
\end{figure}

In Model~II, it is also possible to construct such flows when the number of flavours is far from the aforementioned bound. As shown in Figure~\ref{fig:RGModel2B_3}, taking $y_5$ to be large and negative leads to the effective scaling dimension $\Delta_R^{({\rm eff})}$ being substantially smaller than its UV value $\Delta_R$ in a sizeable region (left panel), which consequently may result in a large IR value of the coupling $\tilde\lambda_{0,{\rm IR}}^{({\rm max})}$ even in the irrelevant case $\Delta_R > 5/2$ (right panel). This is further illustrated in Figure~\ref{fig:RGModel2B_1}, where we plot the coupling $\tilde\lambda_0$ as a function of energy scale $\Lambda$. Conversely, a large positive $y_5$ may suppress $\tilde\lambda_{0,{\rm IR}}^{({\rm max})}$, even when $\Delta_R < 5/2$.

Another mechanism, by which one may obtain large values of $\tilde\lambda_0$ in the IR even when $\Delta_R > 5/2$, is given by taking the scaling dimension of $\mathcal O_\phi$ to be $\Delta_\phi \leq \frac{\Delta}{2\Delta - x_F}$ together with the source $\phi_c \simeq \sqrt{3/\Delta_\phi}$ close to its upper bound (see the discussion around Eq.~\eqref{eq:scalinglargephic}). As can be seen from the left panel of Figure~\ref{fig:RGModel2B_4}, taking $\phi_c$ close to this bound pushes the effective scaling dimension towards its IR value $\Delta_R^{({\rm eff})} = 2$ over a range of energies, such that the maximum IR value of the coupling $\tilde\lambda_{0,{\rm IR}}^{({\rm max})}$ becomes large for any value of $\Delta_R$ (right panel). We illustrate a few examples of RG flows for the coupling $\tilde\lambda_0$ in Figure~\ref{fig:RGModel2B_2}.

Finally, we note the similarity between the right panels of Figures~\ref{fig:RGModel1_2} and~\ref{fig:RGModel2B_4}. In both cases, the dynamics allowing for large IR values of the coupling $\tilde\lambda_{0,{\rm IR}}^{({\rm max})}$ involves significantly deforming the geometry of the background, by pushing either $x_F$ or $\phi_c$ towards their respective upper bounds. In contrast, the dependence of $\tilde\lambda_{0,{\rm IR}}^{({\rm max})}$ on $y_5$, as seen from the right panel of Figure~\ref{fig:RGModel2B_3}, is qualitatively different: (i) the maximum IR value of the coupling does not approach $\tilde\lambda_{0,{\rm IR}}^{({\rm max})} = 1$ as a limiting case, and (ii) it is possible to suppress $\tilde\lambda_{0,{\rm IR}}^{({\rm max})}$ even in the relevant case $\Delta_R < 5/2$.

\section{Spectrum of partially composite fermions}
\label{sec:partiallycompositespectrum}

In this section, we study how the spectrum is modified by coupling the strong sector to an elementary fermion, as in section~\ref{sec:partialcompositeness}. In particular, we will show how the RG flow of $\tilde \lambda_0$ towards the IR may be used to predict when large mixing causes the spectrum to change significantly with respect to that of the isolated strongly coupled sector studied in section~\ref{sec:fermionicspectrum}.

\subsection{Two-point functions and spectrum}

Let us first consider the two-point function of $\mathcal O_R$, computed at finite cutoff $\Lambda$, in the case of $\xi=0$, such that the strongly coupled sector can be considered in isolation. The calculation proceeds along the same lines as outlined in section~\ref{sec:fermionicspectrum}, with a few important differences. We start from the assumption that the GKPW relation is valid at finite cutoff, i.e. that Eq.~\eqref{eq:GKPW-HP} holds at any value of the radial coordinate. As a consequence, we do not add counter-terms to the regularised on-shell action $\mathcal S_{\Psi,{\rm reg}}$ before differentiating with respect to the source. Also, rather than taking the limit of the UV regulator $r_2 \rightarrow \infty$ as in Eq.~\eqref{eq:OORexplicit}, since the cutoff $\Lambda(\mathfrak{r})$ is kept finite, we evaluate the expressions at a finite value of the radial coordinate $r = \mathfrak{r}$. With these considerations, the result for the two-point function is
\beq
\label{eq:OORfinitecutoff}
	\langle \mathcal O_R(q) \overline{\mathcal O}_R(-q) \rangle^{\xi|_\Lambda=0}_{\Lambda(\mathfrak{r})} =  \frac{i \, \delta^2 \mathcal S_{\Psi, \rm reg}}{\delta \overline\psi_L(-q,\mathfrak{r}) \delta \psi_L(q,\mathfrak{r})} = - \mathcal N_\Psi N_L^2 e^{5A} \frac{1}{\slashed q} \frac{\partial_r b}{b} \Big|_{\mathfrak{r}} \,.
\eeq
Note that this expression differs from that of Eq.~\eqref{eq:OORexplicit} by the absence of the term containing $\mathcal F$ that originates from the counter-term action $\mathcal S_{\rm ct}$. We also comment that, since the counter-terms are unimportant for the computation of the spectrum, for $\xi|_\Lambda = 0$, one recovers the same mass spectrum as in section~\ref{sec:fermionicspectrum} in the limit $\Lambda \rightarrow \infty$, when cutoff effects can be neglected.

In order to derive two-point functions for the case of non-zero coupling $\xi$, we write the partition function of the field theory Eq.~\eqref{eq:ZQFT1} in the presence of sources:
\beq
	 Z_{\rm QFT}[\overline J_R, J_R, \overline J_L, J_L; \Lambda(\tilde r); \xi] = \int \mathcal D \overline \chi_L \mathcal D \chi_L \mathcal D {\mathbb M}_\Lambda e^{i \mathcal S[\overline \chi_L,\chi_L,\mathbb M;\Lambda(r_2)] + i \int \dd^4x \left( \overline{\mathcal O}_R J_L + \overline \chi_L J_R + {\rm h.c.} \right)} \,.
\eeq
This has the effect of changing Eq.~\eqref{eq:ZQFT2} into
\beqs
\label{eq:ZQFT2withJs}
	&& Z_{\rm QFT}[\overline J_R, J_R, \overline J_L, J_L; \Lambda(\tilde r); \xi] = \int \mathcal D \overline \psi_L \mathcal D \psi_L Z_{\rm bulk}[\overline \psi_L(\tilde r) + \overline J_L,\psi_L(\tilde r)+J_L;\tilde r] \times \\ \nonumber
	 && \exp \left\{- i \, \mathcal N_\Psi \int \dd^4q \, \overline \psi_L(-q) \frac{i \slashed q}{\xi^2(q,\tilde r)} \psi_L(q) + i  \mathcal N_\Psi^{1/2} \int \dd^4q \left( \frac{1}{\xi(q,\tilde r)} \overline J_R(-q) \psi_L(q) + {\rm h.c.} \right) \right\} \,.
\eeqs
Using Eq.~\eqref{eq:OORfinitecutoff}, this then leads to the following two-point functions:
\beqs
\label{eq:chichiL}
	\langle \chi_L(q) \overline \chi_L(-q) \rangle_{\Lambda(\mathfrak{r})} &=& \frac{-1}{\slashed q - \xi^2(q,\mathfrak{r}) \mathcal N_\Psi^{-1} \langle \mathcal O_R(q) \overline{\mathcal O}_R(-q) \rangle_{\Lambda(\mathfrak{r})}^{\xi|_\Lambda=0}} = \frac{-\slashed q}{q^2 + \xi^2 N_L^2 e^{5A} \frac{\partial_r b}{b}} \Bigg|_\mathfrak{r} \,, \nonumber \\ \nonumber
	\langle \chi_L(q) \overline{\mathcal O}_R(-q) \rangle_{\Lambda(\mathfrak{r})} &=& \frac{- i \, \mathcal N_\Psi^{-1/2} \xi \, \langle \mathcal O_R(q) \overline{\mathcal O}_R(-q) \rangle_\Lambda^{\xi|_\Lambda=0}}{\slashed q - \xi^2(q,\mathfrak{r}) \mathcal N_\Psi^{-1} \langle \mathcal O_R(q) \overline{\mathcal O}_R(-q) \rangle_{\Lambda(\mathfrak{r})}^{\xi|_\Lambda=0}}
	= \frac{i \, \mathcal N_\Psi^{1/2} \xi N_L^2 e^{5A} \frac{\partial_r b}{b}}{q^2 + \xi^2 N_L^2 e^{5A} \frac{\partial_r b}{b}} \Bigg|_\mathfrak{r} \,, \\
	\langle \mathcal O_R(q)\overline{\mathcal O}_R(-q) \rangle_{\Lambda(\mathfrak{r})} &=& \frac{\slashed q \, \langle \mathcal O_R(q) \overline{\mathcal O}_R(-q) \rangle_\Lambda^{\xi|_\Lambda=0}}{\slashed q - \xi^2(q,\mathfrak{r}) \mathcal N_\Psi^{-1} \langle \mathcal O_R(q) \overline{\mathcal O}_R(-q) \rangle_{\Lambda(\mathfrak{r})}^{\xi|_\Lambda=0}}
	= \frac{- \mathcal N_\Psi N_L^2 e^{5A} \frac{\partial_r b}{b} \slashed q}{q^2 + \xi^2 N_L^2 e^{5A} \frac{\partial_r b}{b}} \Bigg|_\mathfrak{r} \,. \ \ \
\eeqs
Note how $\langle \chi_L(q) \overline \chi_L(-q) \rangle_{\Lambda}$ agrees with the expected result from resumming the perturbative series expansion in $\xi$. Again, these expressions for the two-point functions hold for any value of the cutoff $\Lambda(\mathfrak{r})$. Although they depend on the cutoff, as expected, the mass spectrum that can be extracted from their poles is physical and hence RG invariant. We will show explicitly that this is the case for the example of an AdS background in a moment.

In practice, it is convenient to extract the spectrum from Eqs.~\eqref{eq:chichiL} at the UV cutoff $r_{\rm UV}$ for the following reasons. In the derivative expansion of Eq.~\eqref{eq:lambdaexpansion}, we can expect the UV values of the couplings $\tilde \lambda_{i,{\rm UV}}$ to be order one, having been generated in an underlying theory with characteristic scale $\Lambda_{\rm UV}$. Provided that the resonances that we are interested have masses $m \ll \Lambda_{\rm UV}$, we can therefore safely neglect all couplings except the zeroeth order one and put $\tilde \xi_{\rm UV} = \tilde \lambda_{0,{\rm UV}}$ in our computation of the spectrum, from which it follows that one should impose the UV boundary condition
\beq
\label{eq:UVBCPC}
	q^2 + \xi^2 N_L^2 e^{5A} \frac{\partial_r b}{b} \Big|_{r_{\rm UV}} \simeq q^2 + \Lambda^2 \tilde \lambda_0^2 \frac{\Lambda \partial_\Lambda b}{b} \Big|_{\Lambda_{\rm UV}} = 0 \,.
\eeq
In the limit of $\tilde \lambda_{0,{\rm UV}}^2 \rightarrow 0$, this boundary condition reproduces the prescription $b |_{r_{\rm UV}} = 0$, used for computing the spectrum in the strongly coupled sector without a dynamical source. Conversely, in the limit $\tilde \lambda_{0,{\rm UV}}^2 \rightarrow \infty$, one obtains the Neumann boundary condition $\partial_r b |_{r_{\rm UV}} = 0$, which reproduces the prescription given in Ref.~\cite{Contino:2004vy} for a dynamical source. Following the reasoning of appendix~\ref{sec:masslesspoles}, one can show that, for non-zero coupling $\tilde \lambda$, the presence of massless poles is the opposite of what was the case with a non-dynamical source: for the $(+)$ IR boundary condition, a massless pole is present, whereas for $(-)$ IR boundary condition, there is no massless state. Furthermore, we note that in this section, we will compute the decay constant by using the finite cutoff version of Eq.~\eqref{eq:decayconstant}, i.e.
\beq
	f^2 = 2 N_C \frac{e^{2A}}{g_5^2} \frac{\partial_r a}{a} \Big|_{q^2 = 0, r = r_{\rm UV}} \,.
\eeq

Finally, we comment on a relation between the spectra of the strongly coupled sector in isolation and the partially composite spectrum at infinite coupling $\tilde \lambda_{0,{\rm UV}}$. In the range $\frac{3}{2} < \Delta_R < \frac{5}{2}$, these two spectra are identical provided that one makes the replacements $H_\Psi \rightarrow -H_\Psi$ together with $(\pm) \rightarrow (\mp)$ for the IR boundary conditions (as well as use same UV cutoff $\Lambda_{\rm UV}$ in the two computations). This property of the spectra follows by making use of Eq.~\eqref{eq:bRLratio}, which relates the ratio $\partial_r b/b$ for different solutions $b$ of the equation of motion~\eqref{eq:eomb}. For our models in particular, the first of the two replacements implies that one should take $\Delta_R \rightarrow 4 - \Delta_R$ for the scaling dimension of $\mathcal O_R$ in the UV, as well as $y_5 \rightarrow -y_5$ for the Yukawa coupling present in Model~II. We will return to these observations later, when we discuss the spectrum in more detail.

\subsection{Low energy effective actions}

We will now write the low energy effective actions that contain up to and including the lightest massive fermionic resonance. Consider first the case of $(-)$ IR boundary condition. This implies that the strongly coupled theory in isolation is chiral, and contains a massless state. When coupling it to the elementary fermion $\chi_L$, it combines with $\mathcal O_R$ to form a vector, lifting the mass of this state. The denominator appearing in Eqs.~\eqref{eq:chichiL} can be expanded close to the pole of the lightest state (with mass $m$) as
\beq
	q^2 + \xi^2 N_L^2 e^{5A} \frac{\partial_r b}{b} \Big|_\mathfrak{r} \sim (q^2 + m^2) + \cdots \,.
\eeq
We may now write the low energy action
\beq
\label{eq:effectiveactionm}
	\mathcal S_{\rm eff}^{(-)} = \int \dd^4q \, \Big[ - \overline \chi_L(-q) i \slashed q \chi_L(q) - \overline{\mathcal X}_R(-q) i \slashed q \mathcal X_R(q) + \left( f \lambda_\chi \, \overline{\mathcal X}_R(-q) \chi_L(q) + {\rm h.c.} \right) \Big] \,,
\eeq
which reproduces the two-point functions of Eq.~\eqref{eq:chichiL}, close to $-q^2 = m^2$, provided one makes the identification $\lambda_\chi = \frac{m}{f} = \frac{m}{\sqrt{N_C} \tilde f}$ together with
\beq
	\mathcal O_R \longleftrightarrow {\rm sign}(\xi) e^{\frac{5A}{2}} N_L \mathcal N_\Psi^{1/2} \sqrt{\frac{\partial_r b}{b}} \mathcal X_R \,.
\eeq

Similarly, in the case of $(+)$ IR boundary conditions, the low energy effective action can be written as
\beqs
	\mathcal S_{\rm eff}^{(+)} = \int \dd^4q \, \Big[ \hspace{-0.4cm} && - \overline \chi_L(-q) i \slashed q \chi_L(q) - \overline{\mathcal X}_L(-q) i \slashed q \mathcal X_L(q) - \overline{\mathcal X}_R(-q) i \slashed q \mathcal X_R(q) \nonumber\\ &&
	+ \left( m_0 \, \overline{\mathcal X}_R(-q) \mathcal X_L(q) + f \lambda_\chi \, \overline{\mathcal X}_R(-q) \chi_L(q) + {\rm h.c.} \right) \Big] \,,
\label{eq:effectiveactionp}
\eeqs
where $m^2 = m_0^2 + f^2 \lambda_\chi^2$ with $m_0$ the mass of the lightest resonance in the isolated strongly-coupled theory, and, as expected, a massless state is also present.

The coupling $\lambda_\chi$, appearing in the low energy effective actions of Eqs.~\eqref{eq:effectiveactionm} and~\eqref{eq:effectiveactionp}, is related to the IR value $\tilde \lambda_{0,{\rm IR}}$ of the coupling $\tilde \lambda_{0}$ studied in section~\ref{sec:partialcompositeness}, though the two are not identical since $\tilde \lambda_{0,{\rm IR}}$ is normalised in units of $\Lambda_{\rm IR}$ rather than $f$, and also because of the fact that it appears as the lowest order in a derivative expansion which strictly speaking only is valid for $Q = \sqrt{-q^2} \ll \Lambda_{\rm IR}$. Nevertheless, we will see that $\tilde \lambda_{0,{\rm IR}}$ captures well the qualitative features of the spectrum, namely the size of the mixing of the elementary fermion with the strongly coupled sector, and since it is also more convenient to work with in the holographic description, we hence use it, rather than $\lambda_\chi$, to parameterize our results for the partially composite spectra.

\subsection{AdS background}

As in subsections~\ref{sec:AdS} and~\ref{sec:RGAdS}, we introduce a hard-wall IR cutoff at $r = 0$, in order to obtain a mass gap. Furthermore, we parametrize our results for the spectrum in terms of the scaling dimension $[\mathcal O_R] = \Delta_R$ and the IR value of the coupling $\tilde \lambda_{0}(\Lambda_{\rm IR}) = \tilde \lambda_{0,{\rm IR}}$.

\begin{figure}[t]
\begin{center}
\includegraphics[width=\figwidth]{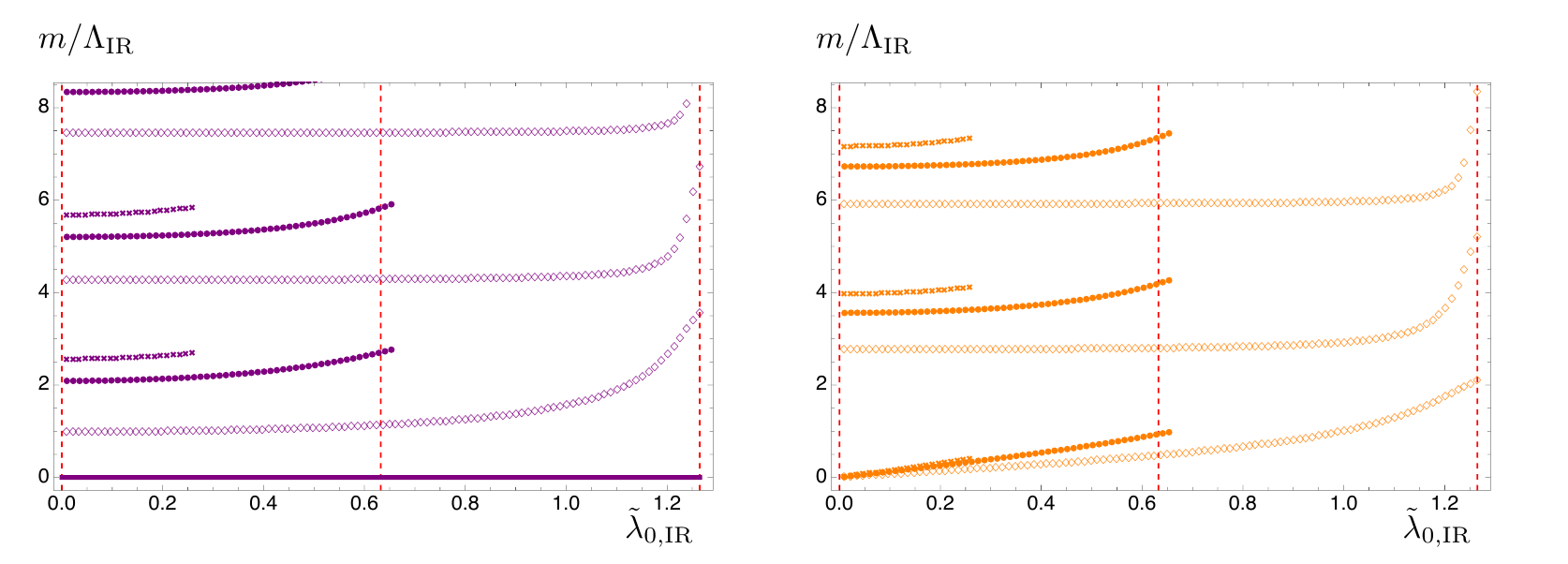}
\caption{AdS background. Fermionic spectrum as a function of $\tilde \lambda_{0,{\rm IR}}$ for $\Delta_R = 1.7, 2.3, 2.6$ (diamonds, dots, crosses). The corresponding IR fixed point values ($\tilde \lambda_{0} = \sqrt{5 - 2\Delta_R}$ when $\Delta_R < 2.5$ and $\tilde \lambda_{0} = 0$ when $\Delta_R \geq 2.5$) are indicated by vertical dashed lines. The left and right panels correspond to imposing $(+)$ and $(-)$ IR boundary conditions, respectively. The solid line in the left panel indicates the presence of a massless state. The ratio between the UV and IR energy scales is equal to $\frac{\Lambda_{\rm UV}}{\Lambda_{\rm IR}} = 10^3$.}
\label{fig:PCAdS_1}
\vspace{0.4cm}
\includegraphics[width=\figwidth]{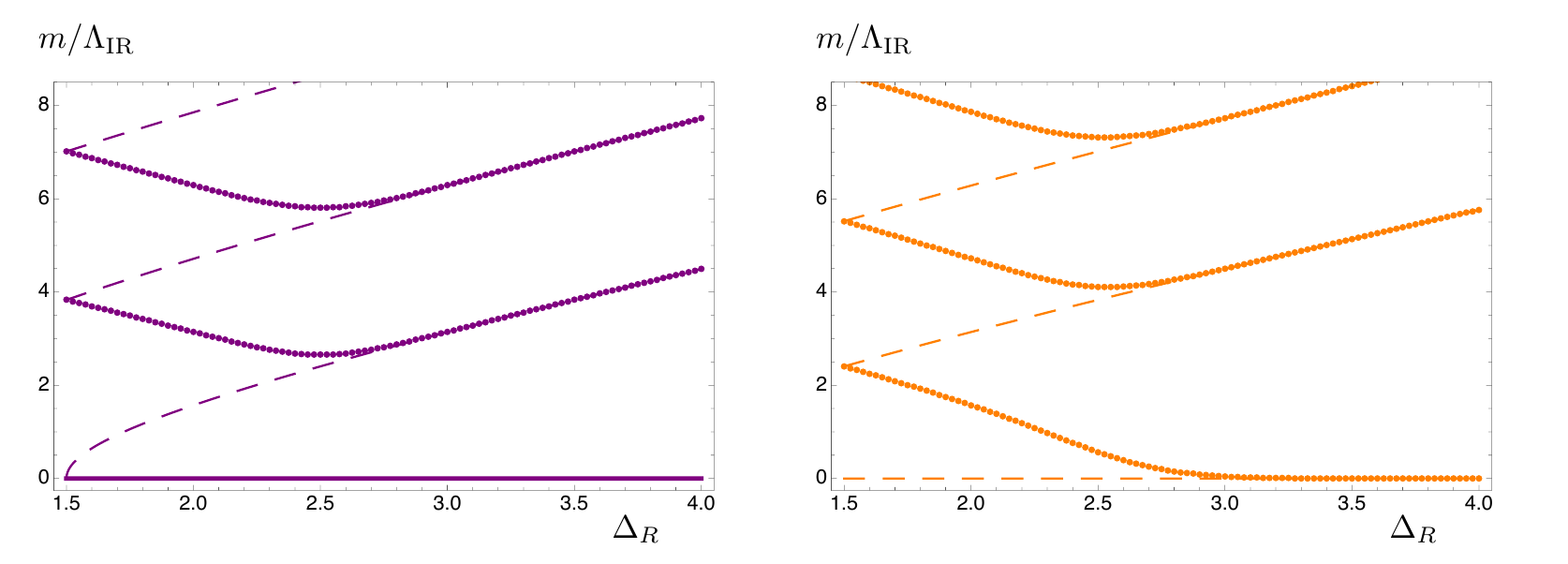}
\caption{AdS background. Fermionic spectrum as a function of $\Delta_R$ for $\tilde \lambda_{0,{\rm IR}} = \tilde \lambda_{0,{\rm IR}}^{({\rm max})}$. The left and right panels correspond to imposing $(+)$ and $(-)$ IR boundary conditions, respectively. The solid line in the left panel indicates the presence of a massless state. The ratio between the UV and IR energy scales is equal to $\frac{\Lambda_{\rm UV}}{\Lambda_{\rm IR}} = 10^3$. The dashed lines indicate the spectrum of the strongly coupled sector in isolation, computed in the limit $\Lambda_{\rm UV} \rightarrow \infty$, and is the same as the one shown in Figure~\ref{fig:AdSmp}.}
\label{fig:PCAdS_2}
\end{center}
\end{figure}

Figure~\ref{fig:PCAdS_1} shows the spectrum as a function of $\tilde \lambda_{0,{\rm IR}}$ for a few different values of $\Delta_R$. The maximum value of the coupling $\tilde \lambda_{0,{\rm IR}}^{\rm (max)}$ depends on $\Delta_R$ and was given by Eq.~\eqref{eq:lambdaIRmax} after making the requirement that the RG flow does not result in a Landau pole at a scale below $\Lambda_{\rm UV}$. When $\Lambda_{\rm UV}$ is large and finite, this implies that one can reach values of $\tilde \lambda_{0,{\rm IR}}$ that are slightly above the IR fixed point. For $(+)$ IR boundary condition, the strongly coupled sector in isolation is vector-like. In the partially composite case, the elementary fermion mixes with the composite states, however there is always a massless state in the theory which will be more or less elementary depending on the strength of $\tilde \lambda_{0,{\rm IR}}$. Conversely, for $(-)$ IR boundary condition, the strongly coupled sector in isolation is chiral-like, and contains an exactly massless state. As the strength of $\tilde \lambda_{0,{\rm IR}}$ is increased, the elementary fermion pairs up with this mode, lifting the mass of the lightest state. We also note that in the irrelevant case, when $\Delta_R > \frac{5}{2}$, the maximum value of $\tilde \lambda_{0,{\rm IR}}$ is suppressed by the factor $\left(\Lambda_{\rm IR}/\Lambda_{\rm UV} \right)^{\Delta_R - \frac{5}{2}}$ appearing in Eq.~\eqref{eq:lambdamaxAdSirrelevant}, such that for large UV cutoff $\Lambda_{\rm UV}$ the low energy spectrum remains almost identical to that of the strongly coupled sector in isolation (with the exception of the presence or not of a massless state).

From Figure~\ref{fig:PCAdS_1}, we can also observe how the spectra of the composite and partially composite cases are related to each other in the range $\frac{3}{2} < \Delta_R < \frac{5}{2}$. The spectrum of the strongly coupled sector in isolation is given by taking $\tilde \lambda_{0,{\rm IR}} \rightarrow 0$, and removing/adding the massless state for $(+)/(-)$ IR boundary condition. This spectrum is identical to that for the maximum value of $\tilde \lambda_{0,{\rm IR}}$, after making the replacement $\Delta_R \rightarrow 4 - \Delta_R$ while also switching the IR boundary condition $(\pm) \rightarrow (\mp)$. In particular, the diamonds ($\Delta_R = 1.7$) of the left panel for $\tilde \lambda_{0,{\rm IR}} = 0$ are the same as the dots ($\Delta_R = 2.3$) of the right panel for $\tilde \lambda_{0,{\rm IR}} = \tilde \lambda_{0,{\rm IR}}^{\rm (max)}$, and vice versa.

Figure~\ref{fig:PCAdS_2} shows the spectrum as a function of the scaling dimension $\Delta_R$ for the maximum value of the coupling $\tilde \lambda_{0,{\rm IR}}$. We have superimposed the spectrum of the strongly coupled sector in isolation, previously displayed in Figure~\ref{fig:AdSmp}, in order to make apparent the effect of turning on the maximum allowed mixing between the elementary and composite sectors. When $\Delta_R > \frac{5}{2}$, the coupling $\tilde \lambda_0$ is irrelevant, and hence for large UV cutoff $\Lambda_{\rm UV}$ the massive spectrum is approximately the same as the case of the composite sector in isolation, with a larger deviation close to the marginal case $\Delta \simeq \frac{5}{2}$. However, in the relevant case $\Delta_R < \frac{5}{2}$, the spectrum changes drastically. In particular, for the $(-)$ IR boundary condition, the previously massless state, present due to the chiral nature of the theory, is lifted as it mixes with the elementary fermion, with a resulting mass that becomes progressively larger as one approaches the free fermion case, $\Delta_R \simeq \frac{3}{2}$.

Before discussing the results for the spectra of Models~I and~II, let us take a slight detour to demonstrate that the mass spectrum indeed is RG invariant; in other words that if one takes into account the RG flow of the couplings, the locations of the poles in the correlators of Eq.~\eqref{eq:chichiL} do not depend on the cutoff $\Lambda$. To show this, we use the analytical solution for $\tilde \xi$ given in Eq.~\eqref{eq:xisolAdS}, and focus on the two-point function $\langle \chi_L(q) \overline \chi_L(-q) \rangle$:
\beq
	\langle \chi_L(q) \overline \chi_L(-q) \rangle_\Lambda = \frac{-\slashed q}{q^2 + \tilde \xi^2 \Lambda^{2} \, \frac{\Lambda \partial_\Lambda b}{b}} \,.
\eeq
In order to proceed we use Eq.~\eqref{eq:dbLdivbL} which we reproduce here for convenience:
\beq
	\frac{\Lambda \partial_\Lambda b}{b} = - \frac{Q}{\Lambda} \frac{J_{\delta_R}(\Lambda^{-1}Q) - c_\pm(Q) Y_{\delta_R}(\Lambda^{-1}Q)}{J_{\delta_R+1}(\Lambda^{-1}Q) - c_\pm(Q) Y_{\delta_R+1}(\Lambda^{-1}Q)} \,,
\eeq
where the integration constant $c_\pm(Q)$ depends on the IR boundary condition and is equal to
\beq
	c_-(Q) = \frac{J_{\delta_R+1}(Q)}{Y_{\delta_R+1}(Q)} \,, \hspace{1cm} c_+(Q) = \frac{J_{\delta_R}(Q)}{Y_{\delta_R}(Q)} \,.
\eeq
A direct calculation, making use of Eq.~\eqref{eq:xisolAdS}, now yields that
\beqs
	&& \langle \chi_L(q) \overline \chi_L(-q) \rangle_\Lambda = \nonumber \\
	&& \frac{\slashed q}{c_\pm(Q) - c_\xi(Q)} \Bigg[ \frac{\pi}{2} \Lambda^{-1} Q^{-1} \left( J_{\delta_R+1}(\Lambda^{-1}Q) - c_\pm(Q) Y_{\delta_R+1}(\Lambda^{-1}Q) \right) \times \\ \nonumber
	&& \hspace{3cm} \left( J_{\delta_R}(\Lambda^{-1}Q) - c_\xi(Q) Y_{\delta_R}(\Lambda^{-1}Q) \right) \Bigg] \,.
\eeqs
The (massive) pole structure follows completely from the very first factor, and leads to the condition $c_\pm(Q) = c_\xi(Q)$, which as advertised does not depend on the cutoff $\Lambda$.

\subsection{Model I}

\begin{figure}[t]
\begin{center}
\includegraphics[width=\figwidth]{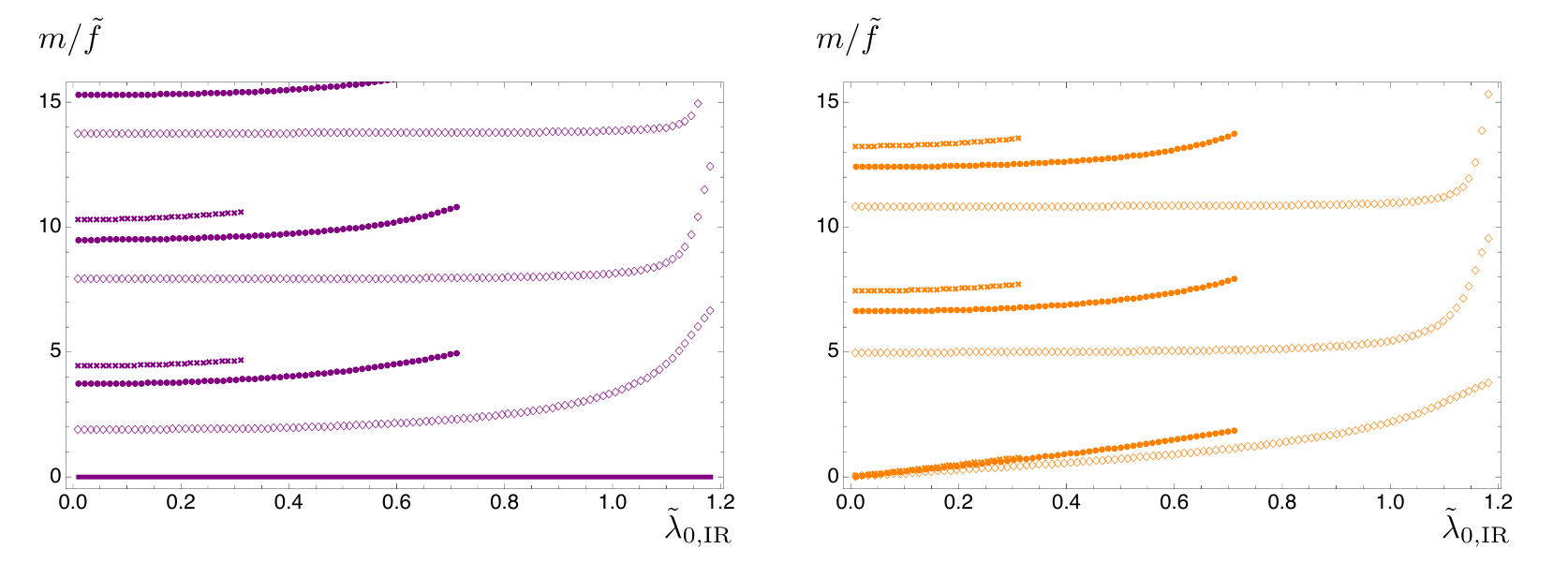}
\caption{Model I. Fermionic spectrum as a function of $\tilde \lambda_{0,{\rm IR}}$ for $\Delta = 2.5$, $x_F = 3$, $g_5 = 8$, and $\Delta_R = 1.7, 2.3, 2.6$ (diamonds, dots, crosses). The left and right panels correspond to imposing $(+)$ and $(-)$ IR boundary conditions, respectively. The solid line in the left panel indicates the presence of a massless state. The ratio between the UV and IR energy scales is equal to $\frac{\Lambda_{\rm UV}}{\Lambda_{\rm IR}} = 10^3$. We used the IR regulator $r_1 = 10^{-12}$ in the numerics.}
\label{fig:PCModel1_1}
\vspace{0.4cm}
\includegraphics[width=\figwidth]{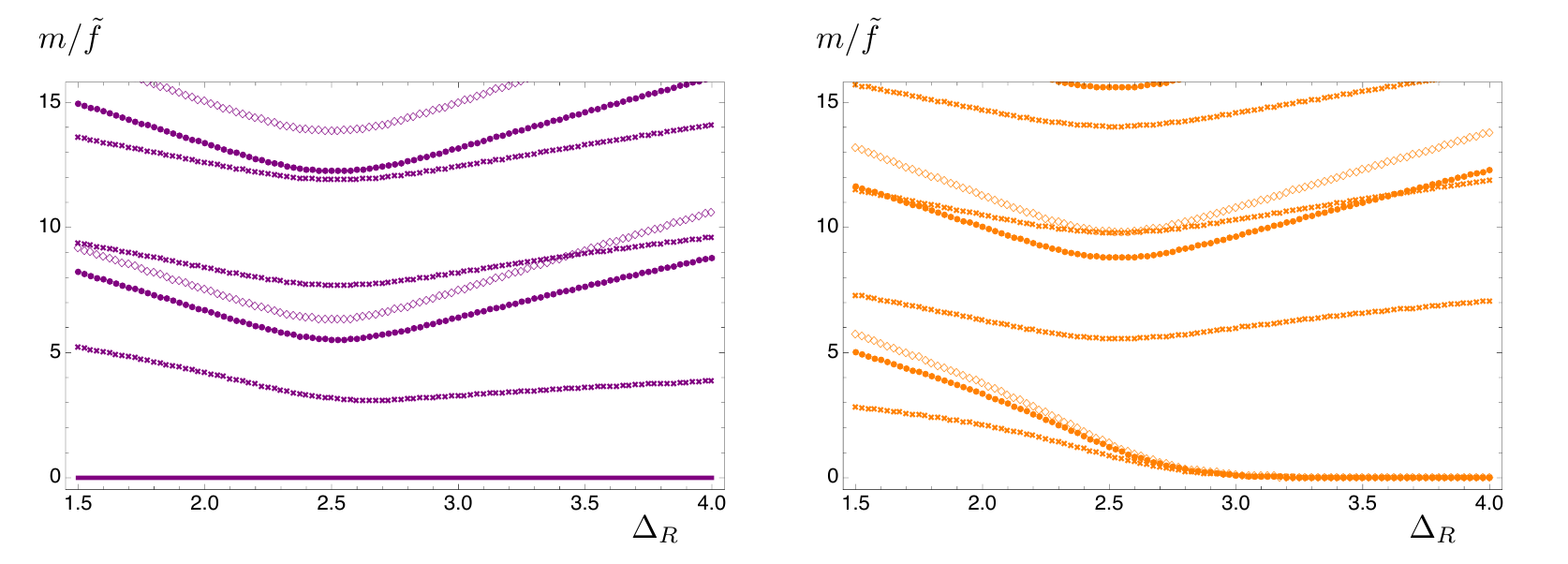}
\caption{Model I. Fermionic spectrum as a function of $\Delta_R$ for $\Delta = 2.5$, $g_5 = 8$, $\tilde \lambda_{0,{\rm IR}} = \tilde \lambda_{0,{\rm IR}}^{({\rm max})}$, and $x_F = 0.5, 2, 4$ (diamonds, dots, crosses). The left and right panels correspond to imposing $(+)$ and $(-)$ IR boundary conditions, respectively. The solid line in the left panel indicates the presence of a massless state. The ratio between the UV and IR energy scales is equal to $\frac{\Lambda_{\rm UV}}{\Lambda_{\rm IR}} = 10^3$. We used the IR regulator $r_1 = 10^{-12}$ in the numerics.}
\label{fig:PCModel1_2}
\end{center}
\end{figure}

In Figure~\ref{fig:PCModel1_1}, we show the spectrum as a function of $\tilde \lambda_{0,{\rm IR}}$ for a few different values of the scaling dimension $\Delta_R$ of $\mathcal O_R$ at the UV fixed point. Notice that these numerical results allow to extract the coupling $\lambda_\chi$, appearing in the low energy effective action of 
Eq.~\eqref{eq:effectiveactionm}, as a function of $\tilde\lambda_{0,\rm IR}$: the lightest massive state in the right panel of the figure is given by $m/\tilde{f} = \lambda_\chi N_C^{1/2}$. Similarly, the parameters in the action of Eq.~\eqref{eq:effectiveactionp} are related to the lightest massive state in the left panel, according to $m/\tilde{f}=(m_0^2/f^2 +\lambda_\chi^2)^{1/2}N_C^{1/2}$. The maximum allowed value of the coupling $\tilde \lambda_{0,{\rm IR}}^{\rm (max)}$ is given in Eq.~\eqref{eq:lambdamaxModelI}, which implies a smaller range of $\tilde \lambda_{0,{\rm IR}}$ when $\Delta_R$ has larger scaling dimension. The resulting spectrum is qualitatively similar to that shown in Figure~\ref{fig:PCAdS_1} for an AdS background. This is consistent with the fact that, since we have chosen moderate values of $x_F = 3$ and $\Delta = 2.5$, the running of $\tilde \lambda_0$ depicted in Figure~\ref{fig:RGModel1_1} only shows small deviations from that of the AdS case in Figure~\ref{fig:RGAdS}.

Figure~\ref{fig:PCModel1_2} shows the spectrum as a function of the scaling dimension $\Delta_R$ for a few different values of $x_F$, and with the maximum allowed value of the coupling $\tilde \lambda_{0,{\rm IR}} = \tilde \lambda_{0,{\rm IR}}^{\rm (max)}$. Comparing with Figure~\ref{fig:Model1F_1}, we see that for $\Delta_R > \frac{5}{2}$, the spectrum remains virtually unchanged with respect to that of the strongly coupled sector in isolation. Conversely, for $\Delta_R < \frac{5}{2}$, the mixing between the elementary and strongly coupled sector becomes large, significantly modifying the spectrum. In particular, the mass of the lightest state, which was previously massless for the $(-)$ IR boundary condition, is lifted and becomes of the same order as the mass gap close to $\Delta_R \simeq \frac{3}{2}$. While all this is qualitatively in agreement with the results for an AdS background, shown in Figure~\ref{fig:PCAdS_2}, the spectrum depends non-trivially on the number of flavours, resulting in a lower mass gap (compared to $\tilde f$) and a more densely packed tower of states, as $x_F$ is increased. We recall that this also was the case for the strongly coupled sector in isolation, as apparent from Figure~\ref{fig:Model1F_2}, when the number of flavours is taken to be close to the upper bound at $x_F = 2\Delta$.

\begin{figure}[t]
\begin{center}
\includegraphics[width=\figwidth]{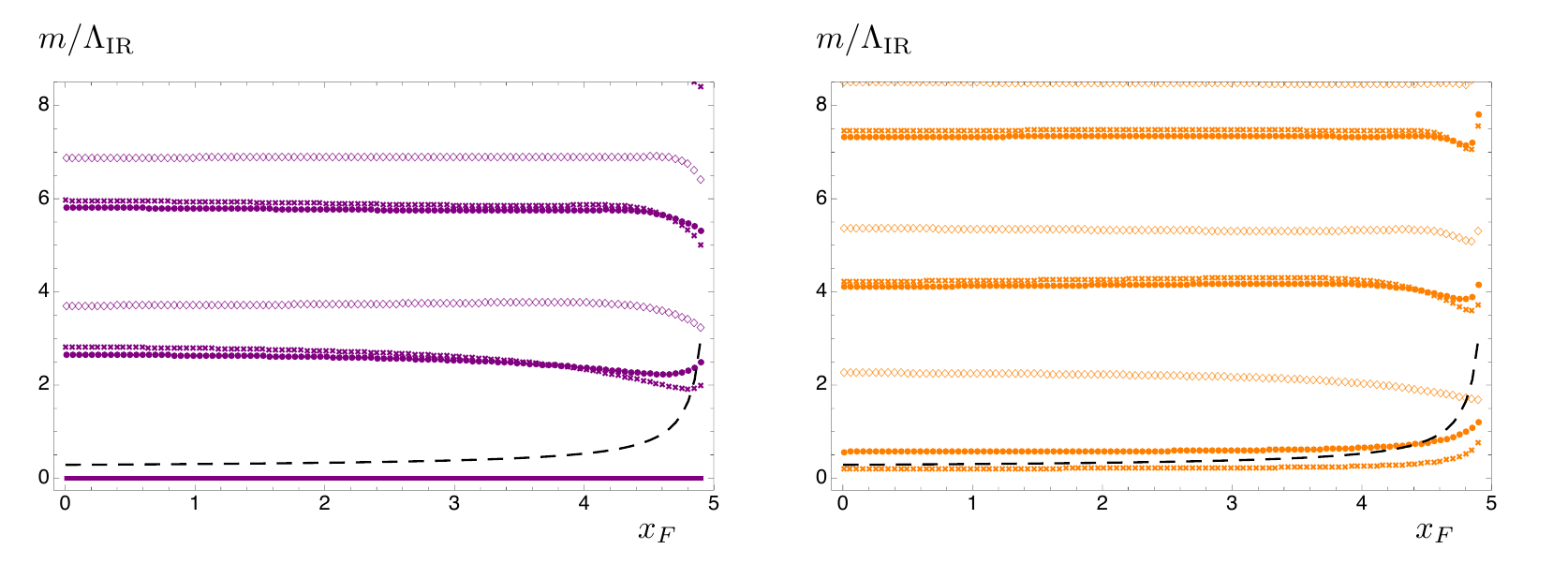}
\caption{Model I. Fermionic spectrum as a function of $x_F$ for $\Delta = 2.5$, $g_5 = 8$, $\tilde \lambda_{0,{\rm IR}} = \tilde \lambda_{0,{\rm IR}}^{({\rm max})}$, and $\Delta_R = 1.6, 2.5, 2.75$ (diamonds, dots, crosses). The left and right panels correspond to imposing $(+)$ and $(-)$ IR boundary conditions, respectively. The solid line in the left panel indicates the presence of a massless state. The decay constant $\tilde f$ is represented by the dashed black line. The ratio between the UV and IR energy scales is equal to $\frac{\Lambda_{\rm UV}}{\Lambda_{\rm IR}} = 10^3$. In the numerics, we used the IR regulator $r_1 = 10^{-12}$ for $x_F \leq 4.5$, while for $x_F > 4.5$ we used $r_1 = 10^{-30}$ in order to minimize the cutoff effects. Even though, for large $x_F \simeq 4.9$, these unphysical effects are visible for the heavy states, we still chose to display the results of the numerical calculation, since we checked that lightest states are less sensitive to the position of the IR regulator.}
\label{fig:PCModel1_3}
\end{center}
\end{figure}

We further explore the flavour dependence of the partially composite spectrum in Figure~\ref{fig:PCModel1_3}. In order to obtain the largest possible effect, we choose $\tilde \lambda_{0,{\rm IR}} = \tilde \lambda_{0,{\rm IR}}^{({\rm max})}$ to be its maximum possible value. As was shown in Figure~\ref{fig:RGModel1_2}, this value of the coupling in the IR can be made to be large close to $x_F \simeq 2\Delta$ even when the scaling dimension $\Delta_R > \frac{5}{2}$. This was also illustrated in Figure~\ref{fig:RGModel1_3}, which depicts the running of $\tilde \lambda_0$ with energy scale for different values of $x_F$. Keeping in mind that, close to the upper bound for $x_F$, the spectrum becomes densely packed in units of $\tilde f$, we instead plot the spectrum in units of the IR scale $\Lambda_{\rm IR}$. The result, shown in Figure~\ref{fig:PCModel1_3}, is consistent with what might be expected from the study of the RG flow of $\tilde \lambda_0$, namely that for $x_F \simeq 2\Delta$ close to its upper bound, it is possible to lift the mass of the light state, present for $(-)$ IR boundary condition, even when $\Delta_R > \frac{5}{2}$. This can be interpreted as that there is significant mixing between the elementary and composite sectors, and shows that despite the coupling $\tilde \lambda_0$ being irrelevant, the details of the IR physics, captured by the bulk geometry near the end of space, may lead to results that deviate to a large extent from models based on an AdS background.

\subsection{Model II}

For Model~II, it is possible to obtain a large IR value of the coupling $\tilde \lambda_0$ even when it is irrelevant, in two additional ways compared to Model~I. As shown in section~\ref{sec:partialcompositeness}, this may happen when either (i) the Yukawa coupling $y_5$ is large and negative, illustrated in Figure~\ref{fig:RGModel2B_1}, or (ii) the scaling dimension of $\mathcal O_\phi$ is $\Delta_\phi \lesssim \frac{\Delta}{2\Delta - x_F}$ and its source $\phi_c \leq \sqrt{3/\Delta}$ is large, illustrated in Figure~\ref{fig:RGModel2B_2}. In our study of the partially composite spectrum of Model~II, we therefore choose to concentrate on these two different cases. Again, we choose $\tilde \lambda_{0,{\rm IR}} = \tilde \lambda_{0,{\rm IR}}^{({\rm max})}$ to be its maximum possible value, in order to obtain the largest possible effect on the spectrum.

\begin{figure}[t]
\begin{center}
\includegraphics[width=\figwidth]{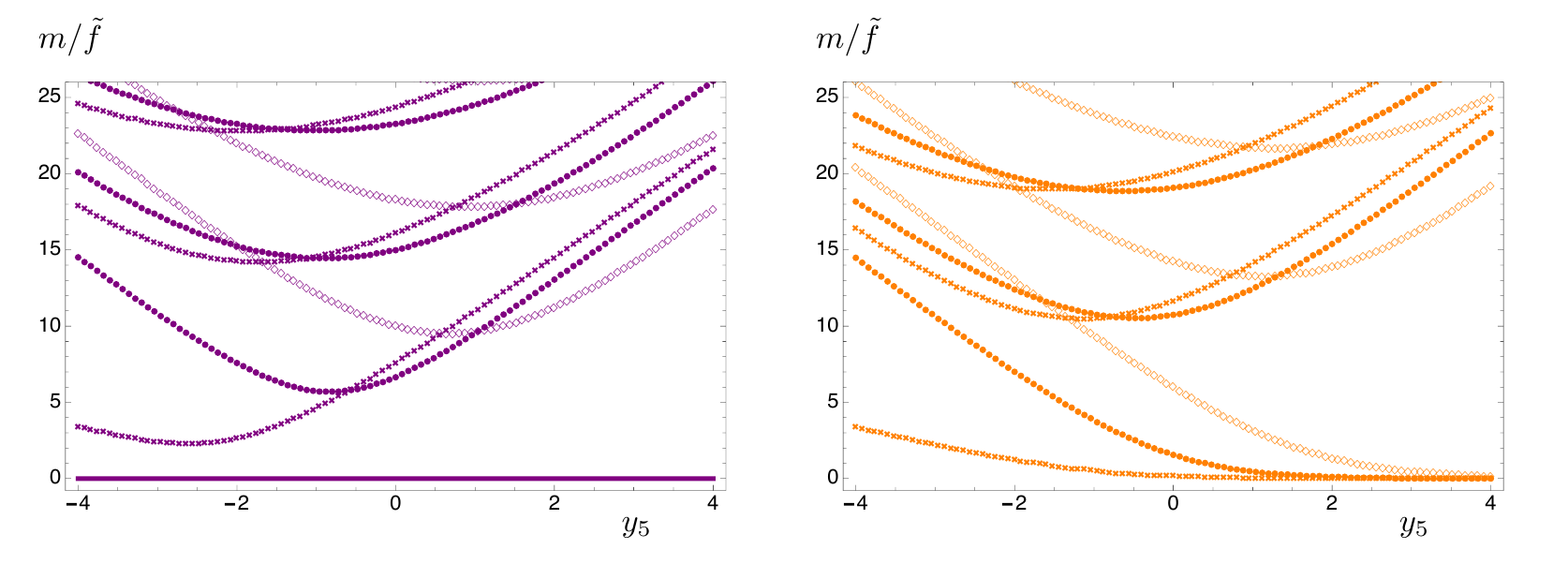}
\caption{Model II. Fermionic spectrum as a function of $y_5$ for $\Delta = 3$, $x_F = 1$, $g_5 = 8$, $\Delta_\phi = 1$, $\phi_c = 1.5$, $\tilde \lambda_{0,{\rm IR}} = \tilde \lambda_{0,{\rm IR}}^{({\rm max})}$, and $\Delta_R = 1.51, 2.5, 3$ (diamonds, dots, crosses). The left and right panels correspond to imposing $(+)$ and $(-)$ IR boundary conditions, respectively. The solid line in the left panel indicates the presence of a massless state. The ratio between the UV and IR energy scales is equal to $\frac{\Lambda_{\rm UV}}{\Lambda_{\rm IR}} = 10^3$. We used the IR regulator $r_1 = 10^{-12}$ in the numerics.}
\label{fig:PCModel2B_1}
\vspace{0.4cm}
\includegraphics[width=\figwidth]{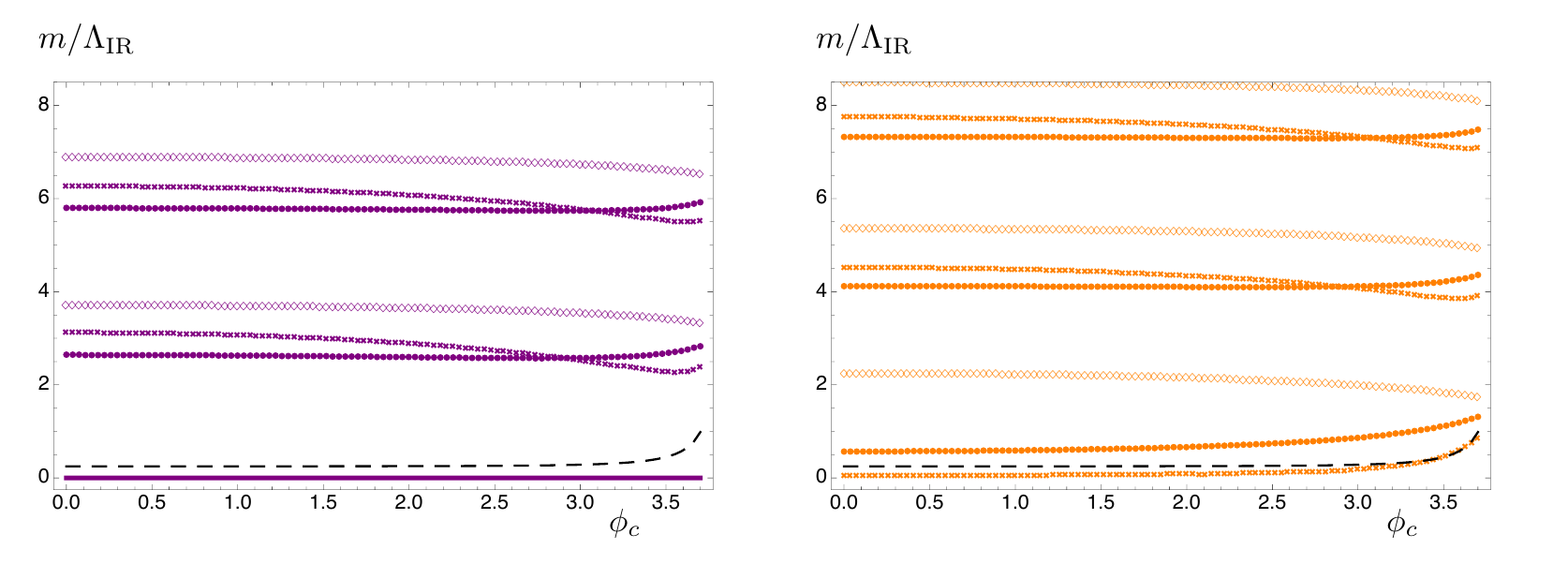}
\caption{Model II. Fermionic spectrum as a function of $\phi_c$ for $\Delta = 3$, $x_F = 1$, $g_5 = 8$, $\Delta_\phi = 0.2$, $y_5 = 0$, $\tilde \lambda_{0,{\rm IR}} = \tilde \lambda_{0,{\rm IR}}^{({\rm max})}$, and $\Delta_R = 1.6, 2.5, 3$ (diamonds, dots, crosses). The maximum allowed value of $\phi_c$ is given by $\sqrt{3/\Delta_\phi} \approx 3.87$. The decay constant $\tilde f$ is represented by the dashed black line. The left and right panels correspond to imposing $(+)$ and $(-)$ IR boundary conditions, respectively. The solid line in the left panel indicates the presence of a massless state. The ratio between the UV and IR energy scales is equal to $\frac{\Lambda_{\rm UV}}{\Lambda_{\rm IR}} = 10^3$. We used the IR regulator $r_1 = 10^{-12}$ in the numerics.}
\label{fig:PCModel2B_2}
\end{center}
\end{figure}

Figure~\ref{fig:PCModel2B_1} shows the resulting spectrum as function of the Yukawa coupling $y_5$. Comparing to Figure~\ref{fig:Model2BF_1}, which shows the spectrum for the strongly coupled sector in isolation, we see that the mass of the light state, present for $(+)$ boundary condition and large negative $y_5$, can be lifted by making $|y_5|$ sufficiently large, even when the scaling dimension $\Delta_R > \frac{5}{2}$ is irrelevant. Similarly, the exactly massless state, present for $(-)$ boundary condition in the isolated strongly coupled sector, can also be lifted by making $y_5$ large and negative, with a smaller effect for larger scaling dimensions. We stress that, contrary to the case represented by taking $x_F \simeq 2 \Delta$, the spectrum does not become densely packed in units of $\tilde f$. We attribute this to the fact that in the expression for the effective scaling dimension $\Delta_R^{\rm (eff)}$ given in Eq.~\eqref{eq:Deff}, it is the factor $H_\Psi = M_\Psi + y_5 \phi$, rather than the warp factor $A$, that is responsible for the large deviation from $\Delta_R$ in the deep IR.

In Figure~\ref{fig:PCModel2B_2}, we show the spectrum as function of the source $\phi_c$. We remind the Reader that when the strongly coupled sector is considered isolation, the spectrum becomes densely packed in units of $\tilde f$ for large $\phi_c$, as can be seen from Figure~\ref{fig:Model2BF_2}. We therefore choose to display the partially composite spectrum in units of $\Lambda_{\rm IR}$. For large values of $\phi_c$ the spectrum shows similar features as that of Model~I with large $x_F \simeq 2\Delta$, displayed in Figure~\ref{fig:PCModel1_3}. In both cases, the IR scale $\Lambda_{\rm IR}$ becomes parametrically small compared to the AdS scale associated with the UV fixed point, leading to multi-scale dynamics in the dual field theory. An important difference, previously discussed, is that the bosonic spectrum, shown in Figure~\ref{fig:SpectrumBosons_1}, approaches a gapped continuum for $x_F$ close to $2\Delta$, whereas for large $\phi_c$ the continuum is no longer gapped: the masses of the lowest resonances approach zero in units of $\tilde f$ both for the fermionic and bosonic states.

\section{Conclusions and Outlook}

We presented holographic models which aim to elucidate the strong dynamics of the composite Higgs scenario in the Veneziano limit. We focused our attention on the fermionic sector, complementing the study carried out for the bosonic sector in~\cite{Elander:2020nyd}. The new results that we presented fall into three categories. First, we calculated the spectrum of fermionic resonances in the strongly coupled sector considered in isolation. Second, we incorporated partial compositeness by coupling this sector to an external elementary fermion, and found a connection to the formalism of holographic Wilsonian RG. Third, we explored the effect of the fermion partial compositeness on the physical spectrum.

Incorporating the Veneziano limit in holographic models necessitates considering background solutions that deviate from AdS to a significant degree,  due to the effect of backreaction of the flavour sector on the bulk geometry. We considerered two models~\cite{Elander:2020nyd} where the backreaction arises due to the dynamics of bulk scalar fields that acquire non-trivial radial profiles on the background. In Model~I, only one such bulk scalar field is present, introduced in order to describe flavour-symmetry breaking. The free parameters are $\Delta$, related to the scaling dimension of the flavour-symmetry breaking operator in the dual field theory, the number of flavours $x_F$, and the bulk gauge coupling $g_5$. In the present study of the fermionic spectrum, $g_5$ only enters into the calculation of the decay constant $\tilde f$, which provides the units in which we present the majority of our results. Model~II contains a second bulk scalar field, associated with the explicit breaking of conformal invariance, leading to the additional parameters $\Delta_\phi$, related to the scaling dimension of the dual flavour-singlet operator, and the source $\phi_c$ which parameterizes the size of the deformation due to such operator.

Common to Models I and II is the scaling dimension $\Delta_R$ of the fermionic operator of the dual field theory, related to the mass of the corresponding bulk fermionic field, as well as the choice of IR boundary condition on the bulk fermionic field, which governs whether the dual strongly-coupled sector is chiral $(-)$ or vector-like $(+)$. For Model~II, we also considered the effect of the Yukawa coupling $y_5$ between the flavour-singlet scalar field and the bulk fermion. We remind the Reader that the masses for flavour singlet and non-singlet fermions are the same, as long as the above parameters are chosen to be the same.

\begin{figure}[t]
\begin{center}
\includegraphics[width=15.2cm]{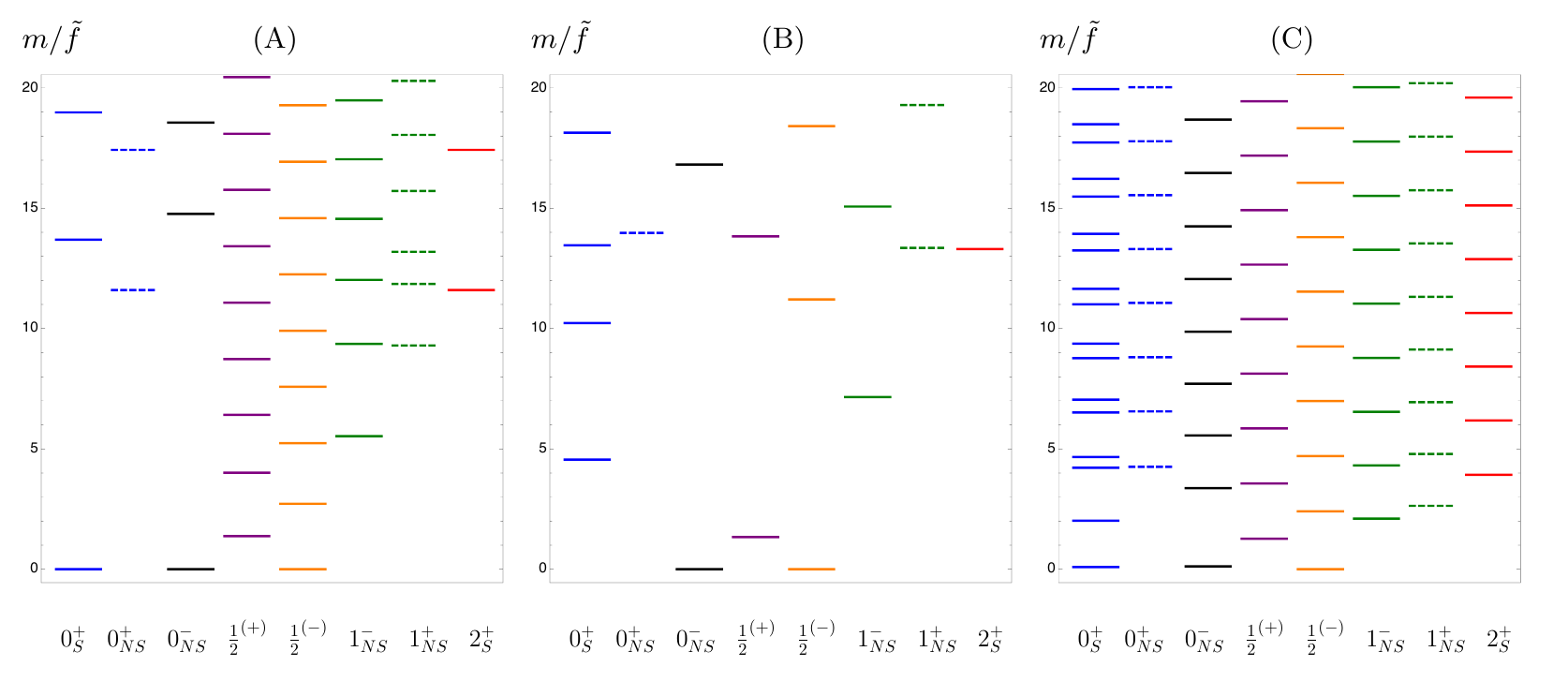}
\caption{Bosonic and fermionic spectra for three especial scenarios described in the text. Scenario (A) corresponds to Model~I with $\Delta = 2.5$, $x_F = 4.6$, $g_5 = 8$, $\Delta_R = 2.5$, $r_1 = 10^{-16}$ (bosons), $r_1 = 10^{-30}$ (fermions), $r_2 = 15$. 
The remaining two scenarios correspond to Model~II with $\Delta = 3$, $x_F = 1$, $g_5 = 8$, $\Delta_R = 2.5$, $r_1 = 10^{-12}$, $r_2 = 15$.
In addition, for scenario (B) we took $\Delta_\phi = 1$, $\phi_c =1.5$, $y_5 = -2$, and for scenario (C) $\Delta_\phi = 0.2$, $\phi_c =3.7$, $y_5 = 0$. 
All plots are normalised in units of the decay constant $\tilde f$. Below the plots, we indicated by $J^P_{S,NS}$ the spin $J$ and parity $P$ of the various bosonic states, as well as whether they transform in singlet ($S$) or non-singlet ($NS$) flavour representations. For the fermionic states, $J=1/2$, the superscripts $(+)$ and $(-)$ refer to the choice of the IR boundary condition, corresponding to the composite sector being vector-like or chiral, respectively. The various sectors from left to right are: singlet scalar (blue, solid lines), non-singlet scalar (blue, dashed lines), non-singlet pseudoscalar (black, solid lines), fermion in the vector-like case (purple, solid lines), fermion in the chiral case (orange, solid lines), non-singlet vector (green, solid lines), non-singlet axial-vector (green, dashed lines), singlet tensor (red, solid lines). The non-singlet scalar spectrum is computed using the invariant $\mathcal I_2$ (for details see \cite{Elander:2020nyd}).
}
\label{fig:SummaryPlot}
\end{center}
\end{figure}

For the case of the composite sector in isolation, and for generic choices of the parameters, we found that the results of the spectrum for both models are qualitatively similar to that of a slice of AdS, obtained in our models as $x_F,\phi_c \rightarrow 0$. In particular, it remains true for our models that, when the spectrum of the dual strongly-coupled sector is vector-like, there may be a light fermionic state when $\Delta_R \simeq 3/2$, corresponding to an approximately free fermion. Additionally, we identified three possible regions of parameter space---illustrated by means of a few examples in Figure~\ref{fig:SummaryPlot}---in which the spectrum deviates substantially from the minimal AdS case, leading to fermionic states parametrically lighter than $4\pi \tilde f$:
\begin{itemize}
	\item[(A)]
		An interesting limiting case is to take the number of flavours $x_F$ close to $2 \Delta$. As one approaches this upper bound on $x_F$, the backreaction of the flavour-symmetry breaking scalar on the bulk geometry becomes large, inducing multi-scale dynamics in the dual field theory. This can be anticipated in the bulk theory, where the IR scale $\Lambda_{\rm IR}$ becomes parametrically small, in units of the AdS curvature in the UV of the geometry. This manifests in a characteristic spectrum, shown for Model~I in Figure~\ref{fig:Model1F_2} for the fermionic sector, and in the left panel of Figure~\ref{fig:SpectrumBosons_1} for the bosonic sector. Close to $x_F \simeq 2\Delta$, the fermionic spectrum approaches that of a continuum (in units of the decay constant $\tilde f$), while the bosonic spectrum becomes that of a gapped continuum, accompanied by two kinds of light states, namely the NGBs, associated with the flavour-symmetry breaking, and the dilaton, associated with the breaking of scale invariance. Therefore, the low energy spectrum is characterised by a dilaton and a number of NGBs, accompanied by a tower of fermionic states, beginning at the scale $m_F \sim \Lambda_{\rm IR} \ll 4\pi \tilde f$. The remaining towers of bosonic states appear only at a 
significantly higher scale governed by the NGB decay constant. Although we illustrated this scenario for Model~I, the spectrum of Model~II shows the same qualitative features close to $x_F \simeq 2 \Delta$, for moderate values of $\phi_c$ and $\Delta_\phi$.
	\item[(B)]
		In Model~II, it is possible to obtain parametrically light fermionic states by dialling the bulk Yukawa coupling $y_5$ to large and negative values. This effect, shown in Figure~\ref{fig:Model2BF_1}, concerns the case when the IR boundary condition for the bulk fermion is chosen to be $(+)$, corresponding to the dual strongly-coupled sector being vector-like. Here, the scale associated with the typical bosonic and fermionic resonances is $m_* \sim 4\pi \tilde f$. However, we interestingly found that an isolated light fermionic state can be obtained, for any value of $\Delta_R$, by making the Yukawa coupling sufficiently large and negative. We were also able to capture this effect in a simple toy model that can be analytically solved, see section~\ref{sec:toymodel}. In addition to this light fermionic state, the low energy spectrum also consists of a number of NGBs accompanied, 
when the singlet operator is close to marginal ($\Delta_\phi \simeq 0$), by a light dilaton (see~\cite{Elander:2020nyd} for further discussion on this point).
	\item[(C)]
		In Model~II, there is an additional limit, in which one obtains a large backreaction on the bulk geometry, and which leads to multiscale dynamics in the dual field theory. For sufficiently small $\Delta_\phi$, one may increase the source $\phi_c$ for the flavour-singlet operator, such that the backreaction of its corresponding bulk scalar field causes the IR scale $\Lambda_{\rm IR}$ to become parametrically small. For large $\phi_c$, the resulting fermionic spectrum, depicted in Figure~\ref{fig:Model2BF_2}, shows similar features as in scenario~(A), namely it approaches that of a continuum (in units of the decay constant $\tilde f$). However,  as can be seen from the right panel of Figure~\ref{fig:SpectrumBosons_1}, the bosonic spectrum no longer contains a gapped continuum, but also approaches a continuum that starts at zero. In this scenario, one therefore finds that the low energy spectrum contains a number of NGBs, accompanied by both fermionic and bosonic towers beginning at the scale $\Lambda_{\rm IR} \ll 4\pi \tilde f$.
\end{itemize}
We remark that out of these three scenarios, the first and the last both rely on large deviations of the bulk geometry, compared to the AdS case, which is the reason that also the bosonic part of the spectrum is significantly affected. Conversely, the mechanism described in scenario~(B) relies on dialling the bulk Yukawa coupling $y_5$, and hence only affects the spectrum of the fermionic sector. We also note that, 
as we chose $\Delta \geq 2$ to realise a spontaneous breaking of the flavour symmetry,  
scenario~(A) requires taking the number of flavours to be large, $x_F\gtrsim 4$, thus  entering a regime in which our models may be less trustable,
see \cite{Elander:2020nyd} as well as the discussion after \eq{FP}.

By making use of the formalism of the holographic Wilsonian RG, we were able to derive a flow equation for an infinite number of couplings, collectively contained in $\tilde \lambda$, between an elementary fermion and the strongly coupled sector. Focusing on the most relevant coupling $\tilde \lambda_0$, obtained at the lowest order in a derivative expansion of $\tilde \lambda$, we studied the dependence of its RG flow on the parameters. For generic choices of the model parameters, the RG flow is qualitatively the same as for the AdS case, with only slight deviations in the deep IR. However, the RG flow can be radically different in all of the three scenarios outlined above. Remarkably, it is possible to induce RG flows leading to sizeable values of the coupling $\tilde \lambda_0$ in the IR, even when the associated operator is irrelevant at the UV fixed point, $\Delta_R > 5/2$. Examples of such special RG flows are shown in Figures~\ref{fig:RGModel1_3}--\ref{fig:RGModel2B_2}.

Finally, we investigated the effect of partial compositeness on the spectrum of fermionic resonances. 
The presence of an exactly massless fermionic state is reversed with respect to the case of a strongly-coupled sector in isolation: when the strong sector is chiral, the coupling to the elementary fermion causes the previously massless state to be lifted, whereas for a vector-like strong sector, there is always a massless state present after partial compositeness is incorporated. In the former case, how much the mass of the light state is lifted depends on the IR value of the coupling $\tilde \lambda_0$, with larger mixing leading to a larger mass, in agreement with the expectations from the analysis of the RG flow (see Figure~\ref{fig:PCModel1_1}). As can be seen from Figure~\ref{fig:PCModel1_2}, for generic values of the parameters, the effect of partial compositeness on the spectrum is simply controlled by the scaling dimension $\Delta_R$ at the UV fixed point: RG flows induced by relevant deformations produce the largest deviations from the low energy spectrum of the strongly-coupled sector in isolation (see Figure~\ref{fig:Model1F_1}). 
Of particular interest is whether the masses of light states can be significantly affected even when $\Delta_R > 5/2$. Our models allow for such dynamics in the following cases:
\begin{itemize}
	\item[(i)]
		When the $(-)$ IR boundary condition is chosen (chiral strongly-coupled sector) one may significantly lift the mass of the light state, provided the model parameters are chosen to be in the regimes corresponding to the scenarios (A--C) described above. This is illustrated in Figures~\ref{fig:PCModel1_3}--\ref{fig:PCModel2B_2}.
	\item[(ii)]
		In Model~II with $(+)$ IR boundary condition (vector-like strongly-coupled sector) the mass of the light state present for large negative $y_5$ in scenario (B) can be significantly lifted provided $|y_5|$ is taken sufficiently large. This is illustrated in Figure~\ref{fig:PCModel2B_1}. In scenarios (A) and (C) the mass of the lightest vector-like fermion was of order (a few) $\Lambda_{\rm IR}$ in the composite sector in isolation: partial compositeness has the effect of enhancing it, most significantly when $\tilde\lambda_{0, \rm IR}$ becomes sizeable.
\end{itemize}
Under these conditions, the operator $\mathcal O_R$ may be considered to be dangerously irrelevant, in the sense that it can have a large effect on the low energy physics, despite its irrelevant nature from the point of view of the UV fixed point. 

Let us draw some phenomenological implications from the above features of the fermionic spectrum. Recall that, for the composite sector in isolation, each resonance corresponds to a full representation of the unbroken flavour group, $Sp(2N_F)$, specifically either a singlet or a two-index representation, see the last column of table  \ref{trili}.  While SM singlet fermions can be light or even massless if sufficiently decoupled, the two-index representations contain several components in exotic SM representations, see \eq{4455} for the minimal case $N_F=5$. As the latter have not been observed, a realistic model requires a (+) IR boundary condition, to remove these components from the chiral content of the theory.\footnote{Alternatively, for $(-)$ IR boundary condition one needs to introduce as many exotic elementary fermions, to pair with their composite partners and thus lift their mass from zero. This scenario is possibly more peculiar, but interestingly it allows to keep massless only desirable components, e.g.~to realise a right-handed top quark that is fully composite.}
Given such large, composite vector-like multiplet, an elementary SM chiral fermion can have a linear coupling to the corresponding component of the multiplet, i.e.~the one carrying the conjugate quantum numbers: this results in a partially composite chiral fermion, as well as a state with mass raised with respect to the other components of the multiplet.\footnote{For simplicity we are neglecting, throughout the discussion, the effect of SM gauge and fermion loops, that may lift significantly the lightest states in the spectrum, in particular in the case of gluon or top-quark loops.} The magnitude of the mass splitting depends crucially on the IR value of such coupling. In particular, the smaller $\Delta_R$, the more relevant $\mathcal O_R$,  resulting in a larger effect.

The generic mass for a composite vector-like fermion is $m_*\sim 4\pi \tilde f$. Figure~\ref{fig:SummaryPlot} illustrates two scenarios (A) and (B) where, instead, the first fermion resonance becomes parametrically lighter. Additionally, in scenario~(C), all resonances become parametrically lighter than $4\pi\tilde f$: the resonance gap is controlled by a flavour-singlet operator, and it is significantly smaller than the scale of flavour-symmetry breaking. The experimental lower bounds lie in the few-TeV (few-hundred-GeV) range for coloured (electroweak) vector-like fermions, while SM singlet fermions could be much lighter.
At the same time one needs to keep $f\equiv \sqrt{N_C} \tilde f \gtrsim 1$ TeV. If light fermions will be discovered, this may imply a stronger lower bound on $f$, and thus no observable deviations e.g.~in SM Higgs couplings. Vice versa, if the latter are observed first, one cannot approach too closely the limiting cases of scenarios~(A--C). Note also that, as a vector-like fermion multiplet becomes lighter, the component mixing with a SM chiral fermion becomes progressively heavier than the others: consequently one might discover exotic states before observing an actual top prime. The light fermion resonances may or may not be accompanied by light bosonic ones (beside the NGBs): as illustrated in Figure~\ref{fig:SummaryPlot}, a light dilaton may or may not be present and, in the case of scenario (C), all bosonic sectors may feature a relatively light state as well.

The above scenarios~(A--C) illustrate how light fermionic modes may emerge in certain bottom-up holographic models, proposed to capture the strongly-coupled dynamics of a class of HC theories in the Veneziano limit. It would be interesting if future lattice simulations could identify regions in the $(N_C,N_F)$ plane leading to spectra with the same qualitative features as in these scenarios. At present, typical lattice simulations of UV-complete gauge theories in the composite-Higgs context (see section~\ref{latt_comp}) are realised for a small number of hypercolours, and are likely to be far from the near-conformal regime where anomalous dimensions could be large. Complementarily, one may ask whether one can build models that implement partial compositeness within the context of top-down gauge-gravity dualities. Such top-down models are rigidly constrained, as their field content and dynamics are dictated by supergravity, and thus allow for greater predictability. In particular, operator scaling dimensions are calculable rather than appearing as free parameters, in contrast with bottom-up models. It would be especially interesting if one could find regimes in which the dynamics is comparable to our scenarios~(A--C), leading to parametrically light fermionic states.

\section*{Acknowledgements}
We thank Kaustubh Agashe and Maurizio Piai for useful discussions. The project leading to this publication has received funding from the Excellence Initiative of Aix-Marseille University - A*MIDEX, a French ``Investissement d'Avenir" programme (ANR-11-LABX-0060 - OCEVU and AMX-19-IET-008 - IPhU).
This project has received support from the European Union’s Horizon 2020 research and innovation programme under the Marie Sk\l odowska-Curie grant agreement No 860881-HIDDeN.

\appendix

\section{Inventory of composite  fermionic operators}\label{zoo}

Given a HC theory with gauge group $Sp(2N_C)$, one can combine the constituent fields into composite, HC-invariant operators, which are suitable to describe the confined phase of the theory.
In table \ref{consti}, we collect the properties of the constituent fields with respect to Lorentz (gauge bosons and Weyl fermions), hypercolour (one- or two-index representations), and the flavour symmetry $G_F=SU(2N_F)\times U(1)$. 
In appendix C of Ref.~\cite{Elander:2020nyd} we classified all possible bosonic operators, while here we present all possible fermionic operators. We only list operators with no (covariant) derivatives, as these can be added straightforwardly, by acting on constituent fields
in all possible ways.

\subsection{Fermion trilinears}\label{Strili}

Let us begin by considering HC-invariant operators made of fermion constituents only. Fermionic operators should include $3+2k$ constituents, for $k=0,1,2,\dots$. The unique invariant tensor with only indexes in the fundamental of $Sp(2N_C)$ is  
$\Omega_{ij}=-\Omega_{ji}$. Therefore,
it is easy to identify all possible contractions of HC indexes,
\be
\psi^T\Omega(\chi\Omega)^{1+2k}\psi~,\quad\quad
{\rm tr}[\chi\Omega\chi\Omega(\chi\Omega)^{1+2k}] ~,
\label{chains}\ee
where each $\psi$ ($\chi$) can be replaced by $\overline\psi$ ($\overline \chi$, $\chi'$, $\overline\chi'$), as they transform in the same HC representation.

Since the canonical scaling dimension of these operators is $9/2 +3k$, one expects that the fermion-trilinear operators ($k=0$) are the most relevant ones, and that they therefore are those with the largest mixing with the SM fermions. Moreover, naively one may expect the mass of a composite fermion to grow with the number of constituents, $M_F \propto (3+2k)m_*$. 
In this case, the fermion-trilinear operators should be sufficient to describe the spectrum of the lightest composite fermions. There are, however, potential exceptions, because (i) composite fermions which are non-generically light are possible, e.g.~zero modes, and
(ii) such light states may belong to  a specific Lorentz and flavour representation, which may not be realised for $k=0$. Keeping this caveat in mind, in the following classification we limit ourself to fermion-trilinear operators only.

\begin{table}[tb]
\renewcommand{\arraystretch}{1.2}
\begin{center}
\begin{tabular}{|c|c|c|c|c|}
\hline
& Lorentz & $Sp(2N_C)$ & $SU(2N_F)$ & $U(1)$  \\
\hline \hline
$A^{\mu}_{ij}$ & $(1/2,1/2)^\mu$ & ${\Yvcentermath1 \tiny \yng(2)}_{\, ij}$ & $\bullet$ & $0$ \\ 
\hline \hline
 $\psi^{\alpha a}_i$ & $(1/2,0)^\alpha$ & ${\Yvcentermath1 \tiny \yng(1)}_{\, i}$ 
 & ${\Yvcentermath1 \tiny \yng(1)}^{\, a}$ & $q_\psi$ \\
\hline
 $\overline{\psi}^{\dot\alpha}_{ai} \equiv \psi^{\dagger\dot\alpha}_{aj} \Omega_{ji}$ & $(0,1/2)^{\dot\alpha}$ & 
 ${\Yvcentermath1 \tiny \yng(1)}_{\, i}$ &
$\overline{\Yvcentermath1 \tiny \yng(1)}_{\, a}$  & $-q_\psi$ \\
\hline\hline
 $\chi^\alpha_{ij}$ ~($\chi'^\alpha_{ij}$) & $(1/2,0)^\alpha$ & ${\Yvcentermath1 \tiny \yng(1,1)}_{\, ij}$ ~(${\Yvcentermath1 \tiny \yng(2)}_{\, ij}$)     
 & $\bullet$ & $q_\chi$ ~($q_{\chi'}$)\\
\hline
 $\overline{\chi}^{\dot\alpha}_{ij}\equiv \Omega_{ik}\chi^{\dag\dot\alpha}_{kl}\Omega_{lj}$ ~($\overline{\chi}'^\alpha_{ij}$) 
 & $(0,1/2)^{\dot\alpha}$ & ${\Yvcentermath1 \tiny \yng(1,1)}_{\, ij}$ ~(${\Yvcentermath1 \tiny \yng(2)}_{\, ij}$)     
 & $\bullet$ & $-q_\chi$ ~($-q_{\chi'}$)\\
\hline
\end{tabular}\end{center}
\caption{The constituent fields of  the HC theory: the gauge bosons $A$, the Weyl fermions $\psi$ and $\chi$ (or $\chi'$) and their conjugates.
In our convention $i,j,\dots$ are $Sp(2N_C)$ indexes, 
$\mu,\nu,\dots$ Lorentz vector indexes, $\alpha,\beta\dots$ and $\dot\alpha,\dot\beta,\dots$ Lorentz spinor indexes,  
$a,b,\dots$ flavour $SU(2N_F)$ indexes. 
Singlets are indicated by a bullet.
The ratio of $U(1)$ charges is determined by the requirement of a vanishing
$U(1)-Sp(2N_C)-Sp(2N_C)$ anomaly: $q_\chi(N_C-1)=-q_\psi N_F$ [or $q_{\chi'}(N_C+1)=-q_\psi N_F$].}
\label{consti}
\end{table}

The interplay between the HC and Lorentz contractions of three fermions is non-trivial, therefore let us analyse these contractions in detail, before presenting the complete list of independent operators.
The $Sp(2N_C)$ contraction of two fundamental representations with a two-index representation reads explicitly
\be
 \psi^T\Omega\chi\Omega\psi = \psi_i \chi_{jk} \psi_l \Omega_{ij}\Omega_{kl}~.
\label{HC121}\ee
With $\chi$ antisymmetric ($\chi'$ symmetric), such contraction is antisymmetric (symmetric) 
under the exchange of HC indexes between the two $\psi$'s.
The contraction of three two-index representations reads
\be 
{\rm tr}(\chi\Omega\chi\Omega\chi\Omega) = \chi_{ij}\chi_{kl}\chi_{mn}\Omega_{jk}\Omega_{lm}\Omega_{ni} ~.
\label{HC222}\ee
With $\chi$ antisymmetric ($\chi'$ symmetric), such contraction is symmetric (antisymmetric) under the exchange of HC indexes between two $\chi$'s. Note that we defined $\overline\psi$ and $\overline\chi$ such that they transform under HC in the same way as $\psi$ and $\chi$, see table~\ref{consti}. 
Therefore Eqs.~(\ref{HC121}) and (\ref{HC222}) hold when one or more fermions are barred as well.

Coming to Lorentz representations, we work with the conventions of \cite{Dreiner:2008tw} for spinor indexes. The combination of three spin-1/2 fermions should contain two spin-1/2 components and one spin-3/2 component. In the case of three left-handed spinors, e.g.~$\psi\chi\psi$, the components in $(1/2,0)^3=2\times (1/2,0) +(3/2,0)$ can be identified as
\be\ba{c}
F_1^\beta \equiv \psi^\alpha \chi^\beta \psi^\gamma \epsilon_{\alpha\gamma}~,\qquad 
F_2^\beta \equiv (\psi^\gamma \chi^\alpha \psi^\beta -\psi^\beta\chi^\gamma\psi^\alpha) \epsilon_{\alpha\gamma}~,\\
G^{\alpha\beta\gamma} \equiv \psi^\alpha \chi^\beta \psi^\gamma - \frac 12 F_1^\beta \epsilon^{\gamma\alpha}+\frac 16 \left(F_2^\alpha \epsilon^{\beta\gamma}- F_2^\gamma \epsilon^{\alpha\beta}\right)~.
\ea\label{LLL}\ee 
The spin-1/2 operator $F_1$ ($F_2$) is symmetric (antisymmetric) under the exchange of the two $\psi$'s,
by taking into account the anticommutation of the spinors. The spin-3/2 operator $G$ satisfies $G^{\alpha\beta\gamma}=-G^{\gamma\beta\alpha}$. Any other product of three left-handed spinors, e.g.~$\chi\chi\chi$, decomposes in the same way.
In the case of two left-handed and one right-handed spinor, e.g.~$\psi\overline{\chi}\psi$, the components in 
$(1/2,0)^2\times (0,1/2)= (0,1/2)+(1,1/2)$ can be identified as
\be
F_3^{\dot\beta} \equiv \psi^\alpha \overline{\chi}^{\dot\beta} \psi^\gamma \epsilon_{\alpha\gamma}~,\qquad 
H_{\mu\nu}^{\dot\beta} \equiv \psi^\alpha \overline{\chi}^{\dot\beta} \psi_\gamma (\sigma_{\mu\nu})_\alpha^{\ \gamma}~.
\ee
The spin-1/2 operator $F_3$ is symmetric under the exchange of the two $\psi$'s.
The operator $H$, despite being Lorentz irreducible, 
contains a spin-1/2 and a spin-3/2 component, both antisymmetric under the exchange of the two $\psi$'s.
Indeed, the spin corresponds to the vector
subgroup of the Lorentz group, $SU(2)_{spin}\subset SU(2)_{left} \times SU(2)_{right}$. Therefore, in order to distinguish the spin components of $H$, one can identify dotted and undotted indexes and develop the same decomposition as in \eq{LLL}.

\begin{table}[tb]
\scriptsize
\renewcommand{\arraystretch}{1.5}
\begin{center}
\begin{tabular}{|c|c|c|c|}
\hline
& Lorentz & $SU(2N_F)$
& $Sp(2N_F)$  \\
\hline \hline
 $F_1^{\beta ab} = \psi^{T \alpha a} \Omega \chi^\beta \Omega \psi^{\gamma b} \epsilon_{\alpha\gamma}$ ~[$\chi\leftrightarrow\chi'$]
 & $(1/2,0)^\beta$ & ${\Yvcentermath1 \tiny \yng(1,1)}^{\, ab}$ ~$\left[{\Yvcentermath1 \tiny \yng(2)}^{\, ab}\right]$
 & ${\Yvcentermath1 \tiny \yng(1,1)}_{\, ab} + \bullet_{aa}$  ~$\left[{\Yvcentermath1 \tiny \yng(2)}_{\, ab}\right]$\\
\hline
 $F_2^{\beta ab} = (\psi^{T \gamma a} \Omega \chi^\alpha \Omega \psi^{\beta b} - \psi^{T \beta a} \Omega \chi^\gamma \Omega \psi^{\alpha b})
 \epsilon_{\alpha\gamma}$ ~[$\chi\leftrightarrow\chi'$] & $(1/2,0)^\beta$ 
 & ${\Yvcentermath1 \tiny \yng(2)}^{\, ab}$ ~$\left[{\Yvcentermath1 \tiny \yng(1,1)}^{\, ab}\right]$
 & ${\Yvcentermath1 \tiny \yng(2)}_{\, ab}$  ~$\left[{\Yvcentermath1 \tiny \yng(1,1)}_{\, ab} + \bullet_{aa}\right]$\\
\hline
$G^{\alpha\beta\gamma ab} = \psi^{T \alpha a} \Omega \chi^\beta \Omega \psi^{\gamma b} - ({\rm spin~1/2~pieces})$ ~[$\chi\leftrightarrow\chi'$]
 & $(3/2,0)^{\alpha\beta\gamma}$ 
 & ${\Yvcentermath1 \tiny \yng(2)}^{\, ab}$ ~$\left[{\Yvcentermath1 \tiny \yng(1,1)}^{\, ab}\right]$
 & ${\Yvcentermath1 \tiny \yng(2)}_{\, ab}$  ~$\left[{\Yvcentermath1 \tiny \yng(1,1)}_{\, ab} + \bullet_{aa}\right]$\\
\hline\hline
 $F^{\beta}_\chi = {\rm tr}\left(\chi^\alpha \Omega \chi^\beta \Omega \chi^\gamma \Omega\right) \epsilon_{\alpha\gamma}$ 
 & $(1/2,0)^\beta$ & $\bullet$ 
 & $\bullet$\\
\hline
 $\left[ G^{\alpha\beta\gamma} = {\rm tr}\left(\chi'^\alpha \Omega \chi'^\beta \Omega \chi'^\gamma \Omega\right) \right]$
 & $\left[(3/2,0)^{\alpha\beta\gamma}\right]$ & $[\bullet]$ 
 & $[\bullet]$\\
\hline\hline
 $F_3^{\dot\beta ab} = \psi^{T \alpha a} \Omega \overline{\chi}^{\dot\beta} \Omega \psi^{\gamma b} \epsilon_{\alpha\gamma}$ ~[$\chi
 \leftrightarrow\chi'$]
 & $(0,1/2)^{\dot\beta}$ & ${\Yvcentermath1 \tiny \yng(1,1)}^{\, ab}$ ~$\left[{\Yvcentermath1 \tiny \yng(2)}^{\, ab}\right]$
 & ${\Yvcentermath1 \tiny \yng(1,1)}_{\, ab} + \bullet_{aa}$  ~$\left[{\Yvcentermath1 \tiny \yng(2)}_{\, ab}\right]$\\
\hline
 $H^{\dot\beta ab}_{\mu\nu} = \psi^{T \alpha a} \Omega  \overline{\chi}^{\dot\beta} \Omega \psi^{b}_{\gamma} (\sigma_{\mu\nu})_\alpha^{~\gamma}$ 
 ~[$\chi \leftrightarrow\chi'$] & $(1,1/2)^{\dot\beta}_{\mu\nu}$ 
 & ${\Yvcentermath1 \tiny \yng(2)}^{\, ab}$ ~$\left[{\Yvcentermath1 \tiny \yng(1,1)}^{\, ab}\right]$
 & ${\Yvcentermath1 \tiny \yng(2)}_{\, ab}$  ~$\left[{\Yvcentermath1 \tiny \yng(1,1)}_{\, ab} + \bullet_{aa}\right]$\\
\hline\hline
 ${F_4^{\dot\gamma}}^a_b = \psi^{T \alpha a} \Omega \chi^\beta \Omega \overline{\psi}^{\dot\gamma}_{b} \epsilon_{\alpha\beta}$ 
 ~[$\chi \leftrightarrow\chi'$] 
 & $(0,1/2)^{\dot\gamma}$ & ${\Yvcentermath1 \tiny \overline{\yng(1)}\yng(1)}_{\, b}^{\, a} + \bullet_a^a$
 & ${\Yvcentermath1 \tiny \yng(2)}_{\, ab} + {\Yvcentermath1 \tiny \yng(1,1)}_{\, ab} + \bullet_{aa}$ \\
\hline
 $H^{\dot\gamma a}_{\mu\nu b} = \psi^{T \alpha a} \Omega \chi^\beta \Omega \overline{\psi}^{\dot\gamma}_{b} 
 (\sigma_{\mu\nu})_\alpha^{~\beta}$
 ~[$\chi \leftrightarrow\chi'$] 
 & $(1,1/2)^{\dot\gamma}_{\mu\nu}$ & ${\Yvcentermath1 \tiny \overline{\yng(1)}\yng(1)}_{\, b}^{\, a} + \bullet_a^a$
 & ${\Yvcentermath1 \tiny \yng(2)}_{\, ab} + {\Yvcentermath1 \tiny \yng(1,1)}_{\, ab} + \bullet_{aa}$ \\
\hline\hline
 $\tilde F_\chi^{\dot\beta } = {\rm tr}\left( \chi^{\alpha} \Omega \overline{\chi}^{\dot\beta} \Omega \chi^{\gamma} \Omega\right) \epsilon_{\alpha\gamma}$
 & $(0,1/2)^{\dot\beta}$ 
 & $\bullet$ & $\bullet$  \\
\hline
 $\left[H^{\dot\beta}_{\mu\nu} = {\rm tr} \left(\chi'^\alpha \Omega  \overline{\chi}'^{\dot\beta} \Omega \chi'_{\gamma} \Omega\right)
 (\sigma_{\mu\nu})_\alpha^{~\gamma}\right]$ 
 & [$(1,1/2)^{\dot\beta}_{\mu\nu}$]
 & [$\bullet$]  & [$\bullet$] \\
\hline
\end{tabular}\end{center}
\caption{The fermion-trilinear operators, and their transformation properties with respect to Lorentz and to the flavour symmetry, before and after SSB. HC indexes (not shown) are contracted according to Eqs.~(\ref{HC121}) and (\ref{HC222}). The case when $\chi_{ij}$ antisymmetric is replaced by $\chi'_{ij}$ symmetric is shown in squared brackets everywhere.} 
\label{trili}
\end{table}

Armed with these tools, one can identify the non-vanishing trilinear operators, that we list in table~\ref{trili}. Let us make some remarks to clarify how the list of independent operators is established,
and how to identify their flavour representation. First of all, 
In the cases $\psi\chi\psi$ and $\psi\overline{\chi}\psi$, the replacement $\chi\leftrightarrow \chi'$ inverts the symmetry in flavour indexes.
In the case $\chi\chi\chi$,  the $F_2$-like contraction vanishes because of antisymmetry under exchange of the two contracted spinors, and the analogue of $G^{\alpha\beta\gamma}$
also vanishes because it is fully antisymmetric in $\alpha,\beta,\gamma=1,2$. In the case $\chi' \chi' \chi'$,
both $F_{1,2}$-like contractions vanish because the two spinors contracted by an $\epsilon$-tensor are antisymmetric in HC.
In the $\chi\overline{\chi}\chi$ ($\chi'\overline{\chi}'\chi'$) case, the $H^{\dot\beta}_{\mu\nu}$-like ($F_3^{\dot\beta}$-like) contraction vanishes by antisymmetry.

Let us discuss the flavour assignment of fermion-trilinear operators. Since each $\psi$ carries an $SU(2N_F)$ index, the $\psi\chi\psi$ operators transform in $SU(2N_F)$ two-index representations (symmetric, antisymmetric or adjoint), which all reduce to $Sp(2N_F)$ two-index symmetric or antisymmetric representations, up to traces that are  flavour singlets. The $\chi\chi\chi$ operators are $SU(2N_F)$ singlets.
The trilinear $U(1)$ charge, not shown in table \ref{trili}, is simply the sum of the three  corresponding constituent $U(1)$ charges, 
shown in table \ref{consti}. Larger operators, those with $k>0$ in \eq{chains},
transform in the same $SU(2N_F)$ representations as the corresponding $k=0$ fermion trilinear, however the non-vanishing components may be different, due to the different 
HC and Lorentz contractions. Of course, the operator $U(1)$ charge also depends on $k$.

The operators with non-trivial SM charges come from the decomposition of the $Sp(2N_F)$ representations 
${\Yvcentermath1 \tiny \yng(2)}_{\,ab}$ or ${\Yvcentermath1 \tiny \yng(1,1)}_{\,ab}$.
In the minimal case $N_F=5$, defined by \eq{1010}, such decomposition under $SU(3)_C\times SU(2)_L\times SU(2)_R \times U(1)_B$
is given by
\be\ba{l}
(55_S)_{Sp(10)} = \left[(1,1,1)_0 + (1,2,2)_0 + (1,1,3)_0 + (1,3,1)_0 + (8,1,1)_0 + (6,1,1)_{2/3} \right.\\
+\left.
(\overline{6},1,1)_{-2/3} + (3,2,1)_{1/3} + (\overline{3},2,1)_{-1/3}+(3,1,2)_{1/3}+(\overline{3},1,2)_{-1/3}\right]_{SU_{3221}}~, \\
(44_A)_{Sp(10)} = \left[2\times (1,1,1)_0 + (1,2,2)_0 + (8,1,1)_0 + (3,1,1)_{-2/3} + (\overline{3},1,1)_{2/3} \right.\\
+\left.(3,2,1)_{1/3} + (\overline{3},2,1)_{-1/3}+(3,1,2)_{1/3}+(\overline{3},1,2)_{-1/3}\right]_{SU_{3221}}~,
\ea
\label{4455}\ee
for the symmetric and antisymmetric case, respectively. As top and bottom quark partners should belong to the $(1/2,0)$ or $(0,1/2)$ Lorentz representation, inspecting table \ref{trili} one finds that the list of possible quark-partner operators is
given by $F_1^{ab},~F_2^{ab},~F_3^{ab}, ~{F_4}^a_b$, as anticipated in \eq{psipsichi}, and analogously 
\eq{psipsichi'} for the $\chi'$ case.

\subsection{Glue plus fermions}\label{SGF}

Let us complete the classification of composite fermionic operators, by considering HC-invariant combinations of gauge and fermion constituents.
The HC gauge-field strength can be written as a two-index symmetric tensor,
\be
F^{\mu\nu}_{ij} \equiv \Omega_{ik} (T^A)^k_j F^{A\mu\nu}~,
\ee
where $A$ runs over the $Sp(2N_C)$ generators $T^A$, which satisfy $\Omega T^A=-(T^A)^T \Omega$, and the contraction with $\Omega$ illustrates that the adjoint of $Sp(2N_C)$ is indeed equivalent to the two-index symmetric representation,
see appendix C in \cite{Elander:2020nyd} for more details. Therefore, $F^{\mu\nu}_{ij}$ transforms as $\chi'_{ij}$ with respect to HC. One can build fermionic operators by combining the same fermion constituents as in \eq{chains} with one or more powers of $F^{\mu\nu}$:
$(\psi^2 \chi^{1+2k} F^{1+n})$  or $(\chi^{3+2k} F^{1+n})$, for $k,n=0,1,2,\dots$, and analogously for $\chi\leftrightarrow \chi'$. These have canonical scaling dimension $\ge 13/2$, therefore they are expected to be highly irrelevant.

However, in the case of an adjoint fermion $\chi'$, there is also the possibility of fermionic operators with a single constituent fermion, $(\chi' F^{1+n})$, with canonical scaling dimension $7/2+2n$. Let us focus on the minimal case $n=0$, where the contraction of HC indexes reads explicitly
\be 
F^{\mu\nu}_{ij} \chi'^\alpha_{kl} \Omega_{jk}\Omega_{li} = {\rm tr} \left(F^{\mu\nu}\Omega \chi'^\alpha \Omega\right)~.
\label{Fchi}\ee 
It contains three irreducible Lorentz components,
\be\ba{l}
\hat F^\beta\equiv {\rm tr} \left(F^{\mu\nu}\Omega \chi'^\alpha \Omega\right) (\sigma_{\mu\nu})_\alpha^{\,\beta} \sim (1/2,0)^\beta ~,\\
\hat G^{\alpha\beta}_{\gamma} \equiv {\rm tr} \left(F^{\mu\nu}\Omega \chi'^\alpha \Omega\right) (\sigma_{\mu\nu})_\gamma^{\,\beta} 
-\dfrac13 \left(\hat F_\gamma\epsilon^{\beta\alpha} +\hat F^\beta\delta_\gamma^{\,\alpha}\right)
\sim (3/2,0)^{\alpha\beta}_{\gamma} ~,\\
\hat H^{\alpha\dot\beta}_{\dot\gamma} \equiv {\rm tr} \left(F^{\mu\nu}\Omega \chi'^\alpha \Omega\right) 
(\overline{\sigma}_{\mu\nu})_{\ \dot\gamma}^{\dot\beta} 
\sim (1/2,1)^{\alpha\dot\beta}_{\dot\gamma} ~.
\ea
\label{hatted}\ee
They are all flavour singlets, and they are conceivably more relevant w.r.t.~to the three-fermion operators (canonical dimension 7/2 versus 9/2), listed in table~\ref{trili}.
It is also conceivable that the lightest flavour-singlet composite fermions are better described as 
bound states of one gluon and one fermion, rather than three-constituent bounds states.

We also note that a composite fermion operator with a smaller scaling dimension should have a more significant mixing with elementary fermions in the same representation.
However, in order to mix the operators in \eq{hatted} with the SM elementary fermions, one would need $\chi'$ to carry SM charges, that is, one should introduce several flavours of 
constituent fermions in the adjoint. This endangers the asymptotic freedom of HC, as well as of the SM gauge interactions, as demonstrated in appendix B of \cite{Elander:2020nyd}.

\section{Fermions in five dimensions and their holographic interpretation} \label{sec:fermions}

In this appendix, we discuss some details about the treatment of fermions in holography~\cite{Henningson:1998cd,Henneaux:1998ch,Contino:2004vy}, in particular pertaining to the computation of two-point functions using the formalism of holographic renormalisation~\cite{Bianchi:2001kw,Bianchi:2001de,Skenderis:2002wp}.

\subsection{Variational problem}
\label{sec:variationalproblem}

Consider a Dirac fermion $\Psi$ in the bulk with the action given by Eq.~\eqref{eq:SPsi} which we repeat here for convenience (in this appendix we suppress the overall normalisation $\mathcal N_\Psi$)
\beqs
	\mathcal S_\Psi &=& - \int \dd^5 x \sqrt{-g} \left[ \frac{1}{2} \left( \overline \Psi \Gamma^M D_M \Psi - \overline{D_M \Psi} \Gamma^M \Psi \right) + H_\Psi \overline \Psi \Psi \right] \nonumber \\
	&& - \sum_{i=1,2} \frac{s_i}{2} \int \dd^4 x \sqrt{-\tilde g} \, \overline \Psi \Psi \Big|_{r_i} \,.
\eeqs
Taking the variation of $\mathcal S_\Psi$ with respect to $\Psi$, we obtain
\beqs
	\delta \mathcal S_\Psi &=& \delta \mathcal S_\Psi^{(1)} + \delta \mathcal S_\Psi^{(2)} \,, \nonumber \\
	\delta \mathcal S_\Psi^{(1)} &=& \int \dd^4x \sqrt{-\tilde g} \left[ \overline \Psi \left( \frac{- s_1 + \gamma^5}{2} \right) \delta \Psi + \delta \overline \Psi \left( \frac{- s_1 - \gamma^5}{2} \right) \Psi \right] \Bigg|_{r_1} \,, \nonumber \\
	\delta \mathcal S_\Psi^{(2)} &=& \int \dd^4x \sqrt{-\tilde g} \left[ \overline \Psi \left( \frac{- s_2 - \gamma^5}{2} \right) \delta \Psi + \delta \overline \Psi \left( \frac{- s_2 + \gamma^5}{2} \right) \Psi \right] \Bigg|_{r_2} \,.
\eeqs
Consider first the UV contribution. Suppose we take $s_2 = -1$. Then
\beq
	\delta \mathcal S_\Psi^{(2)} = \int \dd^4x \sqrt{-\tilde g} \left[ \overline \Psi_R \delta \Psi_L + \delta \overline \Psi_L \Psi_R \right] \Bigg|_{r_2} \,,
\eeq
and when $\Psi_L$ is the source its variation vanishes in the UV, so that $\delta \mathcal S_\Psi^{(2)} = 0$. Conversely, $s_2 = 1$ corresponds to choosing $\Psi_R$ as the source. Because of the invariance of the action under the transformations of Eq.~\eqref{eq:SPsiInvariance}, it is sufficient to consider the case of $\Psi_L$ being the source. Hence, in the following, we put $s_2 = -1$ without loss of generality.

By contrast, in the IR, we do not require either of the variations $\delta \Psi_{L,R}$ to vanish. Choosing $s_1 = -1$ leads to
\beq
	\delta \mathcal S_\Psi^{(1)} = \int \dd^4x \sqrt{-\tilde g} \left[ \overline \Psi_L \delta \Psi_R + \delta \overline \Psi_R \Psi_L \right] \Bigg|_{r_1} \,,
\eeq
and the Dirichlet boundary condition $\Psi_L |_{r_1} = 0$. Conversely, $s_1 = 1$ implies that $\Psi_R |_{r_1} = 0$. We will consider both these cases which we refer to as $(-)$ and $(+)$ IR boundary conditions, respectively. As already discussed in the main body of the paper, the $(-)$ boundary condition implies that the dual strongly-coupled sector is chiral, while the $(+)$ boundary condition describes a vector-like strong sector.

\subsection{Two-point functions}
\label{sec:twopointfunctions}

Two-point functions are obtained by differentiating the on-shell action (supplemented by counter-terms) with respect to the boundary value of the bulk field. To this end, we rewrite the action $\mathcal S_\Psi$, separating it into a bulk part, that vanishes by the equations of motion, and boundary parts. Making use of the identity
\beq
	\omega_M \Gamma^M = \Gamma^M \omega_M - \frac{1}{\sqrt{-g}} \partial_M \left( \sqrt{-g} \, \Gamma^M \right) \,,
\eeq
the action $\mathcal S_\Psi$ can be rewritten as
\beqs
	\mathcal S_\Psi &=& \mathcal S_D + \mathcal S_b \,, \\
	\mathcal S_D &=& - \int \dd^5 x \sqrt{-g} \, \overline \Psi \left[ \Gamma^M D_M + H_\Psi \right] \Psi \,, \\
	\mathcal S_{b} &=& \int \dd^4 x \sqrt{-\tilde g} \, \overline \Psi \, \left( \frac{- s_1 - \gamma^5}{2} \right) \Psi \Big|_{r_1} + \int \dd^4 x \sqrt{-\tilde g} \, \overline \Psi P_R \Psi \Big|_{r_2} \,,
\eeqs
where we used the explicit form of the vielbein $e^M{}_A$ given in Eq.~\eqref{eq:vielbein}. After using the equation of motion for $\Psi$ given by the Dirac equation
\beq
\label{eq:Dirac}
	\left[ \Gamma^M D_M + H_\Psi \right] \Psi = 0 \,,
\eeq
the regularised on-shell action $\mathcal S_{\Psi, \rm reg}$ only receives a contribution from $\mathcal S_b$. Furthermore, after imposing either $\Psi_L |_{r_1} = 0$ ($s_1 = -1$) or $\Psi_R |_{r_1} = 0$ ($s_1 = 1$), only the UV part of $S_b$ contributes on-shell, leaving
\beq
\label{eq:SPsiregL}
	\mathcal S_{\Psi, \rm reg} = \int \dd^4 x \sqrt{-\tilde g} \, \overline \Psi_L \Psi_R \Big|_{r_2} \,.
\eeq

After projecting with $P_{L,R}$, Eq.~\eqref{eq:Dirac} can be written as
\beqs
\label{eq:DiracLR}
	(\partial_r + 2 \partial_r A + H_\Psi ) \Psi_R + e^{-A} \gamma^\mu \partial_\mu \Psi_L &=& 0 \,, \nonumber \\
	- (\partial_r + 2 \partial_r A - H_\Psi ) \Psi_L + e^{-A} \gamma^\mu \partial_\mu \Psi_R &=& 0 \,. 
\eeqs
We define Fourier transforms by the convention
\beq
	\Psi(x) = \int \frac{\dd^4q}{(2\pi)^2} e^{i q_\mu x^\mu} \Psi(q) \,, 
\eeq
and in momentum space we use the notation $\overline \Psi(q,r) \equiv \Psi^\dagger(-q,r) i \gamma^0$. Solving for $\Psi_R$ in Eq.~\eqref{eq:DiracLR}, we then obtain
\beq
\label{eq:PsiRfromL}
	\Psi_R(q,r) = - \frac{i \slashed{q}}{q^2} e^A (\partial_r + 2 \partial_r A - H_\Psi ) \Psi_L(q,r) \,,
\eeq
which after plugging into Eq.~\eqref{eq:SPsiregL} gives
\beq
\label{eq:2ndorderPsiL}
	\mathcal S_{\Psi, \rm reg} = - \int \dd^4 q \, e^{5A} \, \overline \Psi_L(-q,r) \frac{i \slashed{q}}{q^2} (\partial_r + 2 \partial_r A - H_\Psi ) \Psi_L(q,r)  \Big|_{r_2} \,.
\eeq
To this we need to add the counter-term action
\beqs
	\mathcal S_{\Psi, \rm ct} &=& - \int \dd^4 x \sqrt{-\tilde g} \, \overline\Psi_L \Big( \mathcal F(-\tilde g^{\mu\nu} \partial_\mu \partial_\nu) i \Gamma^\sigma \partial_\sigma \Big) \Psi_L \Big|_{r_2} \\ \nonumber
	&=& - \int \dd^4 q \, e^{3A} \, \overline\Psi_L(-q,r) \Big( \mathcal F(e^{-2A} q^2) i \slashed{q} \Big) \Psi_L(q,r) \Big|_{r_2} \,,
\eeqs
to define the subtracted action as $\mathcal S_{\Psi, \rm sub} \equiv \mathcal S_{\Psi, \rm reg} + \mathcal S_{\Psi, \rm ct}$. Here, $\mathcal F$ is a polynomial
\beq
	\mathcal F (e^{-2A} q^2) = \sum_{i=0}^n f_i (e^{-2A} q^2)^i \,,
\eeq
the order $n$ of which depends on the dimension of the operator dual to $\Psi_L$.

It is convenient to rescale $\Psi_L$ as
\beqs
	\Psi_L(q,r) &=& N_L(r) \psi_L(q,r) \,, \\
	N_L(r) &\equiv& \exp \left( -2A(r) + M_\Psi r - \int_r^\infty \dd \tilde r \, h_\Psi(\tilde r) \right) \,,
\eeqs
where we anticipated that it is necessary to include the normalisation factor $N_L(r)$ so that $\psi_L$ becomes the source of the operator $\mathcal O_R$ in the dual field theory.\footnote{Indeed, from Eq.~\eqref{eq:eompsiL} we see that, for asymptotically AdS backgrounds and $\lim\limits_{r \rightarrow \infty} H_\Psi = M_\Psi > - \frac{1}{2}$, the leading mode scales as $\psi_L \sim 1$ in the UV of the geometry (large $r$).} In terms of $\psi_L$, the Dirac equation~\eqref{eq:DiracLR} becomes
\beq
\label{eq:eompsiL}
	\Big[ \partial_r^2 + (\partial_r A + 2 H_\Psi) \partial_r - q^2 e^{-2A} \Big] \psi_L = 0 \,,
\eeq
while the two possible IR boundary conditions, $(-)$ and $(+)$, become Dirichlet $\psi_L|_{r_1} = 0$ $(s_1 = -1)$ or Neumann $\partial_r \psi_L|_{r_1} = 0$ $(s_1 = 1)$, respectively.

After writing $\psi_L(q,r) = b(q,r) \tilde \psi_L(q)$, where $b$ is a scalar, we obtain
\beq
\label{eq:d2SPsi}
	\frac{\delta^2 \mathcal S_{\Psi, \rm sub}}{\delta \overline\psi_L(-q,r_2) \delta \psi_L(q,r_2)} = N_L^2 \, e^{5A}\frac{i}{\slashed{q}} \left( e^{-2A} q^2 \mathcal F(e^{-2A} q^2) + \frac{\partial_r b}{b} \right) \Big|_{r_2} \,.
\eeq
The two-point function of the operator $\mathcal O_R$ is then given by
\beq
\label{eq:OORapp}
	\langle \mathcal O_R(q) \overline{\mathcal O}_R(-q) \rangle = \lim\limits_{r_2 \rightarrow \infty} \bigg\{ \frac{i \, \delta^2 \mathcal S_{\Psi, \rm sub}}{\delta \overline\psi_L(-q,r_2) \delta \psi_L(q,r_2)} \bigg\} \,.
\eeq
We see that the massive poles, and hence the spectrum, can be extracted from the non-trivial solutions that satisfy the UV boundary condition $b|_{r_2} = 0$. We will treat the massless case separately soon.

Finally, we comment on a relation that follows from Eq.~\eqref{eq:eompsiL}. Given a solution $\psi_L(q,r) = b(q,r) \tilde \psi_L(q)$ to Eq.~\eqref{eq:eompsiL}, let us define the ratio
\beq
	x^{\pm}_{H_\Psi} = - \frac{\partial_r b}{b} \,,
\eeq
where the superscript $\pm$ indicates which boundary condition $b$ satisfies in the IR, while the subscript refers to the function $H_\Psi$ appearing in Eq.~\eqref{eq:eompsiL}. Then, it follows that
\beq
\label{eq:bRLratio}
	x^{\pm}_{H_\Psi} = \frac{e^{-2A} q^2}{x^{\mp}_{-H_\Psi}} \,,
\eeq
corresponding to the invariance of \eq{eq:SPsiInvariance}.

\subsection{Massless poles}
\label{sec:masslesspoles}

Whether or not the fermionic correlator of Eq.~\eqref{eq:OORapp} contains massless poles depends on the behaviour of $\frac{\partial_r b}{b}$ in Eq.~\eqref{eq:d2SPsi} at large $r_2$ and small $q^2$: such dependence may or may not lead to the cancellation of the prefactor $\frac{1}{\slashed{q}}$ as $q^2 \to 0$. Hence, let us expand $b$ for small $q^2$ as
\beq
\label{eq:bLsmallq2}
	b(q,r) = \mathcal N(q^2) \left( b^{(0)} + b^{(2)}(r) q^2 + \cdots \right) \,,
\eeq
where $\mathcal N(q^2)$ is an unimportant overall normalisation, and we used the Dirac equation~\eqref{eq:DiracLR} at zero momentum to fix $b^{(0)}$ as having no radial dependence.

Suppose we use the $(+)$ boundary condition. Then, we have that
\beq
	\frac{\partial_r b^+}{b^+} = \frac{\partial_r b^{(2)}}{b^{(0)}} q^2 + \cdots \,,
\eeq
which when plugged into Eq.~\eqref{eq:d2SPsi} and Eq.~\eqref{eq:OORapp} implies that $\langle \mathcal O_R(q) \overline{\mathcal O}_R(-q) \rangle^+$ has no massless pole. On the other hand, the $(-)$ boundary condition implies that $b^{(0)} = 0$, and hence
\beq
	\frac{\partial_r b^-}{b^-} = \frac{\partial_r b^{(2)}}{b^{(2)}} + \cdots \,,
\eeq
where $b^{(2)}(r)$ satisfies the equation of motion
\beq
	\big[ \partial_r^2 + (\partial_r A + 2 H_\Psi) \partial_r \big] b^{(2)} = 0 \,,
\eeq
supplemented by the IR boundary condition $b^{(2)}(r_1) = 0$. In an asymptotically AdS background, this leads to the UV expansion
\beq
	b^{(2)}(r) = b^{(2)}_1 + b^{(2)}_2 \left( e^{-(1+2M_\Psi) r} + \cdots \right)
\eeq
where $b^{(2)}_1$ and $b^{(2)}_2$ are integration constants. Which in turn implies that
\beq
	\frac{\partial_r b^-}{b^-} = -\frac{b^{(2)}_2}{ b^{(2)}_1} (1+ 2M_\Psi) e^{-(1+2M_\Psi) r} + \cdots
\eeq
for large $r$, such that the exponential factor cancels against the factor $N_L^2 e^{5A}$ in Eq.~\eqref{eq:d2SPsi}. Consequently, Eq.~\eqref{eq:OORapp} implies that $\langle \mathcal O_R(q) \overline{\mathcal O}_R(-q) \rangle^-$ has a massless pole, except for special cases (e.g. when $M_\Psi = -1/2$ such that the scaling dimension $\Delta_R = 3/2$ is that of a free fermion, see next subsection).

\subsection{Application to AdS background}
\label{AdSapp}

It is instructive to compute the fermionic two-point functions for the case of an AdS background. Hence, we take $A(r) = r$ and $H_\Psi = M_\Psi$ constant, such that Eq.~\eqref{eq:eompsiL} becomes
\beq
	\Big[ \partial_r^2 + (1 + 2 M_\Psi) \partial_r - q^2 e^{-2r} \Big] b = 0 \,
\eeq
with the solution (for $q^2 \leq 0$)
\beq
	b = e^{-\alpha r} \Big[ c_1(Q) Y_\alpha(e^{-r} Q) + c_2(Q) J_\alpha(e^{-r} Q) \Big] \,, \quad \alpha \equiv M_\Psi + \frac{1}{2} \,, \quad Q \equiv \sqrt{-q^2} \,,
\eeq
where $c_1$ and $c_2$ are integration constants.

After imposing the two possible IR boundary conditions $b|_{r_1} = 0$ $(-)$ or $\partial_r b|_{r_1} = 0$ $(+)$, one obtains
\beqs
\label{eq:dbLdivbL}
	\frac{\partial_r b^-}{b^-} &=&
Q \, e^{-r} \frac{J_{\alpha -1}\left(e^{-r} Q\right) Y_{\alpha
   }\left(e^{-r_1} Q\right)-J_{\alpha }\left(e^{-r_1} Q\right)
   Y_{\alpha -1}\left(e^{-r} Q\right)}{J_{\alpha }\left(e^{-r_1}
   Q\right) Y_{\alpha }\left(e^{-r} Q\right)-J_{\alpha }\left(e^{-r}
   Q\right) Y_{\alpha }\left(e^{-r_1} Q\right)}
	\,, \\
	\frac{\partial_r b^+}{b^+} &=&
Q \,e^{-r} \frac{J_{\alpha -1}\left(e^{-r} Q\right) Y_{\alpha
   -1}\left(e^{-r_1} Q\right)-J_{\alpha -1}\left(e^{-r_1} Q\right)
   Y_{\alpha -1}\left(e^{-r} Q\right)}{J_{\alpha -1}\left(e^{-r_1}
   Q\right) Y_{\alpha }\left(e^{-r} Q\right)-J_{\alpha }\left(e^{-r}
   Q\right) Y_{\alpha -1}\left(e^{-r_1} Q\right)}
	\,.
\eeqs
It is straightforward to verify that these expressions are consistent with the relation given in Eq.~\eqref{eq:bRLratio}.

Recalling that the counter-terms are polynomial in $q^2$, the massive poles of $\langle \mathcal O_R(q) \overline{\mathcal O}_R(-q) \rangle^-$ are given by the zeroes of $b^-(Q,r_2)$ in the limit $r_2 \rightarrow \infty$. One can show that these are given by those $Q$ for which
\beq
	J_\alpha(e^{-r_1} Q) = 0 \,.
\eeq
Similarly, the massive poles of $\langle \mathcal O_R(q) \overline{\mathcal O}_R(-q) \rangle^+$ can be obtained from the zeroes of $b^+(Q,r_2)$ in the limit $r_2 \rightarrow \infty$, which leads to
\beq
	J_{\alpha-1}(e^{-r_1} Q) = 0 \,.
\eeq
We showed the resulting spectrum as a function of $\Delta_R = 2 + M_\Psi$ in Figure~\ref{fig:AdSmp}. 

For the computation of the two-point function, including the cancellation of potential UV divergencies, we consider separately the two cases corresponding to non-integer or integer values of $\alpha$. In the former case, one can show that the counterterm coefficients $f_k$ may be chosen to be\footnote{Note that no counterterms are needed when $\alpha < 1$.}
\beq
	f_k = (-1)^k \Theta(\alpha - 1) g_k^{(\alpha)} \,, \hspace{1cm} k = 0, \cdots, \lfloor \alpha \rfloor \,,
\eeq
where $\lfloor \alpha \rfloor$ is the largest integer smaller or equal to $\alpha$ and the coefficients $g_k^{(\alpha)}$ are defined through the expansion
\beq
	\frac{J_{1-\alpha}(z))}{J_{-\alpha}(z)} \equiv \sum_{k \geq 0} g_k^{(\alpha)} z^{2k+1} \,.
\eeq
The resulting two-point functions become equal to
\beq
	\langle \mathcal O_R(q) \overline{\mathcal O}_R(-q) \rangle^\pm = \pm \frac{2}{\slashed{q}} \left( \frac{Q}{2} \right)^{2\alpha} \frac{\Gamma(1-\alpha)}{\Gamma(\alpha)} \frac{J_{-\alpha + \frac{1}{2} \pm \frac{1}{2}}(e^{-r_1} Q)}{J_{\alpha - \frac{1}{2} \mp \frac{1}{2}}(e^{-r_1} Q)} \,,
\eeq
which confirms the above statements regarding the position of the massive poles. It is also easy to verify by expanding for small $Q$ that, as expected, $\langle \mathcal O_R(q) \overline{\mathcal O}_R(-q) \rangle^-$ has a massless pole while $\langle \mathcal O_R(q) \overline{\mathcal O}_R(-q) \rangle^+$ does not. Finally note that in the limit $\alpha \rightarrow 0$, corresponding to the free fermion case $\Delta_R = \frac{3}{2}$, both two-point functions vanish identically.

For integer $\alpha = n \geq 1$, one has to take into account the presence of logarithmic divergencies. After choosing the counterterm coefficients to be
\begin{align}
	f_k &= (-1)^{k+1} h_k^{(n)} \,, \hspace{1cm} k = 0, \cdots, n - 2 \,, \\
	f_{n - 1} &= (-1)^n \left[ \frac{1}{2^{2(n-1)} \kappa_n (n - 1)!} \left(r - \log \left( \frac{\mu}{2} \right) \right) + h_{n-1}^{(n)} \right] \,,
\end{align}
where $\mu$ is a constant, $\kappa_0 = -2\gamma_E$, $\kappa_n = (n - 1)!$ ($n \geq 1$), and the coefficients $h_k^{(n)}$ are defined through the expansion
\beq
	\frac{Y_{n -1}(z) - \frac{2}{\pi} \log \left( \frac{z}{2} \right) J_{n - 1}(z)}{Y_{n}(z) - \frac{2}{\pi} \log \left( \frac{z}{2} \right) J_{n}(z)} \equiv \sum_{k \geq 0} h_k^{(n)} z^{2k+1} \,,
\eeq
one obtains the two-point functions
\beq
	\langle \mathcal O_R(q) \overline{\mathcal O}_R(-q) \rangle^\pm = \frac{2}{\slashed{q}} \left( \frac{Q}{2} \right)^{2n} \frac{1}{\kappa_n (n - 1)!} \left[ \pi \frac{Y_{n - \frac{1}{2} \mp \frac{1}{2}}(e^{-r_1} Q)}{J_{n - \frac{1}{2} \mp \frac{1}{2}}(e^{-r_1} Q)} -2 \log \left( \frac{Q}{\mu} \right) \right] \,.
\eeq
As can be seen, we have traded the UV scale corresponding to a finite $r_2$ for the renormalisation scale $\mu$.

\section{Comparison with bosonic spectrum}
\label{sec:bosonicspectrum}

\begin{figure}[t]
\begin{center}
\includegraphics[width=\figwidth]{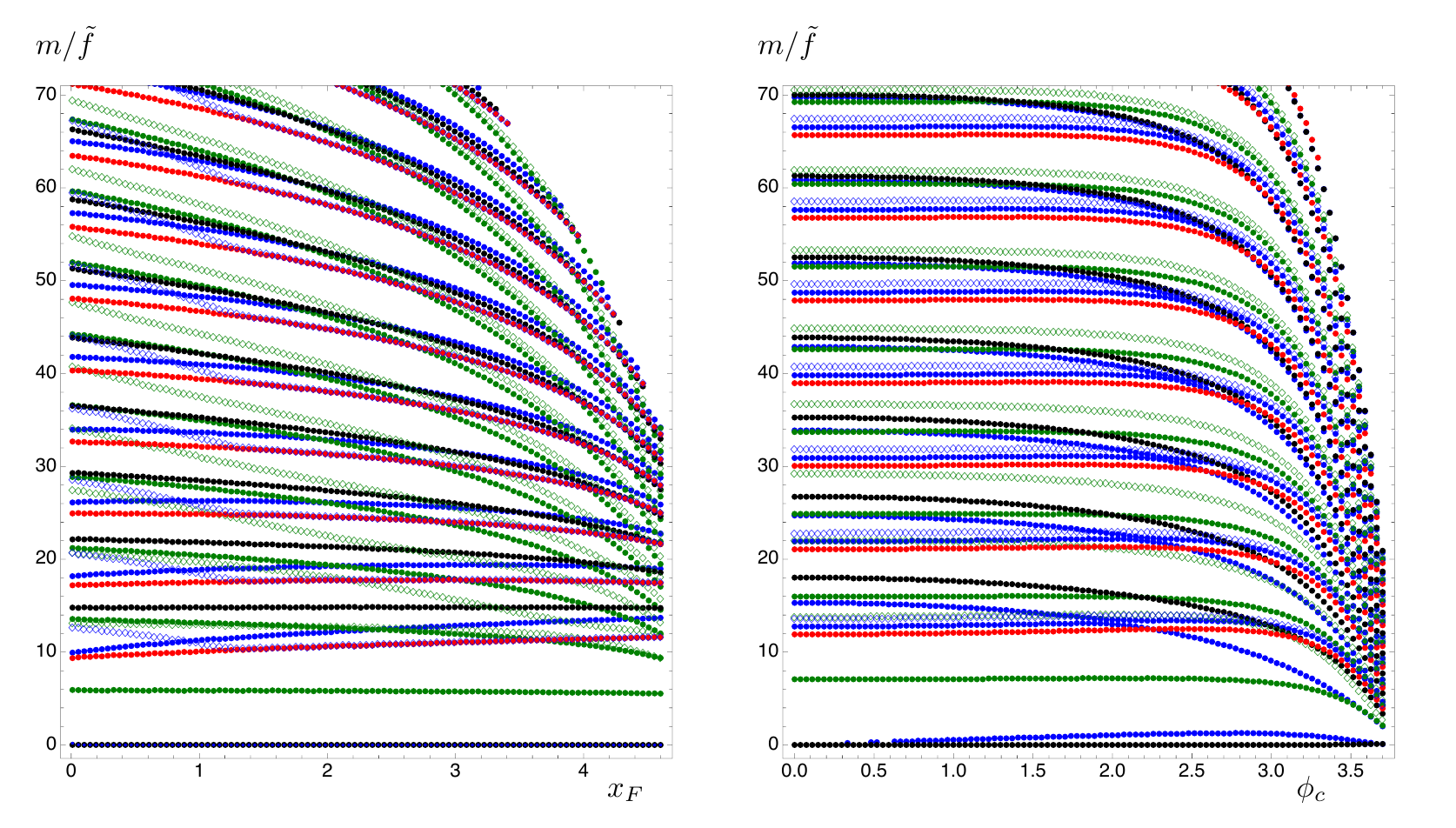}
\caption{Bosonic spectrum for Model~I (left panel) and Model~II (right panel), as a function of $x_F$ and $\Delta_\phi$, respectively. In Model~I, we used $\Delta = 2.5$, $g_5 = 8$, $r_1 = 10^{-10}$, and $r_2 = 15$. In Model~II, we used $x_F =1$, $\Delta = 3$, $g_5 = 8$, $\Delta_\phi = 0.2$, $r_1 = 10^{-12}$, and $r_2 = 15$. Both plots are normalised in units of the decay constant $\tilde f$. The colour coding for the spectrum is: singlet scalar (blue), non-singlet scalar (blue diamonds), tensor (red), pseudoscalar (black), vector (green), axial-vector (green diamonds); for further details see \cite{Elander:2020nyd}. In Model~I, since $\Delta > 2$, there is also a massless dilaton in the spectrum. In both panels, it is understood that heavier states are present in the right- and left-upper corners.}
\label{fig:SpectrumBosons_1}
\end{center}
\end{figure}

In order to make the comparison between the fermionic and bosonic spectra of Models~I and~II, we here present two additional plots shown in Figure~\ref{fig:SpectrumBosons_1}, complementing the results previously obtained in~\cite{Elander:2020nyd}. We remind the Reader that what we call Model~II in the current paper is referred to as Model~IIB in~\cite{Elander:2020nyd}. Furthermore, the spectrum is computed using the invariant $\mathcal  I = \mathcal I_2$, which is a choice that only affects the results for the non-singlet scalar (for details, see~\cite{Elander:2020nyd}). The left panel of Figure~\ref{fig:SpectrumBosons_1} shows the bosonic spectrum for Model~I, as one approaches the upper bound on the number of flavours $x_F \leq 2 \Delta$. Conversely, the right panel shows the bosonic spectrum for Model~II, as one approaches the upper bound on $\phi_c < \sqrt{3/\Delta_\phi}$ for a value of $\Delta_\phi < \frac{\Delta}{2\Delta - x_F}$. In both limits, the spectrum begins to form a continuum, i.e. the spacing between the heavy states approaches zero in units of $\tilde f$. However, there is an important difference in that for the former case the mass gap remains of the same order as $x_F$ is varied, resulting in a gapped continuum, while for the latter case, the continuum that forms for large values of $\phi_c$ starts at zero. We also note the presence of light states in both cases, namely the NGBs and the dilaton, which are both massless for Model~I, while in Model~II, increasing the source $\phi_c$ results in lifting the mass of the dilaton. The latter effect is small due to the choice of $\Delta_\phi = 0.2$ being close to marginal.

\section{Holographic Wilsonian RG}
\label{sec:hWRGapp}

For completeness, we give here a derivation of the flow equation given by Eq.~\eqref{eq:flowxi}, following the arguments outlined in~\cite{Faulkner:2010jy} for bosons, and generalised to fermions in~\cite{Laia:2011wf,Elander:2011vh}. We start by observing that Eq.~\eqref{eq:ZQFTUV} is invariant under change of the cutoff surface parametrized by $\mathfrak{r}$. Recalling Eq.~\eqref{eq:SPsiFiniteCutoff}, the actions appearing in Eq.~\eqref{eq:ZQFTUV} are
\begin{align}
	\mathcal S_\Psi[r_1,\mathfrak r;s_1,-1] = & - \int_{r_1}^{\mathfrak{r}} \dd r \int \dd^4 x \sqrt{-g} \left[ \frac{1}{2} \left( \overline \Psi \Gamma^M D_M \Psi - \overline{D_M \Psi} \Gamma^M \Psi \right) + H_\Psi \overline \Psi \Psi \right] \nonumber \\ & - \frac{s_1}{2} \int \dd^4 x \sqrt{-\tilde g} \, \overline \Psi \Psi \Big|_{r_1} + \frac{1}{2} \int \dd^4 x \sqrt{-\tilde g} \, \overline \Psi \Psi \Big|_{\mathfrak{r}}
\end{align}
and
\begin{align}
	\mathcal S_{\rm UV}[\mathfrak{r},r_{\rm UV}] = - \int \dd^4 q \sqrt{-\tilde g} \, \overline \Psi_L(-q,\mathfrak{r}) f(q,\mathfrak{r}) \Psi_L(q,\mathfrak{r}) \,,
\end{align}
where, for ease of presentation, we have defined (see Eq.~\eqref{eq:SUVdefinition})
\beq
\label{eq:fdefinition}
	f(q,\mathfrak{r}) \equiv \frac{i \slashed{q}}{e^{4A} N_L^2 \xi^2} \,,
\eeq
and we omitted the factor $\mathcal N_\Psi$ which only appears as an overall constant, irrelevant in the following. In the large-$N_C$ limit, these actions are to be evaluated on the classical solutions, and the RG invariance of Eq.~\eqref{eq:ZQFTUV} hence implies that
\beq
\label{eq:RGinvariance}
	\partial_\mathfrak{r} \big( \mathcal S_\Psi[r_1,\mathfrak r;s_1,-1] + \mathcal S_{\rm UV}[\mathfrak{r},r_{\rm UV}] \big) = 0 \,.
\eeq
In addition, the variational problem demands that
\beqs
\label{eq:PsiRBCs}
	0 &=& \delta \mathcal S_\Psi[r_1,\mathfrak r;s_1,-1] + \delta \mathcal S_{\rm UV}[\mathfrak{r},r_{\rm UV}] \nonumber \\
	&=& \int \dd^4q \sqrt{-\tilde g} \, \bigg\{ \Big[ \overline \Psi_R(-q,\mathfrak{r}) - \overline \Psi_L(-q,\mathfrak{r}) f(q,\mathfrak{r}) \Big] \delta \Psi_L(q,\mathfrak{r}) \nonumber \\
	&& \qquad\qquad\qquad + \delta \overline \Psi_L(-q,\mathfrak{r}) \Big[ \Psi_R(q,\mathfrak{r}) - f(q,\mathfrak{r}) \Psi_L(q,\mathfrak{r}) \Big] \bigg\} \,,
\eeqs
leading to the boundary conditions relating $\Psi_R$ ($\overline \Psi_R$) to $\Psi_L$ ($\overline \Psi_L$):
\begin{align}
	\overline \Psi_R(-q,\mathfrak{r}) &= \overline \Psi_L(-q,\mathfrak{r}) f(q,\mathfrak{r}) \nonumber \\
	\Psi_R(q,\mathfrak{r}) &= f(q,\mathfrak{r}) \Psi_L(q,\mathfrak{r}) \,.
\end{align}
Together, Eqs.~\eqref{eq:RGinvariance} and~\eqref{eq:PsiRBCs} imply a flow equation for $f$, as follows. We have that
\begin{align}
	& \hspace{-0.8cm} \partial_\mathfrak{r} \mathcal S_\Psi[r_1,\mathfrak r;s_1,-1] = \nonumber \\ =& \int \dd^4 x \, \bigg\{ \frac{1}{2} \partial_r \left( \sqrt{-\tilde g} \, \overline \Psi \Psi \right) - \sqrt{-g} \left[ \frac{1}{2} \left( \overline \Psi \Gamma^M D_M \Psi - \overline{D_M \Psi} \Gamma^M \Psi \right) + H_\Psi \overline \Psi \Psi \right] \bigg\} \bigg|_{\mathfrak{r}} \nonumber \\
	=& \int \dd^4 q \, e^{4A} \bigg\{ \left( 2 \partial_\mathfrak{r} A - H_\Psi \right) \Big[ \overline \Psi_R(-q,\mathfrak{r}) \Psi_L(q,\mathfrak{r}) + \overline \Psi_L(-q,\mathfrak{r}) \Psi_R(q,\mathfrak{r}) \Big] \nonumber \\ 
	& \hspace{2cm} + \overline \Psi_R(-q,\mathfrak{r}) \partial_\mathfrak{r} \Psi_L(q,\mathfrak{r}) + \partial_\mathfrak{r} \overline \Psi_L(-q,\mathfrak{r}) \Psi_R(q,\mathfrak{r}) \nonumber \\
	& \hspace{2cm} - e^{-A} \Big[ \overline \Psi_L(-q,\mathfrak{r}) (i \slashed{q}) \Psi_L(q,\mathfrak{r}) + \overline \Psi_R(-q,\mathfrak{r}) (i \slashed{q}) \Psi_R(q,\mathfrak{r}) \Big] \bigg\} \\
	=& \int \dd^4 q \, e^{4A} \bigg\{ \left( 4 \partial_\mathfrak{r} A - 2 H_\Psi \right) \overline \Psi_L(-q,\mathfrak{r}) f(q,\mathfrak{r}) \Psi_L(q,\mathfrak{r}) \nonumber \\
	& \hspace{2cm} + \overline \Psi_L(-q,\mathfrak{r}) f(q,\mathfrak{r}) \partial_\mathfrak{r} \Psi_L(q,\mathfrak{r}) + \partial_\mathfrak{r} \overline \Psi_L(-q,\mathfrak{r}) f(q,\mathfrak{r}) \Psi_L(q,\mathfrak{r}) \nonumber \\
	& \hspace{2cm} - e^{-A} \Big[ \overline \Psi_L(-q,\mathfrak{r}) (i \slashed{q}) \Psi_L(q,\mathfrak{r}) + \overline \Psi_L(-q,\mathfrak{r}) f(q,\mathfrak{r})(i \slashed{q}) f(q,\mathfrak{r}) \Psi_L(q,\mathfrak{r}) \Big] \bigg\} \,, \nonumber
\end{align}
where we used Eq.~\eqref{eq:PsiRBCs} to obtain an expression in terms of $\Psi_L$ and $\overline \Psi_L$. Furthermore,
\begin{align}
	\partial_\mathfrak{r} \mathcal S_{\rm UV}[\mathfrak{r},r_{\rm UV}] =& - \int \dd^4 q \, e^{4A} \bigg\{ 4 \partial_\mathfrak{r} A \overline \Psi_L(-q,\mathfrak{r}) f(q,\mathfrak{r}) \Psi_L(q,\mathfrak{r}) + \overline \Psi_L(-q,\mathfrak{r}) \partial_\mathfrak{r} f(q,\mathfrak{r}) \Psi_L(q,\mathfrak{r}) \nonumber \\
	& \hspace{1cm} + \partial_\mathfrak{r} \overline \Psi_L(-q,\mathfrak{r}) f(q,\mathfrak{r}) \Psi_L(q,\mathfrak{r}) + \overline \Psi_L(-q,\mathfrak{r}) f(q,\mathfrak{r}) \partial_\mathfrak{r} \Psi_L(q,\mathfrak{r}) \bigg\} \,,
\end{align}
so that Eq.~\eqref{eq:RGinvariance} becomes
\begin{align}
\label{eq:functionaleq}
	0 &= \partial_\mathfrak{r} \big( \mathcal S_\Psi[r_1,\mathfrak r;s_1,-1] + \mathcal S_{\rm UV}[\mathfrak{r},r_{\rm UV}] \big) \nonumber \\ &= - \int \dd^4 q \, e^{4A} \overline \Psi_L(-q,\mathfrak{r}) \bigg\{ \partial_\mathfrak{r} f(q,\mathfrak{r}) + 2 H_\Psi f(q,\mathfrak{r}) \nonumber \\ & \hspace{4.3cm} + e^{-A} i \slashed{q}  + f(q,\mathfrak{r}) \left(e^{-A} i \slashed{q} \right) f(q,\mathfrak{r}) \bigg\} \Psi_L(q,\mathfrak{r}) \,.
\end{align}
One now observes that Eq.~\eqref{eq:functionaleq} needs to hold for {\it any} classical solution. In particular, after imposing an IR boundary condition on $\Psi_L$, one can think of the space of such solutions as being parametrized by the value that $\Psi_L$ takes at the coordinate $r = \mathfrak{r}$, which implies that
\beq
	\partial_\mathfrak{r} f = - 2 H_\Psi f - e^{-A} i \slashed{q}  - f \left(e^{-A} i \slashed{q} \right) f \,.
\eeq
Using the definition of $f(q,\mathfrak{r})$ given in Eq.~\eqref{eq:fdefinition}, one finally obtains the RG flow equation for $\xi$:
\beq
	\partial_\mathfrak{r} \xi^2 = - N_L^{-2} e^{-5A} q^2 + N_L^2 e^{3A} \xi^4 \,.
\eeq

\end{document}